\documentclass[12pt,a4]{article}
\topmargin -10mm \textwidth 165mm \textheight 220mm \evensidemargin 0mm \oddsidemargin 0mm \parskip=\medskipamount \def\baselinestretch{1.4} \arraycolsep 2pt
\renewcommand{\baselinestretch}{1.25}
\usepackage{amsthm,amsmath,latexsym,amssymb,amsfonts,amscd}
\usepackage{graphics,lscape,fancyhdr,array,stmaryrd,euscript}
\usepackage{empheq}
\usepackage{sidecap}
\usepackage{ifpdf}
\usepackage{verbatim}
\usepackage{color}
\usepackage{relsize}
\usepackage{tikz-cd}
\numberwithin{equation}{section}
\usepackage{hyperref,setspace}

\usepackage[numbers,sort&compress]{natbib}
\usepackage[nottoc]{tocbibind}
\usepackage{skak}
\newcommand{\pl}{\partial}

\newcommand{\leading}{\text{lead}}
\newcommand{\subleading}{\text{sub}}
\newcommand{\be}{\begin{equation}}
\newcommand{\ee}{\end{equation}}

\newcommand{\D}{\text{D}}

\newcommand{\fdu}[2]{{}_{#1}{}^{#2}\,}

\newcommand{\besubeqs}{\begin{subequations}}
\newcommand{\esubeqs}{\end{subequations}}

\newcommand{\self}{\text{self}}
\newcommand{\tri}{\text{ver}}
\newcommand{\zb}{{\bar{z}}}

\newcommand{\pvec}{{\boldsymbol{p}}}
\newcommand{\qvec}{{\boldsymbol{q}}}
\newcommand{\kvec}{{\boldsymbol{k}}}
\newcommand{\pb}{{\bar{p}}}

\newcommand{\jb}{{\bar{j}}}

\newcommand{\tr}{{\text{Tr}}}

\newcommand{\pfrac}[1]{{\frac{\pl}{\pl #1}}}

\newcommand{\deltas}[1]{{\delta\left(#1\right)}}
\newcommand{\PP}{{\mathbb{P}}}

\newcommand{\PPb}{{\overline{\mathbb{P}}}}

\begin{document}
\pagenumbering{gobble}
\graphicspath{ {./pics/} }
%%%%%%%%%%%%%%%%%%%%%%%%%%%%%%%%%%%%%%%%%%%%%%%%%%%%%%%%%%
\begin{center}
{\Large\bfseries 
More on Quantum Chiral Higher Spin Gravity
\vspace{0.4cm}
} \\

\vskip 0.04\textheight

Evgeny Skvortsov,${}^{\symking,\dagger}$ Tung Tran,${}^{\symknight,\symking}$ Mirian Tsulaia ${}^{\symqueen}$

\vskip 0.04\textheight

{\em ${}^{\symknight}$ Arnold Sommerfeld Center for Theoretical Physics\\
Ludwig-Maximilians University Munich\\
Theresienstr. 37, D-80333 Munich, Germany}\\

\vspace{5pt}
{\em$^{\symking}$  Albert Einstein Institute, \\
Am M\"{u}hlenberg 1, D-14476, Potsdam-Golm, Germany}

\vspace{5pt}
{\em$^{\dagger}$ Lebedev Institute of Physics, \\
Leninsky ave. 53, 119991 Moscow, Russia}\\

\vspace{5pt}
{\em$^{\symqueen}$ Okinawa Institute of Science and Technology, \\ 1919-1 Tancha, Onna-son, Okinawa 904-0495, Japan}

\vskip 0.02\textheight

{\bf Abstract }

\end{center}
\begin{quotation}
Chiral Higher Spin Gravity is unique in being the smallest higher spin extension of gravity and in having a simple local action both in flat and (anti)-de Sitter spaces. It must be a closed subsector of any other higher spin theory in four dimensions, which makes it an important building block and benchmark. Using the flat space version for simplicity, we perform a thorough study of quantum corrections in Chiral Theory, which strengthen our earlier results \href{https://arxiv.org/abs/1805.00048}{arXiv:1805.00048 [hep-th]}. Even though the interactions are naively non-renormalizable, we show that there are no UV-divergences in two-, three- and four-point amplitudes at one loop thanks to the higher spin symmetry. We also give arguments that the AdS Chiral Theory should exhibit similar properties. It is shown that Chiral Theory admits Yang-Mills gaugings with $U(N)$, $SO(N)$ and $USp(N)$ groups, which is reminiscent of the Chan-Paton symmetry in string theory.
\end{quotation}

\pagestyle{empty}
\newpage
\setcounter{page}{1}
\pagestyle{plain}
\renewcommand{\baselinestretch}{0.75}\normalsize
\tableofcontents
\renewcommand{\baselinestretch}{1.25}\normalsize
\newpage

%%%%%%%%%%%%%%%%%%%%%%%%%%%%%%%%%%%%%%%%%%%%%%%%%%%%%%%%%%
\section{Introduction and Main Results}
%%%%%%%%%%%%%%%%%%%%%%%%%%%%%%%%%%%%%%%%%%%%%%%%%%%%%%%%%%
\pagenumbering{arabic}
\setcounter{page}{2}
We report on the recent progress in addressing the Quantum Gravity problem from the Higher Spin Gravity (HiSGRA) vantage point. The model we consider is Chiral Higher Spin Gravity that exists both in flat  \cite{Metsaev:1991mt,Metsaev:1991nb,Ponomarev:2016lrm} and anti-de Sitter space \cite{Metsaev:2018xip,Skvortsov:2018uru}. The results of this paper extend considerably the ones of  \cite{Skvortsov:2018jea} and confirm that Chiral Theory does not have UV-divergences even though the two-derivative graviton self-interaction as well as infinitely many vertices involving higher spin fields are naively non-renormalizable when taken one by one. For simplicity we perform the calculations in the Minkowski Chiral Theory where Weinberg and Coleman-Mandula theorems leave no room for nontrivial S-matrix for HiSGRA. Nevertheless, this is an important consistency check and we do not expect the structure of UV-divergences be affected by the cosmological constant. 

The general idea behind HiSGRA is to look for extensions of gravity with massless higher spin fields, $s>2$, that would make the graviton to be a part of a much larger multiplet of gauge fields. The multiplet is usually infinite and so is the gauge symmetry. It is expected that the infinite-dimensional higher spin symmetry imposes sufficiently strong constraints on interactions and, in particular, restricts counterterms.  This expectation is justified, for example, by the fact that higher spin symmetry completely fixes the holographic S-matrix, i.e. there are unique higher spin invariant holographic correlation functions \cite{Maldacena:2011jn,Boulanger:2013zza,Alba:2013yda,Alba:2015upa}. In fact, the correlation functions are directly given by invariants of a higher spin algebra \cite{Colombo:2012jx,Didenko:2013bj,Didenko:2012tv,Bonezzi:2017vha}. Other quantum tests of holographic higher spin theories include one-loop determinants  \cite{Gopakumar:2011qs,Tseytlin:2013jya,Giombi:2013fka,Giombi:2014yra,Beccaria:2014jxa,Beccaria:2014xda,Beccaria:2015vaa,Gunaydin:2016amv,Bae:2016rgm,Skvortsov:2017ldz} and one-loop corrections to the four-point function via AdS unitarity cuts \cite{Ponomarev:2019ltz,Ponomarev:2019ofr}. 

While the checks of the quantum consistency of HiSGRA alluded to above are encouraging, they are either indirect or do not sufficiently probe the structure of interactions. The only model with propagating massless higher spin fields where direct computations are possible at the moment is Chiral HiSGRA \cite{Ponomarev:2016lrm}, which is heavily based on the earlier works by Metsaev \cite{Metsaev:1991mt,Metsaev:1991nb}. Chiral Theory is the smallest extension of gravity with massless higher spin fields. It exists both in flat and (anti)-de Sitter spaces \cite{Metsaev:2018xip,Skvortsov:2018uru}, which makes it a unique model of this kind. Chiral Theory must be a closed subsector in any other higher spin theory in four dimensions with the same free spectrum, which makes it an important building block. The specific structure of interactions allows Chiral Theory to escape from all no-go-type results both in flat, see e.g. \cite{Weinberg:1964ew,Coleman:1967ad,Bekaert:2010hp,Roiban:2017iqg}, and (anti)-de Sitter spaces \cite{Bekaert:2015tva,Maldacena:2015iua,Sleight:2017pcz,Ponomarev:2017qab}.

It is interesting that the holographic S-matrix of the $AdS_4$ Chiral Theory is nontrivial \cite{Skvortsov:2018uru} and is related to Chern-Simons Matter Theories, which should be confronted with its triviality in flat space. It seems that the interactions fine-tuned by the higher spin symmetry result in a perfect annihilation of all terms in physical amplitudes when the space-time is flat  or very close to flat (for example, one should find the same result for high energy scattering in the interior of (anti)-de Sitter space). When the space-time is curved the higher derivative nature of the interactions becomes important and there is no perfect cancellation anymore, which results in a nontrivial holographic $S$-matrix. Probing the UV structure of interactions in flat space is important for the quantum consistency in (anti)-de Sitter space as well. Had we found any UV-divergence in the Minkowski Chiral Theory, its (A)dS version would have suffered from the same problem. Therefore, our preliminary conclusion is that the AdS Chiral Theory does not have UV-divergences. In addition, the quantum consistency of Chiral Theory is an important test of the more general $4d$ (holographic) higher spin theories which have to have it as a subsector.  

One of the crucial ideas behind Chiral HiSGRA \cite{Metsaev:1991mt,Metsaev:1991nb,Ponomarev:2016lrm,Skvortsov:2018jea,Skvortsov:2018uru} was to stick to the light-cone or light-front approach, which was applied to the higher spin problem for the first time in \cite{Bengtsson:1983pd,Bengtsson:1983pg}. It was already in 1983 that some evidence for existence of higher spin theories was obtained in \cite{Bengtsson:1983pd}: 'Our conclusion is that the higher-spin theories are likely to exist, at least as classical field theories, although they may not have a manifestly covariant form'. Due to Weinberg's and Coleman-Mandula theorems the $S$-matrix approach is not applicable in flat space. The light-cone approach is the most general approach to local dynamics, which can be used both in flat and (anti)-de Sitter spaces. It goes well with understanding gauge symmetry as redundancy of description. 

Technically, the idea of the light-cone approach is to construct the generators of the space-time symmetry algebra directly in terms of local physical degrees of freedom. In particular, Chiral Theory results from checking the same equations, which are a part of the Poincare algebra,
\begin{align}\tag{$\ast$}
[J^{a-},J^{c-}]&=0\,, &
[J^{a-},P^{-}]&=0\,,\label{stringeq}
\end{align}
as it is done in string theory in the light-cone gauge \cite{Goddard:1973qh}. One difference with string theory is that we first look for the classical realization of the algebra via Poisson brackets. Then, the Hamiltonian $H=P^-$ gives a classical action $S$ that we invoke to compute quantum corrections. Another difference is that we do not have any prior knowledge of how the theory looks like and what the spectrum of states is. One can put in at least one massless higher spin field with certain minimal self-interaction. The Lorentz algebra implies that one needs an infinite multiplet comprising massless fields of all spins with very specific interactions in order to fulfil \eqref{stringeq}, \cite{Metsaev:1991mt,Metsaev:1991nb,Ponomarev:2016lrm}. In particular, the graviton must belong to the multiplet. Chiral Theory is the most minimal solution of this problem in the sense of having the least possible number of fields, which is still infinite, and the least number of interactions. One remarkable property of Chiral Theory is that the interactions truncate at cubic terms. 

Another way to approach the higher spin problem is to start 
with string theory --- a natural candidate for
a consistent theory of quantum gravity. This theory contains an infinite number of massive higher spin fields,
and these fields are crucial for making the quantum theory finite. Therefore, in order to formulate a HiSGRA on flat space, one can try to find some form of a symmetric phase of string theory, by taking its high energy (low tension)
limit, for example. This limit \cite{Gross:1988ue}, being opposite to the low energy (supergravity) limit, is still not completely understood even in the simplest case of the bosonic string theory.\footnote{See, however, \cite{Eberhardt:2018ouy} for the tensionless limit of strings on $AdS_3$.} One possible approach, which eventually leads to nontrivial interactions, is as follows. As the first step one takes $\alpha^\prime \rightarrow \infty$ in the free equations \cite{Sagnotti:2003qa}
(see also \cite{Sorokin:2018djm} for a recent work in this direction and \cite{Bonelli:2003kh,Lindstrom:2003mg,Bakas:2004jq,Moeller:2005ez,Lee:2015wwa,Bagchi:2016yyf}  
for other works on the high energy limit of string theory),
thus obtaining a consistent gauge invariant formulation of massless 
fields. As the second
step one promotes the original linear gauge symmetries and field
equations to nonlinear ones \cite{Fotopoulos:2007nm, Sagnotti:2010at, Fotopoulos:2010ay} and this way one can reproduce nontrivial cubic interaction vertices obtained using other methods \cite{Metsaev:1993ap, Metsaev:2005ar,Metsaev:2007rn,Manvelyan:2010jr,Manvelyan:2010je}. However, the most  difficult problems when considering
interacting massless higher spin fields arise at the level of quartic interactions. These problems
manifests themselves either in a form of nonlocal terms in quartic vertices and four-point functions or by a failure of various consistency checks
for the symmetries of four-point scattering amplitudes
\cite{Sagnotti:2010at, Fotopoulos:2010ay,Bekaert:2010hp,Dempster:2012vw, Roiban:2017iqg}. To summarize, a consistent HiSGRA is still to be obtained this way.

Both string theory and Chiral HiSGRA require infinitely many higher spin fields (massive or massless) for consistency. Another stringy feature of Chiral HiSGRA is that one can extend it to a class of theories where all fields are charged with respect to spin-one fields in a way that is reminiscent of the Chan-Paton approach. The $SO(N)$-case was studied already in \cite{Metsaev:1991nb}, see also \cite{Konstein:1989ij} for an earlier important result within a different approach. Here we extend it to $U(N)$ and $USp(N)$. For $SO(N)$ and $USp(N)$ cases the representations that fields take values in depend on whether the spin is even or odd, which is again similar to string theory \cite{Marcus:1982fr} and \cite{Konstein:1989ij}. Our findings indicate that higher spin fields are essential for quantization of gravity and replacing massive fields with massless ones allows us to find nontrivial toy models that are much smaller and simpler than string theory, which should be helpful for understanding the quantum gravity problem.

The outline of the paper is as follows. In section \ref{sec:hisgra}
we begin by presenting the action of Chiral Theory. In section \ref{sec:Feynamnrules} we collect the Feynman rules, which are used in the subsequent sections to compute quantum corrections. In section \ref{sec:trees} we compute all tree-level amplitudes and show that, in accordance with Weinberg's theorem they, vanish on-shell, which is a result of a highly nontrivial cancellation between all Feynman diagrams due to coupling conspiracy. In section \ref{sec:bubbles} we compute the vacuum diagrams. We shown that the vacuum loop diagrams vanish identically either due to the coupling conspiracy or due to the fact that the regularized number of effective degrees of freedom vanishes. In section \ref{sec:leggedloop} we compute
the loop diagrams with external legs and demonstrate that they do not have UV-divergences and are also proportional to the total number of effective degrees of freedom, hence, can be made to vanish. We conclude with section \ref{sec:conclusions} that contains 
a summary of our results and discussion of possible future developments. A crash course on the light-cone approach as well as some useful technical details are collected in the Appendices. In particular, in Appendix \ref{app:gauging} we study in detail the Chan-Paton gauging of the theory. In particular, we show that the closure of the Poincare algebra in the light-cone gauge
admits three types of gauge groups: $U(N)$, $SO(N)$ and
$USp(N)$.

%%%%%%%%%%%%%%%%%%%%%%%%%%%%%%%%%%%%%%%%%%%%%%%%%%%%%%%%%%
\section{Classical Chiral Higher Spin Gravity}
\label{sec:hisgra}
%%%%%%%%%%%%%%%%%%%%%%%%%%%%%%%%%%%%%%%%%%%%%%%%%%%%%%%%%%
We begin directly with the action of Chiral Theory. The action follows from the Hamiltonian $H=P^-$ that together with the other generators obey the Poincare algebra. A short summary of the light-front approach can be found in Appendix \ref{app:LC}. 

One important feature of the four-dimensional world is that a massless spin-$s$ field has two degrees of freedom and effectively it looks like two scalar fields representing helicity $\pm s$ states. Usually, in the covariant formulation a massless spin-$s$ particle is described by a rank-$s$ tensor $\Phi_{a_1...a_s}(x)$. Upon imposing the light-cone gauge and integrating out auxiliary fields one is left with two helicity eigen states $\Phi^{\pm s}(x)$. We would like to study possible interactions between such states. It is convenient to work with the Fourier transformed fields 
\begin{align}
    \Phi^{\lambda}_{\pvec}\equiv\Phi^\lambda(\pvec)&: && \lambda= \pm s\,.
\end{align}
Throughout the paper we shall work in momentum space and four-momentum $\pvec$ is split as\footnote{Since $p^+$ is present in many expressions the shorthand notation $\beta$ for $p^+$ appears to be  very handy.} $\pvec=(\beta\equiv p^+,p^-,p,\bar{p})$. The action of Chiral Theory reads
\begin{align}
\begin{aligned}\label{eq:chiralaction}
S=-\sum_{\lambda\geq0}\int (\pvec^2)\mathrm{Tr}[\Phi^{\lambda}(\pvec)^\dag \Phi^\lambda(\pvec)] +\sum_{\lambda_{1,2,3}}\int C_{\lambda_1,\lambda_2,\lambda_3} V(\pvec_1,\lambda_1;\pvec_2,\lambda_2;\pvec_3,\lambda_3)\,.
\end{aligned}
\end{align}
Let us now discuss all the ingredients of this action. It consists of the canonical kinetic term, where we sum over all spins, and specific cubic interactions. The fields are assumed to take values in some matrix algebra, to be specified below, and hence we use the trace $\mathrm{Tr}$ to form a singlet. As is well-known, given any three helicities there is a unique cubic vertex or cubic amplitude.\footnote{One important exception is when $\lambda_1+\lambda_2+\lambda_3=0$. In this case the only allowed vertex is the scalar cubic self-interaction, $\lambda_1=\lambda_2=\lambda_3=0$. } In the light-cone gauge such a vertex has the form \cite{Metsaev:1991mt,Metsaev:1991nb}
\small
\begin{equation}\label{eq:generalvertex}
    V(\pvec_1,\lambda_1;\pvec_2,\lambda_2;\pvec_3,\lambda_3)=\frac{\PPb^{\lambda_1+\lambda_2+\lambda_3}}{\beta_1^{\lambda_1}\beta_2^{\lambda_2}\beta_3^{\lambda_3}}\tr[\Phi^{\lambda_1}_{\pvec_1}\Phi^{\lambda_2}_{\pvec_2}\Phi^{\lambda_3}_{\pvec_3}]\delta^4(\pvec_1+\pvec_2+\pvec_3)\,,
\end{equation}
\normalsize
where $\lambda_1+\lambda_2+\lambda_3\geq0$ and
\begin{align} \label{PP1}
    \PPb=\frac13\left[ (\beta_1-\beta_2)\bar{p}_3+(\beta_2-\beta_3)\bar{p}_1+(\beta_3-\beta_1)\bar{p}_2\right]\,.
\end{align}
The complex conjugate of the above gives the vertices for $\lambda_1+\lambda_2+\lambda_3\leq0$. Note that the only admissible vertex with $\lambda_1+\lambda_2+\lambda_3=0$ is the scalar self-interaction. It is straightforward to establish a dictionary between the light-cone approach and the spinor helicity formalism. To this end \cite{Chakrabarti:2005ny,Chakrabarti:2006mb,Ananth:2012un,Akshay:2014qea,Bengtsson:2016jfk,Ponomarev:2016cwi}, let us introduce two-component spinors   
\begin{equation}
\label{9a11}
|i] = \frac{2^{1/4}}{\sqrt{\beta_i}}\left(\begin{array}{c}
 \bar q_i \\
  -\beta_i
\end{array}\right)=2^{1/4}\left(\begin{array}{c}
 \bar q_i \beta_i^{-1/2} \\
 - \beta^{1/2}_i
\end{array}\right)\,.
\end{equation}
The contractions can be expressed as
\begin{equation}
\label{9a1}
[i  j] = \sqrt{\frac{2}{\beta_i\beta_j}}\PPb_{ij}\,, \qquad\quad \langle ij \rangle = \sqrt{\frac{2}{\beta_i\beta_j}}\PP_{ij}\,,
\end{equation}
where $\PPb_{km}=\pb_k\beta_m-\pb_m\beta_k$ and similarly for $|i\rangle$. Then the kinematical factor in the cubic vertex (\ref{eq:generalvertex}) has the standard form \cite{Benincasa:2007xk,Benincasa:2011pg}
\begin{align}\
\frac{\PPb^{\lambda_1+\lambda_2+\lambda_3}}{\beta_1^{\lambda_1}\beta_2^{\lambda_2}\beta_3^{\lambda_3}} \sim 
    [12]^{\lambda_1+\lambda_2-\lambda_3}[23]^{\lambda_2+\lambda_3-\lambda_1}[13]^{\lambda_1+\lambda_3-\lambda_2}\,,
\end{align}
where the momentum conservation has to be used to replace $\PPb$ with any of $\PPb_{12}$, $\PPb_{23}$, $\PPb_{31}$.  The light-cone approach provides an off-shell extension \cite{Brandhuber:2007vm,Ponomarev:2016cwi,Ponomarev:2017nrr}
of the on-shell three-point amplitudes. Therefore, the cubic vertices are the canonical ones, but written in the light-cone gauge. 

The ingredients above are kinematical. The dynamical input is in the coupling constants $C_{\lambda_1,\lambda_2,\lambda_3}$. For example, the action of Yang-Mills Theory up to the cubic terms would require $C^{+1,+1,-1}=C^{-1,-1,+1}=ig_{YM}$ and $C_{\lambda_1,\lambda_2,\lambda_3}=0$ for all other combinations. Similarly, the Einstein-Hilbert action up to the cubic terms is reproduced by $C^{+2,+2,-2}=C^{-2,-2,+2}=l_{p}$, where $l_p$ is the Planck length, and $C_{\lambda_1,\lambda_2,\lambda_3}=0$ for all other triplets. Chiral Theory requires 
\begin{equation}\label{eq:magicalcoupling}
    C_{\lambda_1,\lambda_2,\lambda_3}=\frac{\kappa\,(l_p)^{\lambda_1+\lambda_2+\lambda_3-1}}{\Gamma(\lambda_1+\lambda_2+\lambda_3)}
\end{equation}
that is a unique solution of the Poincare algebra relations provided at least one higher spin field is present together with a nontrivial self-interaction. The explicit expressions for the generators of the Poincare algebra can be found in \cite{Metsaev:1991mt,Metsaev:1991nb,Ponomarev:2016lrm,Ponomarev:2016jqk,Ponomarev:2017nrr} and Appendix \ref{app:LC}. 

The constant $l_p$ can be associated with the Planck length since the chiral half of the Einstein-Hilbert two-derivative cubic vertex belongs to the action, $C^{+2,+2,-2}=\kappa\, l_p$. The chiral half of the Goroff-Sagnotti \cite{Goroff:1985th} counterterm 
\begin{equation}\label{Div}
\int \sqrt{g}\, R_{\mu\nu\rho\sigma}
R^{\rho\sigma\lambda\tau} R_{\lambda\tau}^{~~~\mu\nu}\,,
\end{equation}
corresponds to $C^{+2,+2,+2}=\kappa\, (l_p)^5/5!$. Note that the number of derivatives in the covariant description corresponds to the total power of $\PPb$ in the light-cone gauge. In general we see infinitely many higher derivative interactions present in the action. Naively, it is not power-counting renormalizable. Nevertheless, we will show that there are no UV-divergences. 

The action does stop at the cubic order and no higher order corrections are required to make it consistent. Formally, there is one more dimensionless coupling $\kappa$ that does not play any role in the present paper, but is important for making contact between Chern-Simons Matter theories and $AdS_4$ Chiral Theory \cite{Skvortsov:2018uru}. Specific form \eqref{eq:magicalcoupling} of the coupling constants  discriminates between helicities: if the sum of helicities entering the vertex is zero or negative, the coupling vanishes, while all positive sums are allowed. Therefore, the theory is chiral and violates parity. It is close in spirit to self-dual Yang-Mills theory, which in the light-cone gauge also looks like half of the Yang-Mills' cubic action with higher order terms erased \cite{Chalmers:1996rq}.

The last optional ingredient is that fields $\Phi^{\lambda}_{\pvec}$ can be extended to carry color degrees of freedom to which we shall refer as Chan-Paton factors, the terminology borrowed from the string theory. In practice, this means that each $\Phi^{\lambda}$ takes values in the algebra of matrices:
\begin{align}
    \Phi^{\lambda}(\pvec)\equiv \Phi_a^{\lambda}(\pvec)T^{a, \lambda} \equiv (\Phi^{\lambda}_{\pvec})^A_{\ B}\,.
\end{align}
The reason why we call them Chan-Paton factors, see also \cite{Konstein:1989ij} for the first occurrence in the higher spin context, (with technical details left to Appendix \ref{app:gauging}) is that, similarly to what happens in open string theory \cite{Marcus:1982fr}, only three options for gauge groups are allowed: (i) $U(N)$ gauging: fields are (anti)-Hermitian matrices; (ii) $SO(N)$ gauging, studied in \cite{Metsaev:1991nb}: even spins are symmetric matrices, while odd spins are anti-symmetric  matrices; (iii) $USp(N)$ gauging, where the symmetry is the opposite as compared to the $SO(N)$ case. The most minimal Chiral Theories can be obtained as particular cases: $U(1)$-gauging leads to a theory with all integer spins in the spectrum, each in one copy. $SO(1)$-gauging leads to even spins only, each in one copy. In what follows we work with the $U(N)$-case by default. 

%%%%%%%%%%%%%%%%%%%%%%%%%%%%%%%%%%%%%%%%%%%%%%%%%%%%%%%%%%
\section{Feynman rules}
\label{sec:Feynamnrules}
%%%%%%%%%%%%%%%%%%%%%%%%%%%%%%%%%%%%%%%%%%%%%%%%%%%%%%%%%%
Using the results of the previous section and of Appendix \ref{app:gauging}, we can write down the Feynman rules for Chiral Theories with Chan-Paton factors. The propagator is found to be
\begin{align}
   \parbox{3.cm}{\includegraphics[scale=0.22]{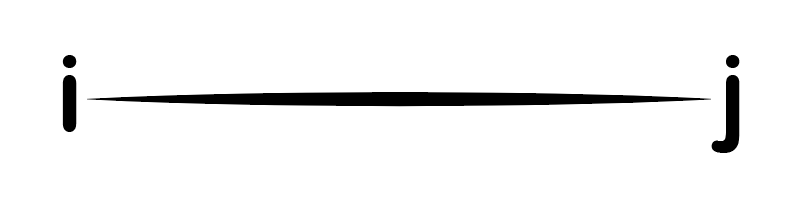}}= \parbox{3.6cm}{\includegraphics[scale=0.15]{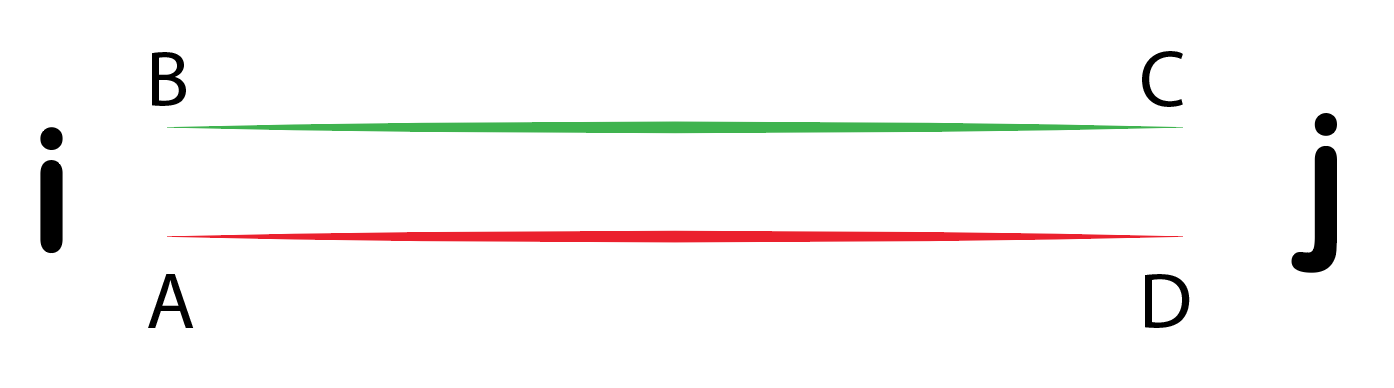}} = \frac{\delta^{\lambda_i+\lambda_j,0}\delta^4(\pvec_i+\pvec_j)}{\pvec_i^2}\,\Xi_{\text{gauge}}\,,
\end{align}
where $\Xi_{\text{gauge}}$ is the part that comes from the double line notation. For $U(N)$ gauging, which is the easiest case, we find that\footnote{Note that the somewhat strange sign factor is due to the fact that odd spins correspond to anti-Hermitian matrices, while even spins to Hermitian ones. Therefore, the kinetic term, which has $\tr[\Phi^\dag \Phi]$, is always Hermitian and positive definite.}
    \begin{equation}
        \Xi_{U(N)}=(-)^{\lambda_i}\delta^C_{\ B}\delta^A_{\ D}\,.
    \end{equation}
And, for $SO(N)/USp(N)$ gauging, one finds
    \begin{align}
        \Xi_{SO(N)}&=\frac{\delta_{AC}\delta_{BD}+(-)^{\lambda_i}\delta_{BC}\delta_{AD}}{2}\,,\\
        \Xi_{USp(N)}&=\frac{C_{AC}C_{BD}+(-)^{\lambda_i+1}C_{BC}C_{AD}}{2}\,.
    \end{align}
Computations for $SO(N)/USp(N)$-valued fields are a bit more subtle compared to the $U(N)$ case. Lastly, the vertex for all cases can be presented in the 't Hooft double line notation as
\begin{align}
   \parbox{3.cm}{\includegraphics[scale=0.25]{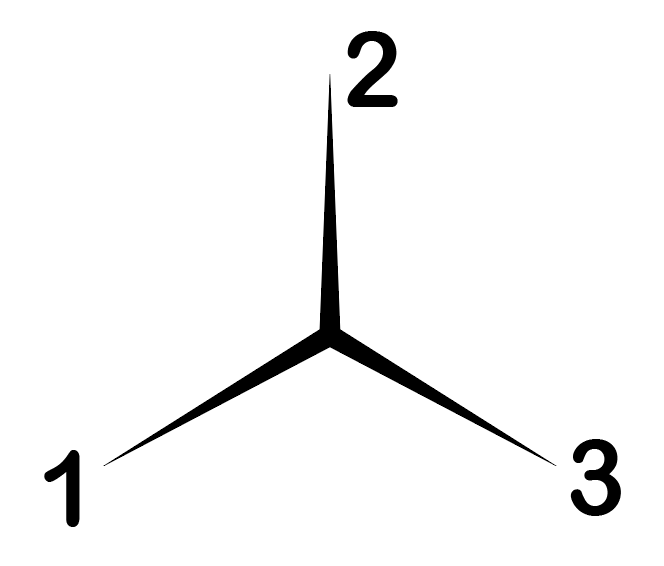}}= \parbox{3.2cm}{\includegraphics[scale=0.26]{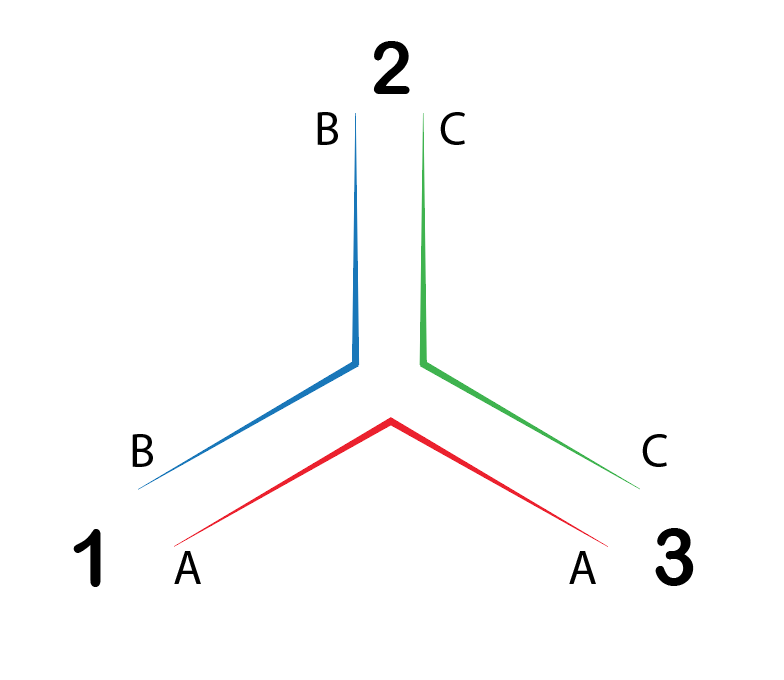}} = \delta^4(\pvec_1+\pvec_2+\pvec_3)\tr[\Phi^{\lambda_1}_{\pvec_1}\Phi^{\lambda_2}_{\pvec_2}\Phi^{\lambda_3}_{\pvec_3}]\frac{\PPb^{\lambda_1+\lambda_2+\lambda_3}}{\beta_1^{\lambda_1}\beta_2^{\lambda_2}\beta_3^{\lambda_3}}\,,
\end{align}
where the $\tr$ is the trace over implicit $U(N)$, $SO(N)/USp(N)$ indices.

%%%%%%%%%%%%%%%%%%%%%%%%%%%%%%%%%%%%%%%%%%%%%%%%%%%%%%%%%%
\section{Tree Amplitudes}
\label{sec:trees}
%%%%%%%%%%%%%%%%%%%%%%%%%%%%%%%%%%%%%%%%%%%%%%%%%%%%%%%%%%
In this section we compute all tree level amplitudes in Chiral Theory. We will show that all of them vanish on-shell, which is a result of a highly nontrivial cancellation among all Feynman diagrams. The triviality of the S-matrix, $S=1$, follows from the Weinberg low energy theorem.  

The proof proceeds by induction. First, we explicitly compute $4$-, $5$,- and $6$-point amplitudes with one off-shell leg. These amplitudes turn out to have a very compact form which suggests a general result for the $n$-point amplitude. Following Berends-Giele method \cite{Berends:1987me}, the $n$-point amplitude can be obtained by taking one cubic vertex and attaching two of its legs to various $(n-k)$- and $k$-point amplitudes for all possible $k$. This trick allows us to avoid explicit summation over all Feynman graphs. In order to carry out this procedure it is necessary to know all lower order amplitudes with one off-shell leg. The result of the recursion gives us a $(n+1)$-point amplitude with one leg being again off-shell.

Finally, we find that all amplitudes are proportional to $\pvec^2$ of the off-shell leg and therefore vanish on-shell. To simplify the calculations even further we work with the Chiral Theory extended by $U(N)$ Chan-Paton factors since one has to compute color-ordered sub-amplitudes only.

%%%%%%%%%%%%%%%%%%%%%%%%%%%%%%%%%%%%%%%%%%%%%%%%%%%%%%%%%%
\subsection{Four-point Amplitude}
\label{sec:Fourpoint}
%%%%%%%%%%%%%%%%%%%%%%%%%%%%%%%%%%%%%%%%%%%%%%%%%%%%%%%%%%
On-shell three-point amplitudes for massless spinning fields vanish due to kinematical reasons, see e.g. \cite{Benincasa:2007xk}. The scalar cubic self-coupling is absent due to the higher spin symmetry. Therefore, the simplest amplitude that may not be zero is the four-point one. Below we demonstrate the calculations for the case of the $U(N)$ Chan-Paton symmetry. The cases of $SO(N)$ and $USp(N)$ gauge groups can be treated in a similar way. An $n$-point amplitude can be represented as
\begin{align}
    A_n(\pvec_1,\lambda_1;...;\pvec_n,\lambda_n)&= \sum_{S_n/Z_n} \mathrm{Tr}[T_{\sigma(1)}...T_{\sigma(n)}] \hat{A}_n(\pvec_{\sigma_1},\lambda_{\sigma_1};...;\pvec_{\sigma_n},\lambda_{\sigma_n})\,,
\end{align}
which is a sum over $(n-1)!$ permutations and  $\sigma_1,...,\sigma_n$ denotes various permutations of $1,...,n$. The elementary blocks, sub-amplitudes $\hat{A}_n$, should be computed using the color-ordered Feynman rules. In the case of four-point function the sub-amplitude consists of two graphs: 
\begin{align*}
   \parbox{2.2cm}{\includegraphics[scale=0.3]{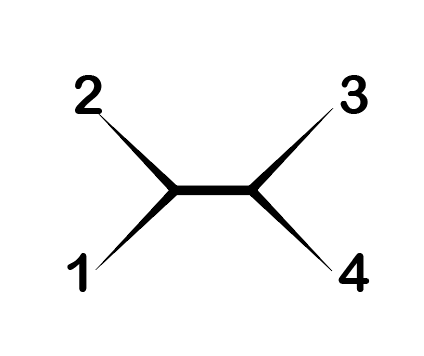}}+\parbox{2.2cm}{\includegraphics[scale=0.3]{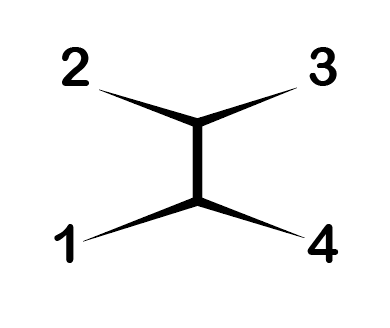}}\equiv \parbox{2.8cm}{\includegraphics[scale=0.24]{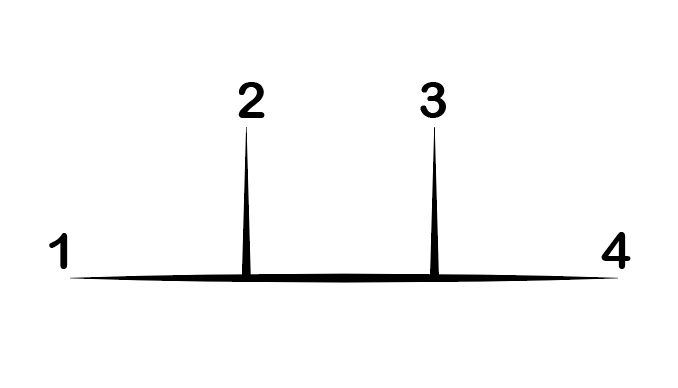}}+\parbox{2.8cm}{\includegraphics[scale=0.24]{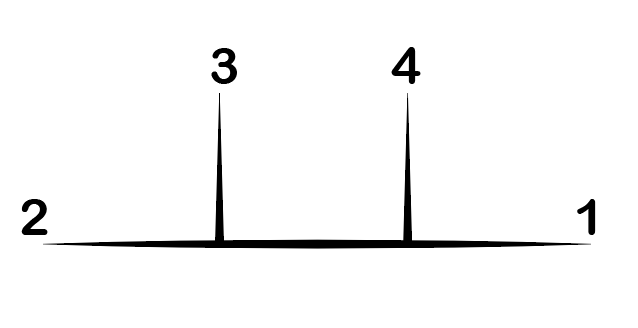}}
\end{align*}
The sum of these diagrams gives, see also \cite{Ponomarev:2016lrm,Skvortsov:2018jea}, 
\begin{equation}
    A_4(1234)=\frac{\delta(\sum_i\pvec_i)}{\Gamma(\Lambda_4-1)\prod_{i=1}^4\beta_i^{\lambda_i}}\Big[\frac{\PPb_{12}\PPb_{34}(\PPb_{12}+\PPb_{34})^{\Lambda_4-2}}{(\pvec_1+\pvec_2)^2}+\frac{\PPb_{23}\PPb_{41}(\PPb_{23}+\PPb_{41})^{\Lambda_4-2}}{(\pvec_2+\pvec_3)^2}\Big]
\end{equation}
where $\Lambda_4=\lambda_1+...+\lambda_4$. In what follows we drop an overall momentum-conserving $\delta$-function. 

It is important to notice that the sum over intermediate helicities is bounded both from above and from below due to the specific form of the  coupling constants (\ref{eq:magicalcoupling}). This is no longer so if we add up the chiral and anti-chiral vertices together with the idea to look for the more general higher spin theory. 

Next we use various kinematic identities from (\ref{eq:Bianchilike}) to (\ref{eq:magicidentity}) for $\PPb$ that are collected in the Appendix \ref{app:kinematics}.  Let us assume that the first momenta is off-shell, $\pvec_1^2\neq0$. Then,
\begin{equation}
    A_4(1234)=\frac{\alpha_4^{\Lambda_4-2}}{\Gamma(\Lambda_4-1)\prod_{i=1}^4\beta_i^{\lambda_i-1}}\frac{\beta_3\, \pvec_1^2}{4\beta_1\PP_{23}\PP_{34}}\,,
\end{equation}
where $\alpha_4=\PPb_{12}+\PPb_{34}=\PPb_{23}+\PPb_{41}$ is cyclic invariant. It is obvious that the total amplitude vanishes when all momenta are on-shell.

%%%%%%%%%%%%%%%%%%%%%%%%%%%%%%%%%%%%%%%%%%%%%%%%%%%%%%%%%%
\subsection{Five-point Amplitude}
\label{sec:Fivepoint}
%%%%%%%%%%%%%%%%%%%%%%%%%%%%%%%%%%%%%%%%%%%%%%%%%%%%%%%%%%
In the case of five-point amplitude we have five diagrams, which are cyclic permutations of a single comb diagram: 
\begin{align}
 \includegraphics[scale=0.35]{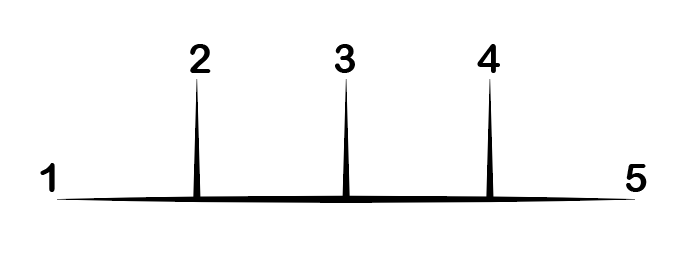}
\end{align}
However, according to our general discussion we can equivalently
represent the five-point amplitude as a sum of three
diagrams 
\begin{align*}
   \parbox{2.8cm}{\includegraphics[scale=0.3]{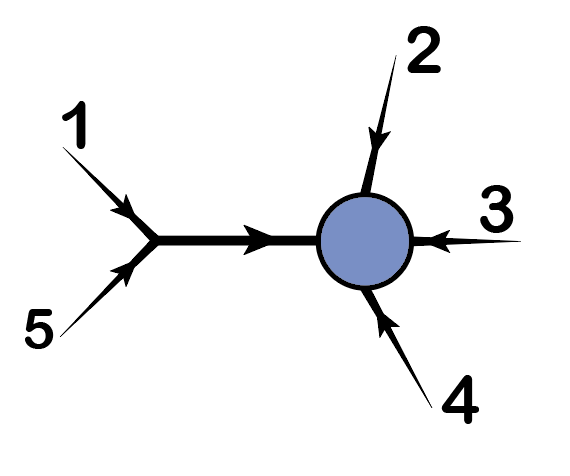}}\,\, + \parbox{3.8cm}{\includegraphics[scale=0.3]{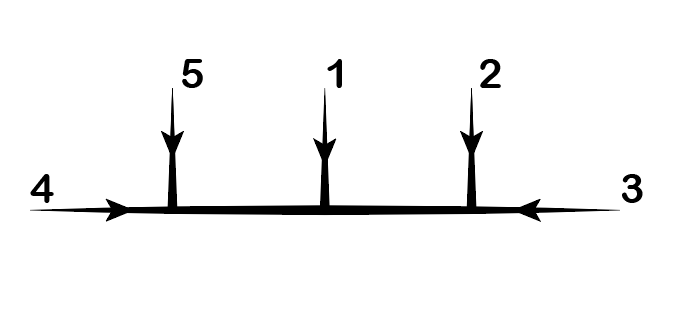}} \,\, + \parbox{3.8cm}{\includegraphics[scale=0.3]{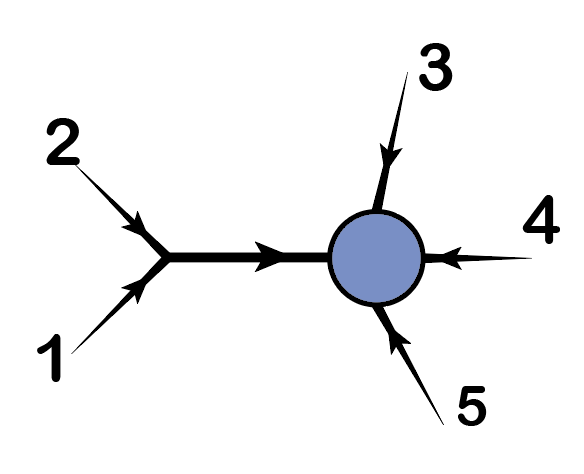}}
\end{align*}
Let us keep again the four-momentum of the first particle off-shell.
Using the results of the previous subsection for the four-point amplitude as well as the form of the cubic vertex we have for the 
first diagram
\begin{eqnarray}\label{eq:5ptsubexample-1-2}
{A}^{I}_5(12345)&=&\frac{1}{\Gamma(\Lambda_5 -2)\prod_{i=1}^5\beta_i^{\lambda_i}}\frac{ \PPb_{51} \PPb_{34}
\PPb_{23}
(\PPb_{51}+\PPb_{23}+\PPb_{24}+\PPb_{34})^{\Lambda_5-3}}{s_{23}\,s_{34}}
\\ \nonumber
&=&\frac{1}{4\Gamma(\Lambda_5 -2)\prod_{i=1}^5\beta_i^{\lambda_i}}\frac{ \PPb_{51} 
(\PPb_{51}+\PPb_{23}+\PPb_{24}+\PPb_{34})^{\Lambda_5-3}
\beta_2 \beta_3^2 \beta_4}{\PP_{23}
\PP_{34}}\,,
\end{eqnarray}
where $\Lambda_5=\lambda_1+...+\lambda_5$ and $s_{ij}=(\pvec_i+\pvec_j)^2$.
We also have used (\ref{C3}) to obtain the second line
in (\ref{eq:5ptsubexample-1-2}). With the help of the cubic vertex one obtains for the second diagram
\begin{eqnarray}\label{eq:5ptsubexample}
{A}_5^{II}(12345)&=&\frac{1}{\Gamma (\Lambda_5 -2)\prod_{i=1}^5\beta_i^{\lambda_i}}\frac{ \PPb_{45}\PPb_{23}(\PPb_{41}+\PPb_{51})(\PPb_{45}+\PPb_{23}+
\PPb_{41}+\PPb_{51})^{\Lambda_5-3}}{s_{23}\,s_{45}}  \\ \nonumber
&=&\frac{1}{4\Gamma (\Lambda_5 -2)\prod_{i=1}^5\beta_i^{\lambda_i}}\frac{ (\PPb_{41}+\PPb_{51})(\PPb_{45}+\PPb_{23}+
\PPb_{41}+\PPb_{51})^{\Lambda_5-3}\beta_2 \beta_3 \beta_4 \beta_5}{\PP_{23}\,\PP_{45}}\,.
\end{eqnarray}
Finally the third diagram can be obtained from the first one through the cyclic permutation of the indices
\begin{eqnarray}\label{eq:5ptsubexample-1-1}
{A}^{III}_5(12345)&=&\frac{1}{\Gamma(\Lambda_5 -2)\prod_{i=1}^5\beta_i^{\lambda_i}}\frac{ \PPb_{12} \PPb_{45}
\PPb_{34}
(\PPb_{12}+\PPb_{34}+\PPb_{35}+\PPb_{45})^{\Lambda_5-3}}{s_{34}\,s_{45}} \\ \nonumber
&=&
\frac{1}{4\Gamma(\Lambda_5 -2)\prod_{i=1}^5\beta_i^{\lambda_i}}\frac{ \PPb_{12} 
(\PPb_{12}+\PPb_{34}+\PPb_{35}+\PPb_{45})^{\Lambda_5-3}
\beta_3 \beta_4^2 \beta_5}{\PP_{34}
\PP_{45}}\,.
\end{eqnarray}
Let us notice that factors that are raised to power
$(\Lambda_5-3)$ are all equal to
$\alpha_5=\PPb_{12}+\PPb_{13}+\PPb_{23}+\PPb_{45}$.
Adding the contributions from three sub-amplitudes we get
\begin{equation} \label{5Am-1}
   {A}_5(12345) = {\mathcal{C}_5}
   (\PPb_{51} \PP_{45}\beta_2 \beta_3 +
  (\PPb_{41}+ \PPb_{51}) \PP_{34}\beta_2 \beta_5+
  \PPb_{12} \PP_{23}\beta_4 \beta_5)\,,
\end{equation}
where
\begin{equation}
    \mathcal{C}_5=\frac{\alpha_5^{\Lambda_5-3}}
    {4\Gamma(\Lambda_5-2)
    \PP_{23}\PP_{34}\PP_{45} \beta_1 \beta_2 \beta_5 
    \prod_{i=1}^5\beta_i^{\lambda_i-1}}\,.
\end{equation}
Now we shall perform a step which will be used for all higher 
point tree level amplitudes. Namely we shall transform
the last term in  (\ref{5Am-1}) using equation (\ref{C1})
to get
\begin{eqnarray} \label{5Am-1-2}
   {A}_5(12345)& =& {\mathcal{C}_5} (-\frac{\pvec_1^2}{2}
  \beta_2 \beta _3 \beta_4 \beta_5 - 
   \PPb_{14} \PP_{43}\beta_2 \beta_5 
   -\PPb_{15} \PP_{53}\beta_2 \beta_4
    \\ \nonumber
   &+&\PPb_{51} \PP_{45}\beta_2 \beta_3 +
  (\PPb_{41}+ \PPb_{51}) \PP_{34}\beta_2 \beta_5)\,.
\end{eqnarray}
Now, collecting the terms proportional to 
$\PPb_{51}$ and $\PPb_{41}$ we see that they vanish by virtue of the Bianchi-like identities (\ref{eq:Bianchilike}). Therefore we are left  only
with the first term in (\ref{5Am-1-2}), which is proportional to
$\pvec_1^2$. Therefore, we finally get the five-point amplitude with one off-shell leg
\begin{equation}\label{eq:5ptcyclic}
    A_5(12345)
    =-\frac{\alpha_5^{\Lambda-3}}{8\Gamma (\Lambda_5-2)\prod_{i=1}^5 \beta_i^{\lambda_i-1}} \frac{\beta_3\beta_4\,\pvec_1^2}{\beta_1\PP_{23}\PP_{34}\PP_{45}}\,.
\end{equation}
Again, it vanishes on-shell.

%%%%%%%%%%%%%%%%%%%%%%%%%%%%%%%%%%%%%%%%%%%%%%%%%%%%%%%%%%
\subsection{Six-Point Amplitude}
\label{sec:sixpoint}
%%%%%%%%%%%%%%%%%%%%%%%%%%%%%%%%%%%%%%%%%%%%%%%%%%%%%%%%%%
In order to prepare for computations of general $n$-point
tree-level amplitudes and demonstrate the pattern 
let us consider explicitly the six-point contributions. 
Again, we keep the four-momentum of the first particle off-shell.
The total amplitude is a sum of four sub-amplitudes:
\begin{align*}
   \parbox{3.2cm}{\includegraphics[scale=0.2]{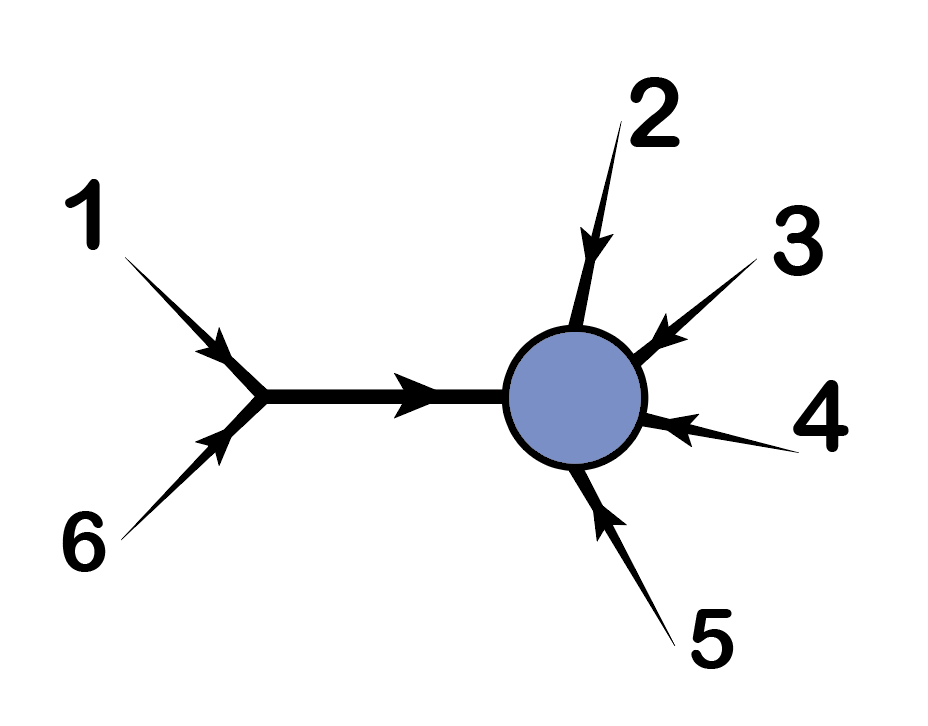}} \,\,\,\,
   +
   \parbox{3.2cm}{\includegraphics[scale=0.2]{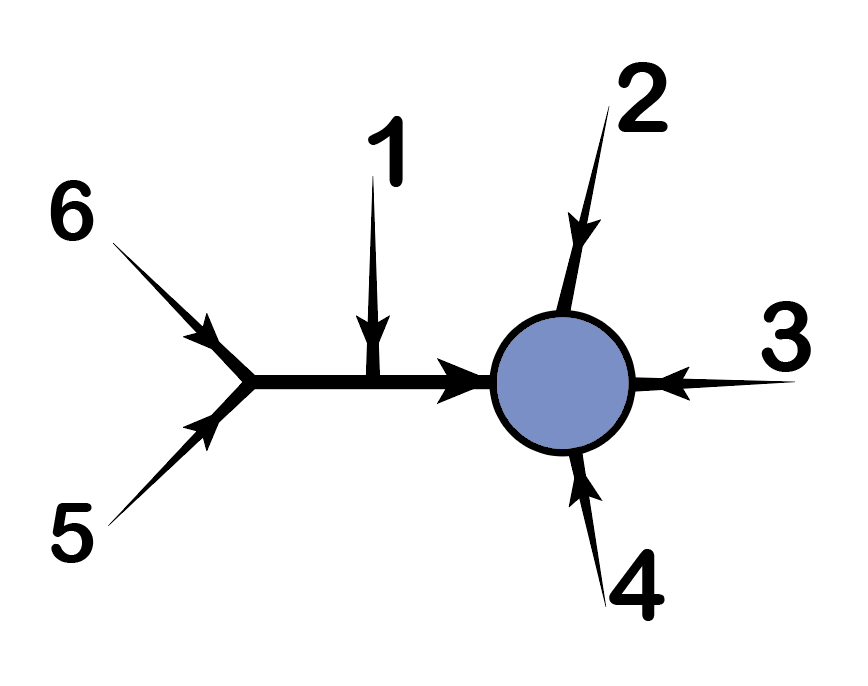}}+ \parbox{2.8cm}{\includegraphics[scale=0.2]{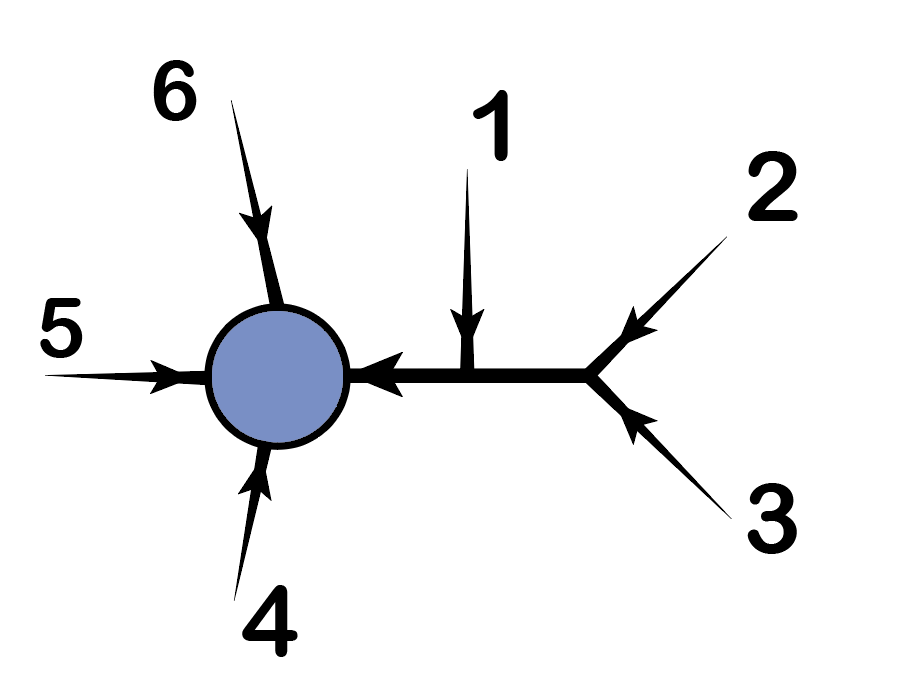}}+
   \parbox{2.8cm}{\includegraphics[scale=0.2]{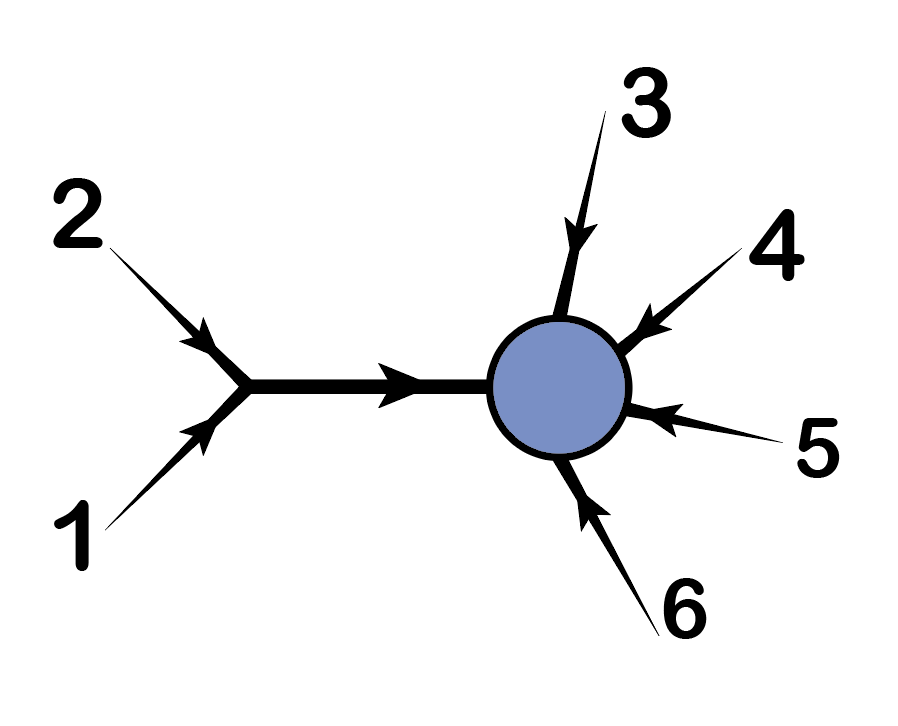}}
\end{align*}
Using the results of previous subsection
for four- and five-point amplitudes
as well as the explicit form of the cubic vertex we obtain
{\allowdisplaybreaks
\besubeqs
\begin{align}
{A}^{I}(123456)&=-\frac{(\PPb_{23} +\PPb_{24}+ \PPb_{34}+ \PPb_{45} + \PPb_{35}+ \PPb_{25}+ \PPb_{61})^{\Lambda_6-4}}{8\Gamma(\Lambda_6-3)\prod_{i=1}^6\beta_i^{\lambda_i-1}}
\frac{\PPb_{61}}{\PP_{23} \PP_{34} \PP_{45}} \frac{\beta_3 \beta_4}{\beta_6 \beta_1}
\,,\\
{A}^{II}(123456)&=-\frac{(\PPb_{56} +\PPb_{51}+ \PPb_{61}+ \PPb_{23} + \PPb_{24}+ \PPb_{34})^{\Lambda_6-4}}{8\Gamma(\Lambda_6-3)\prod_{i=1}^6\beta_i^{\lambda_i-1}}
\frac{(\PPb_{51}+ \PPb_{61})}{\PP_{23} \PP_{34} \PP_{56}} \frac{\beta_3 }{ \beta_1}
\,,\\
{A}^{III}(123456)&=-\frac{(\PPb_{12} +\PPb_{13}+ \PPb_{23}+ \PPb_{45} + \PPb_{46}+ \PPb_{56})^{\Lambda_6-4}}{8\Gamma(\Lambda_6-3)\prod_{i=1}^6\beta_i^{\lambda_i-1}}
\frac{(\PPb_{12}+ \PPb_{13})}{\PP_{23} \PP_{45} \PP_{56}} \frac{\beta_5 }{ \beta_1}
\,,\\
{A}^{IV}(123456)&=-\frac{(\PPb_{34} +\PPb_{35}+ \PPb_{45}+ \PPb_{56} + \PPb_{46}+ \PPb_{36}+ \PPb_{12})^{\Lambda_6-4}}{8\Gamma(\Lambda_6-3)\prod_{i=1}^6\beta_i^{\lambda_i-1}}
\frac{\PPb_{12}}{\PP_{34} \PP_{45} \PP_{56}} \frac{\beta_4 \beta_5}{\beta_1 \beta_2}\,,
\end{align}\esubeqs}%
\noindent where $\Lambda_6 = \lambda_1+ \cdots + \lambda_6$.
As in the previous cases, the terms with power $\Lambda_6-4$ all have the same base 
\begin{equation}
    \alpha_6 = \PPb_{12}+ \PPb_{13}+ \PPb_{14}+ \PPb_{23}+ \PPb_{24} +
    \PPb_{34} + \PPb_{56}\,.
\end{equation}
Next, let us add the expressions for the sub-amplitudes together. We get
\begin{eqnarray} \notag
{A}(123456)&=& \mathcal{C}_6 (\PPb_{61}\PP_{56} \beta_2 \beta_3 \beta_4
+ (\PPb_{61}+ \PPb_{51})\PP_{45} \beta_2 \beta_3 \beta_6
+ (\PPb_{61}+ \PPb_{51} + \PPb_{41})\PP_{34} \beta_2 \beta_5 \beta_6 \\ 
&+&\PPb_{12}\PP_{23} \beta_4 \beta_5 \beta_6)\,,\notag
\end{eqnarray}
where
\begin{equation}
 \mathcal{C}_6=   -\frac{\alpha_6^{\Lambda_6-4}}{8\Gamma(\Lambda_6-3)\prod_{i=1}^6\beta_i^{\lambda_i-1}}
\frac{1}{ \PP_{23}\PP_{34} \PP_{45} \PP_{56}} \frac{1}{\beta_1 \beta_2 \beta_6}\,.
\end{equation}
Now, following our general strategy, we transform
the last term, which corresponds to the fourth diagram,
according to equation (\ref{C1}) to get
\begin{eqnarray} \label{6pt-1} \nonumber
{A}(123456)&=& \mathcal{C}_6(\PPb_{61}\PP_{56} \beta_2 \beta_3 \beta_4
+ (\PPb_{61}+ \PPb_{51})\PP_{45} \beta_2 \beta_3 \beta_6
+ (\PPb_{61}+ \PPb_{51} + \PPb_{41})\PP_{34} \beta_2 \beta_5 \beta_6 \\ 
&-&(\PPb_{14}\PP_{43} \beta_2 \beta_5 \beta_6
+\PPb_{15}\PP_{53} \beta_2 \beta_4 \beta_6
+ \PPb_{16}\PP_{63} \beta_2 \beta_4 \beta_5) \\ 
&-& \frac{1}{2} \pvec_1^2 \beta_2  \beta_3 \beta_4 \beta_5)\,.\nonumber
\end{eqnarray}
Next, we shall proceed as follows. Consider first the terms proportional to $\PPb_{61}$. The contributions from  the first and from the second sub-diagrams, i.e. from the first two terms
in (\ref{6pt-1}), combine to
\begin{equation} \label{6-1-2}
    \PPb_{56}\beta_2 \beta_3 \beta_4 + \PPb_{45} \beta_2
    \beta_3 \beta_6 = \PPb_{46} \beta_2 \beta_3 \beta_5
\end{equation}
due to the Bianchi identities. The right hand side of (\ref{6-1-2})
adds up to the contribution from the third diagram, i.e. with the third term in (\ref{6pt-1}), to give
\begin{equation} \label{6-1-3}
    \PPb_{34}\beta_2 \beta_5 \beta_6 + \PPb_{46} \beta_2
    \beta_3 \beta_5 = \PPb_{36} \beta_2 \beta_4 \beta_5
\end{equation}
and the right hand side of (\ref{6-1-3}) cancels the contribution
from the fourth sub-diagram. Repeating this procedure for the terms proportional to $\PPb_{51}$ and $\PPb_{41}$ one can see that they all cancel out and we are left only with the term proportional to the off-shell momentum $\pvec_1^2$. Therefore, one finally gets for the six-point amplitude
\begin{equation}
    A(123456)=\frac{\alpha_6^{\Lambda_6-4}}{16\Gamma(\Lambda_6-3)
    \prod_{i=1}^6\beta_i^{\lambda_i-1}}\frac{\beta_3\beta_4\beta_5\,\pvec_1^2 }{\beta_1\PP_{23}\PP_{34}\PP_{45}\PP_{56}}\,,
\end{equation}
which vanishes on-shell, as expected. Let us note that the same amplitude can be computed in a slightly alternative way, which is given in Appendix \ref{app:sixpoint}.

%%%%%%%%%%%%%%%%%%%%%%%%%%%%%%%%%%%%%%%%%%%%%%%%%%%%%%%%%%
\subsection{Recursive Construction}
\label{sec:npoint}
%%%%%%%%%%%%%%%%%%%%%%%%%%%%%%%%%%%%%%%%%%%%%%%%%%%%%%%%%%
Given the results of the previous subsections, it is easy to guess the $n$-point amplitude with one off-shell leg
\begin{equation}\label{eq:npointrecursive}
    A_n(1...n)=\frac{(-)^n\,\alpha_n^{\Lambda_n-(n-2)}\beta_3...\beta_{n-1}\,\pvec_1^2}{2^{n-2}\Gamma(\Lambda_n-(n-3))\prod_{i=1}^{n}\beta_i^{\lambda_i-1}\beta_1\PP_{23}...\PP_{n-1,n}}\,, 
    \end{equation}
    \begin{equation} \label{alpha-n}
    \alpha_n=\sum_{i<j}^{n-2}\PPb_{ij}+\PPb_{n-1,n}\,,
\end{equation}
where $\Lambda_n=\lambda_1+...+\lambda_n$. Below we shall prove by induction that (\ref{eq:npointrecursive}) is indeed the correct answer. The $n$-point amplitude can be represented diagramatically as follows
\begin{align*}
   \parbox{2.8cm}{\includegraphics[scale=0.3]{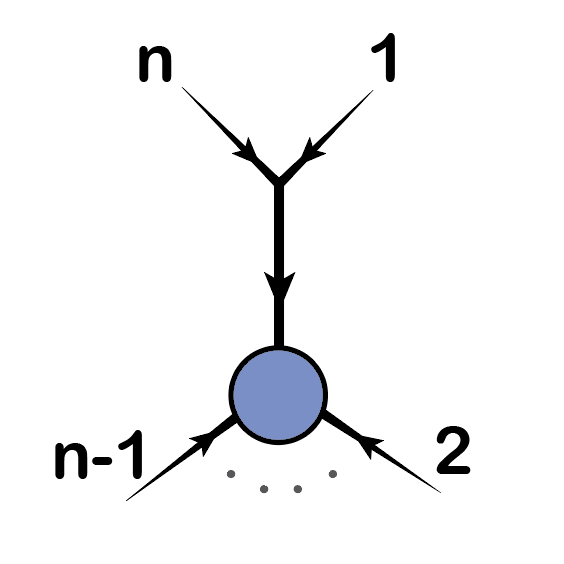}}\,\, + \parbox{3.8cm}{\includegraphics[scale=0.3]{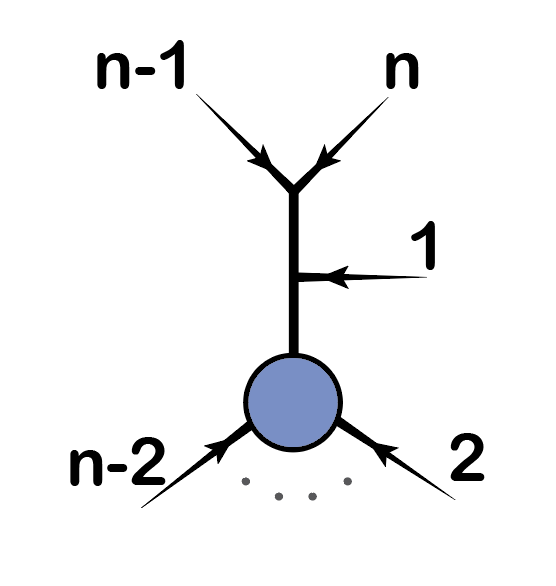}} \,\, + \parbox{3.8cm}{\includegraphics[scale=0.3]{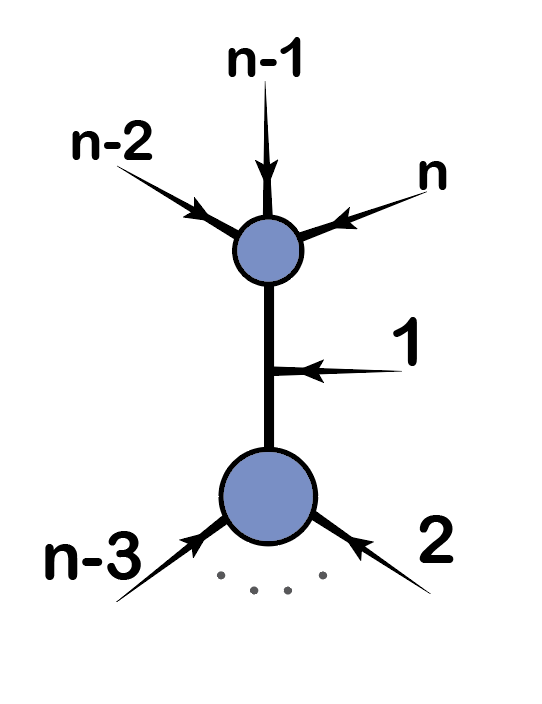}} + ...
\end{align*}
\begin{align*}
...+   \parbox{2.8cm}{\includegraphics[scale=0.3]{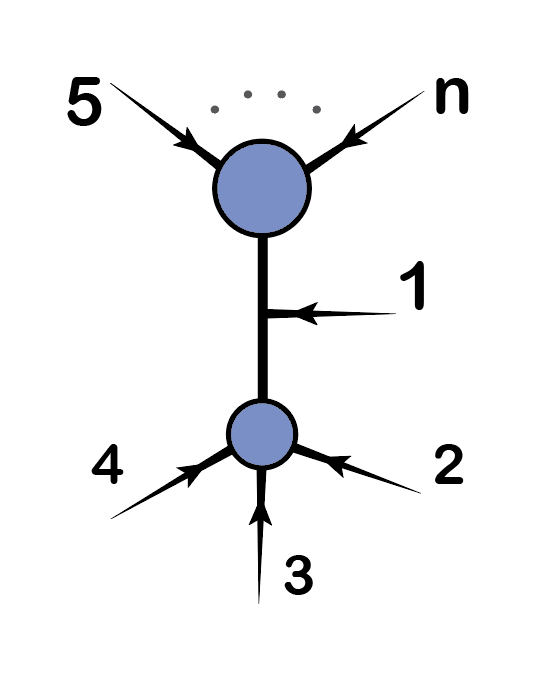}}\,\, + \parbox{3.8cm}{\includegraphics[scale=0.3]{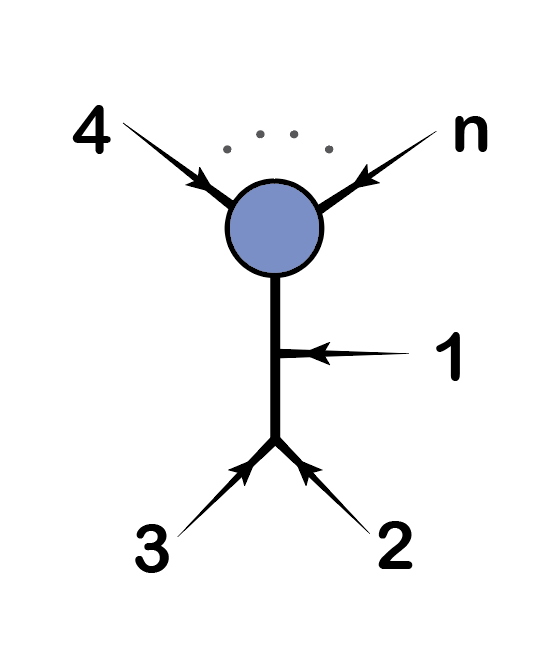}} \,\, + \parbox{3.8cm}{\includegraphics[scale=0.3]{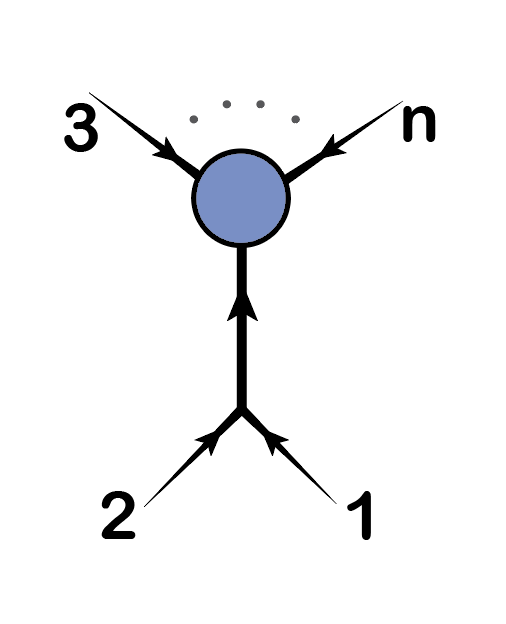}}
\end{align*}
First, let us prove by induction that the factor  $\alpha_n$
has the form (\ref{alpha-n}) and is common for all diagrams. The overall $\Gamma$-function for the $n$-point amplitude follows directly from our previous calculations and therefore we shall not consider it below. Since we have already checked the cases of 4-, 5- and 6-point amplitudes we proceed to the induction step.

Consider the first diagram. The corresponding $\alpha_n^I$-factor is equal to
\begin{equation}\label{alp-1}
    \alpha^I_n
    = \PPb_{n,1} + (\PPb_{n-1,m} + \PPb_{23}+...+\PPb_{2,n-2}
    + \PPb_{34}+...+\PPb_{3,n-2}+...+ \PPb_{n-4,n-2}+ \PPb_{n-3,n-2})\,,
\end{equation}
where the momentum on the internal line has index $m$. Now using the momentum conservation
\begin{equation} \label{alp-2}
   \PPb_{n-1,m}=-  \PPb_{n-1,2}-...-\PPb_{n-1,n-2}
\end{equation}
we see that (\ref{alp-1}) coincides with (\ref{alpha-n}).
Next, let us demonstrate that this factor is the same for all
sub-diagrams. Consider the second diagram, whose
$\alpha^{II}$-factor reads
\begin{equation} \label{alp-3}
\alpha_n^{II}=
     \PPb_{n-1,n} + \PPb_{p,1}+
    (\PPb_{n-2,m} + \PPb_{23}+..+\PPb_{2,n-3}
    + \PPb_{34}+...+\PPb_{3,n-3}+...+ \PPb_{n-5,n-3}+ \PPb_{n-4,n-3})\,.
\end {equation}
Similarly in the equation above  the subscript $p$ corresponds to internal momentum that exits $(n-1, n)$ part of the diagram
and the subscript $m$ corresponds to the internal momentum that enters $(2,..., n-2)$ part of the diagram. Using relation (\ref{alp-2}) as well as
\begin{equation}
     \PPb_{n-2,m}=-  \PPb_{n-2,2}-...-\PPb_{n-2,n-3},
\qquad\quad
  \PPb_{p,1}= \PPb_{n,1} + \PPb_{n-1,1}  
\end{equation}
one can see that the difference $\alpha^I_n-\alpha_n^{II}$ is indeed zero. The proof that the $\alpha_n$-factor is equal to \eqref{alpha-n} for all sub-diagrams with $((n-k,n),1,(2,n-k-1))$
partition of external momenta is completely analogous.

Now let us prove that the $n$-point amplitude has the required form
(\ref{eq:npointrecursive}). Again we proceed with the induction step.
The sum of the $n$-point diagrams has the form: 
\begin{eqnarray}\nonumber
{ A}(1,2,...,n) &=& \mathcal{C}_n^\prime \left ( \frac{\PPb_{n,1}}{\PP_{23}...
\PP_{n-2,n-1} \underline{\PP_{n-1,n}}}
\frac{ \beta_3...\beta_{n-2} \underline{\beta_{n-1}}}{\beta_n} +
\frac{\PPb_{n,1}+\PPb_{n-1,1}}{\PP_{23}...\underline{\PP_{n-2,n-1}}
\PP_{n-1,n}}
\beta_3...\underline{\beta_{n-2}\beta_{n-1}} \right. \\ 
&+& \frac{\PPb_{n1}+\PPb_{n-1,1} + \PPb_{n-2,1}}{\PP_{23}...\underline{\PP_{n-3,n-2}}
\PP_{n-2,n-1}\PP_{n-1,n}}
\beta_3...\underline{\beta_{n-3} \beta_{n-2}} \beta_{n-1} + ...
 \\ \nonumber
&+&\frac{\PPb_{n,1}+... \PPb_{n,5}}{\PP_{23}...\underline{\PP_{4,5}}
...\PP_{n-1,n}}
\beta_3 \underline{\beta_{4} \beta_{5}}...\beta_{n-1}
\\ \nonumber
&+&\left. \frac{\PPb_{n,1}+... + \PPb_{n,4}}
{\PP_{23} \underline{\PP_{34}} 
...\PP_{n-1,n}} 
\underline{\beta_{3} \beta_{4}}... \beta_{n-1} +
\frac{\PPb_{12}}{\underline{\PP_{23}}... 
\PP_{n-1,n}}
\frac{ \underline{\beta_3}...\beta_{n-2} {\beta_{n-1}}}{\beta_2} 
\right )\,,
\end{eqnarray}
where the underlined expression is omitted and
\begin{equation}
\mathcal{C}_n^\prime=\frac{(-)^{n-1}\,\alpha_n^{\Lambda_n-(n-2)}}
    {2^{n-3}\Gamma(\Lambda_n-(n-3)) \beta_1 
    \prod_{i=1}^{n}\beta_i^{\lambda_i-1}}    \,.
\end{equation}
Adding these terms together and extracting the common denominator
\begin{equation}
    \frac{1}{\beta_2 \beta_n \PP_{23}\PP_{34}...\PP_{n-1,1}} 
\end{equation}
 we get
\begin{eqnarray} \label{npt-i} \nonumber 
{ A}(1,2,...,n)&=& \mathcal{C}_n( \PPb_{n,1}\PP_{n-1,n} \beta_2...\underline{\beta_{n-1} \beta_n}
+(\PPb_{n1}+ \PPb_{n-1,1})\PP_{n-2,n-1} \beta_2...\underline{\beta_{n-2} \beta_{n-1}} \beta_n 
\\ \nonumber
&+&(\PPb_{n,1}+ \PPb_{n-1,1} + \PPb_{n-2,1})\PP_{n-3,n-2}
\beta_2...\underline{\beta_{n-3} \beta_{n-2}} \beta_{n-1}\beta_n
+...\\
&+&(\PPb_{n,1}+ \PPb_{n-1,1} + ...+ \PPb_{51})\PP_{45}
\beta_2 \beta_3 \underline{\beta_4 \beta_5}... \beta_n
\\ \nonumber
&+&
(\PPb_{n,1}+ \PPb_{n-1,1} + ...+ \PPb_{41})\PP_{34}
\beta_2  \underline{\beta_3 \beta_4}... \beta_n 
\\ \nonumber
&+& \PPb_{12} \PP_{23}  \underline{\beta_2 \beta_3}... \beta_n)\,,
\end{eqnarray}
where
\begin{equation}
    \mathcal{C}_n=
    \frac{(-)^{n-1}\,\alpha_n^{\Lambda_n-(n-2)}}
    {2^{n-3}\Gamma(\Lambda_n-(n-3)) \PP_{23}...\PP_{n-1,n}\beta_1 \beta_2 \beta_n
    \prod_{i=1}^{n}\beta_i^{\lambda_i-1}}\,.
  \end{equation}
Now, as we have done in the cases of five- and six-point functions, we transform the last term in
(\ref{npt-i}) as
\begin{eqnarray}
\PPb_{12} \PP_{23}  \beta_4... \beta_{n-1}\beta_n&=&
-\frac{1}{2} \pvec_1^2 \beta_2 \beta_3... .\beta_n 
- \PPb_{14} \PP_{43} \beta_2 \underline{\beta_3 \beta_4} ... \beta_n - ...
\\ \nonumber
&-& \PPb_{1,n-1} \PP_{n-1,3} \beta_2 \underline{\beta_3}...
\underline{\beta_{n-1}} \beta_n
-\PPb_{1,n} \PP_{n,3}\beta_2 \underline{\beta_3}...
\underline{\beta_{n}} \,.
\end{eqnarray}
Further, we collect the terms proportional to
$\PPb_{n,1}$ in (\ref{npt-i}). They have the form
\begin{eqnarray} \label{npt-x}
&&\PP_{n-1,n} \beta_2...\underline{\beta_{n-1} \beta_{n}}\nonumber
+
\PP_{n-2,n-1} \beta_2...\underline{\beta_{n-2} \beta_{n-1}} \beta_n
 \\ 
&& +
\PP_{n-3,n-2}
\beta_2...\underline{\beta_{n-3} \beta_{n-2}} \beta_{n-1}\beta_n
\\ \nonumber
&&
...+\PP_{45}
\beta_2 \beta_3 \underline{\beta_4 \beta_5}... \beta_n
+
\PP_{34}
\beta_2  \underline{\beta_3 \beta_4}... \beta_n 
 \\ \nonumber
&&-
\PP_{n,3}\beta_2 \underline{\beta_3}...
\underline{\beta_{n}}
\end{eqnarray}
Now we shall use the Bianchi identities. First, we apply the Bianchi identity to the first line in 
(\ref{npt-x}) to obtain $\PPb_{n,n-2}\beta_2 ...\underline{\beta_{n-2}} \beta_{n-1} {\underline \beta_n}$.
Then we add this expression to the second line in 
(\ref{npt-x}) and then
apply the Bianchi identity again. Proceeding this way 
we see that the sum of terms proportional 
to $\PPb_{n,1}$ vanishes.
Next, we repeat the same procedure for the terms proportional
to $\PPb_{n-1,1}$ in (\ref{npt-i})
and obtain that their sum is equal to zero as well, and so on.
Finally, we see that all the terms except for the one which is proportional to $\pvec_1^2$ cancel out. Collecting the intermediate results together we find the final expression for the 
$n$-point tree amplitude to be (\ref{eq:npointrecursive}), as conjectured.

The final conclusion here is that all $n$-point amplitudes with one off-shell leg have a remarkably simple form and vanish on-shell. Hence, at the tree-level Chiral Theory is consistent with the numerous no-go theorems like Weinberg's low energy theorem and Coleman-Mandula theorem that imply $S=1$ once at least one massless higher spin particle is in the game. From the explicit calculations above it is clear that (i) it is important to have all spins in the spectrum without any upper/lower bounds and gaps; (ii) the coupling constants must have a very particular dependence on spins, $C_{\lambda_1,\lambda_2,\lambda_3}\sim 1/\Gamma(\lambda_1+\lambda_2+\lambda_3)$. This situation was referred to as coupling conspiracy \cite{Skvortsov:2018jea}. The fact that the tree-level amplitudes vanish on-shell indicates that there should not be any nontrivial cuts of the loop diagrams and, hence, the loop corrections are expected to have a better UV-behaviour. 

%%%%%%%%%%%%%%%%%%%%%%%%%%%%%%%%%%%%%%%%%%%%%%%%%%%%%%%%%%
\section{Vacuum Bubbles}
\label{sec:bubbles}
%%%%%%%%%%%%%%%%%%%%%%%%%%%%%%%%%%%%%%%%%%%%%%%%%%%%%%%%%%
It is easy to show that all vacuum corrections vanish in accordance with the naive expectation that vacuum partition function for higher spin gravities should be one, $Z=1$, which indicates that the total regularized number of degrees of freedom vanishes. This is in accordance with similar findings both in flat and AdS spaces \cite{Gopakumar:2011qs,Tseytlin:2013jya,Giombi:2013fka,Giombi:2014yra,Beccaria:2014jxa,Beccaria:2014xda,Beccaria:2015vaa,Gunaydin:2016amv,Bae:2016rgm,Skvortsov:2017ldz}.

%%%%%%%%%%%%%%%%%%%%%%%%%%%%%%%%%%%%%%%%%%%%%%%%%%%%%%%%%%
\subsection{Determinants}
\label{sec:oneloopbubble}
%%%%%%%%%%%%%%%%%%%%%%%%%%%%%%%%%%%%%%%%%%%%%%%%%%%%%%%%%%
The simplest vacuum corrections probe the spectrum of a theory via determinants of the kinetic operators. First, let us consider the free higher spin theory in four-dimensional flat space \cite{Beccaria:2015vaa}. The action is the sum over all spins of the kinetic terms of massless fields:
\begin{align}
    S&= \sum_{s} \int d^4x\, \phi_{a_1...a_s} \square \phi^{a_1...a_s}\,, && \delta \phi_{a_1...a_s}=\pl_{a_1}\xi_{a_2...a_s}+\text{perm.}\,,
\end{align}
where we have already partially gauged fixed the action, so that both the fields and the gauge parameters are transverse and traceless. The partition function is
\begin{align}
    \parbox{2.0cm}{\includegraphics[scale=0.34]{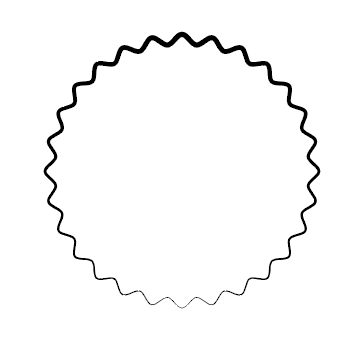}}&:  &
   Z_{\text{1-loop}}&= \frac{1}{\det^{1/2}_{0}|-\partial^2|}\prod_{s>0}\frac{\det^{1/2}_{s-1,\perp}|-\partial^2 |}{\det^{1/2}_{s,\perp}|-\partial^2|}=\frac{1}{(z_0)^{\frac{1}{2}}}\prod_{s>0}\frac{(z_{s-1})^{\frac{1}{2}}}{(z_s)^{\frac{1}{2}}}\,,
\end{align}
where the determinants are of the Laplacian $-\pl^2$ defined on symmetric traceless transverse tensors, see e.g. \cite{Tseytlin:2013jya,Beccaria:2015vaa}. The numerator in the formula corresponds to ghosts, i.e. to pure gauge degrees of freedom. The determinant of a free scalar field stays aside since it is not a gauge field.

On one hand it is tempting to choose a regularization for the infinite product such that the ghost of the spin-$s$ field cancels the spin-$(s-1)$ contribution in the denominator. This would give $Z_{\text{1-loop}}=1$, as a result. On the other hand it is the same problem as determining the value of the infinite sum $1-1+1-...$. Indeed, for theories with infinitely many fields a prescription of how to sum over the spectrum has to be given
by hand and this is one of the instances where higher spin gravity reveals its 'stringy' nature. However unlike string theory,
where summation goes over relevant Riemann surfaces, we do not have any geometric understanding of how the sum over spins needs to be done.

Therefore, we have to come up with some plausible idea of what the total number of degrees of freedom is. The prescription of \cite{Beccaria:2015vaa} that gives $Z = 1$ instructs us to count degrees of freedom as follows 
\begin{align}\label{pdof}
    \nu_0=\sum_{\lambda} 1= 1+2 \sum_{\lambda>0}\lambda=1+2\zeta(0)=0\,,
\end{align}
where $1$ is for the scalar field and $2$ per each massless field. Although this regularization seems to be ad hoc,
the success \cite{Giombi:2013fka,Giombi:2014yra,Gunaydin:2016amv,Giombi:2016pvg,Bae:2016rgm,Skvortsov:2017ldz} of the zeta-function regularization \cite{Dowker:1975tf,Hawking:1976ja} in the study of
determinants of higher spin theories on AdS background provides a strong support for \eqref{pdof}.

Let us recall that the kinetic operators of massless spinning fields on $AdS_d$ have spin-dependent mass-like terms and the naive cancellation, as above, is not possible. The determinants can be computed via spectral zeta-function \cite{Camporesi:1991nw,Camporesi:1992wn,Camporesi:1992tm,Camporesi:1993mz,Camporesi:1994ga,Camporesi:1995fb} and the spin sums can be taken with the help of zeta-function. One can perform the one-loop computations for various spectra of fields and on various backgrounds (Euclidian, thermal and global $AdS_d$). The final result is highly nontrivial and is consistent with the AdS/CFT expectations. Therefore, the zeta-function regularization seems to be well-tested, which justifies \eqref{pdof}.

%%%%%%%%%%%%%%%%%%%%%%%%%%%%%%%%%%%%%%%%%%%%%%%%%%%%%%%%%%
\subsection{Higher Vacuum Loops}
\label{sec:2loopbubble}
%%%%%%%%%%%%%%%%%%%%%%%%%%%%%%%%%%%%%%%%%%%%%%%%%%%%%%%%%%
The two-loop diagram vanishes due to the chirality of interactions: assuming some combination of helicities $\lambda_{i=1,2,3}$ assigned
to the left vertex of
\begin{align*}
    \parbox{2.0cm}{\includegraphics[scale=0.34]{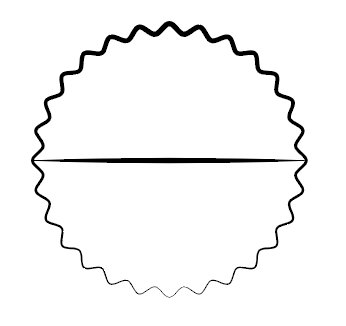}}=0
\end{align*}
we find the opposite triplet, i.e. $-\lambda_{i=1,2,3}$, entering the right vertex. However, $1/\Gamma[\Lambda]$ and $1/\Gamma[-\Lambda]$ factors coming from the product of the two couplings cannot both be nonzero. Hence, the diagram vanishes. The same arguments as above show that the three-loop diagrams also vanish: there is no such assignment of helicities that makes all $1/\Gamma[...]$-factors nonzero at the same time.
\begin{align*}
    \parbox{2.0cm}{\includegraphics[scale=0.34]{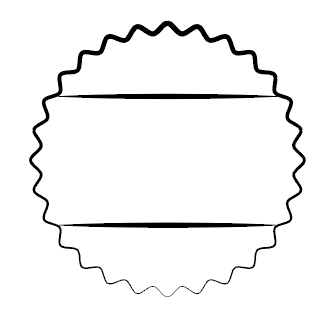}}&=0&  &
    \parbox{2.0cm}{\includegraphics[scale=0.34]{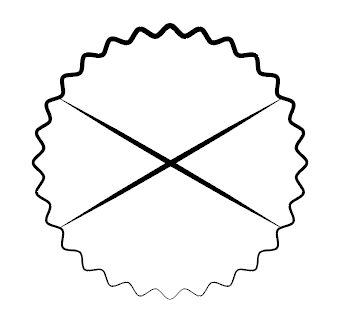}}=0
\end{align*}
It is easy to see that this is true for all loops. Indeed, the total helicity --- the sum over all ends of the propagators must be zero since there are no external legs and the propagator connects helicities of opposite sign. The same sum can be represented as a sum over triplets of helicities entering the vertices. In order for a vacuum diagram to be nonzero each triplet must have positive total helicity, otherwise the coupling constant is zero. Therefore, in this case we shall have a finite sum of positive numbers that equals zero, which is impossible. Therefore, all vacuum diagrams with more than one loop vanish identically.

%%%%%%%%%%%%%%%%%%%%%%%%%%%%%%%%%%%%%%%%%%%%%%%%%%%%%%%%%%
\section{Loops with Legs}
\label{sec:leggedloop}
We shall discuss the behaviour of $n$-legged loop diagrams by examining the tadpole, self-energy, vertex correction and the four-point amplitude at one loop. Then, we give a general argument for multi-loop amplitudes. An important thing to remember is that vanishing of tree-level amplitudes should eliminate all $\log$-divergences that would lead to cuts otherwise. In the higher spin case it always makes sense to check explicitly if an argument developed for low-spin theories works for higher spin ones as well. We also would like to see if there are any power divergences and how slightly different regularizations work. 
%%%%%%%%%%%%%%%%%%%%%%%%%%%%%%%%%%%%%%%%%%%%%%%%%%%%%%%%%%

%%%%%%%%%%%%%%%%%%%%%%%%%%%%%%%%%%%%%%%%%%%%%%%%%%%%%%%%%%
\subsection{Tadpole}
\label{sec:tadpole}
%%%%%%%%%%%%%%%%%%%%%%%%%%%%%%%%%%%%%%%%%%%%%%%%%%%%%%%%%%
The light-cone approach is not suitable for the computation of one-point functions, like tadpole. Nevertheless, 
tadpoles for the external lines with non-zero helicity must vanish by Lorentz invariance. A tadpole for the scalar field also vanishes due to the absence of the relevant vertex in the action. Lastly, if the external helicity is zero and the internal one is some $\mu$, then at the vertex we still have $\Gamma(0+\mu-\mu)^{-1}=0$. Therefore, 
\begin{align*}
    \parbox{1.5cm}{\includegraphics[scale=0.21]{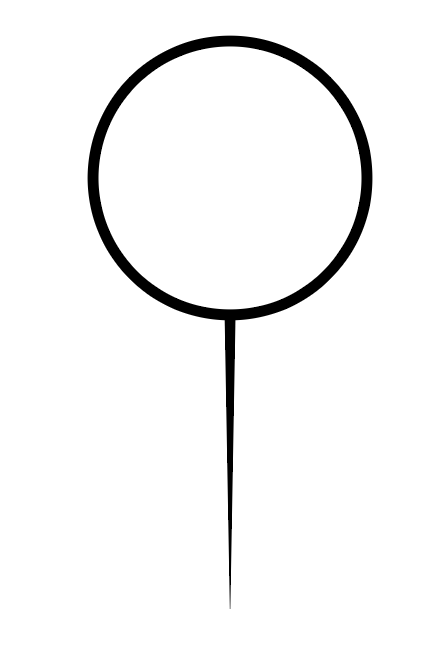}}=0
\end{align*}

%%%%%%%%%%%%%%%%%%%%%%%%%%%%%%%%%%%%%%%%%%%%%%%%%%%%%%%%%%
\subsection{Self-energy}
\label{sec:selfenergy1loop}
%%%%%%%%%%%%%%%%%%%%%%%%%%%%%%%%%%%%%%%%%%%%%%%%%%%%%%%%%%
We recall that the $U(N)$-version of Chiral Theory is studied for concreteness. All general conclusions below are also true for the other cases, which can be treated in a similar way. For a given $N$ we can first have a look at the planar diagrams, which are simpler. For the self-energy diagram, there are contributions from planar and non-planar diagrams:
\begin{align*}
   \parbox{3.7cm}{\includegraphics[scale=0.21]{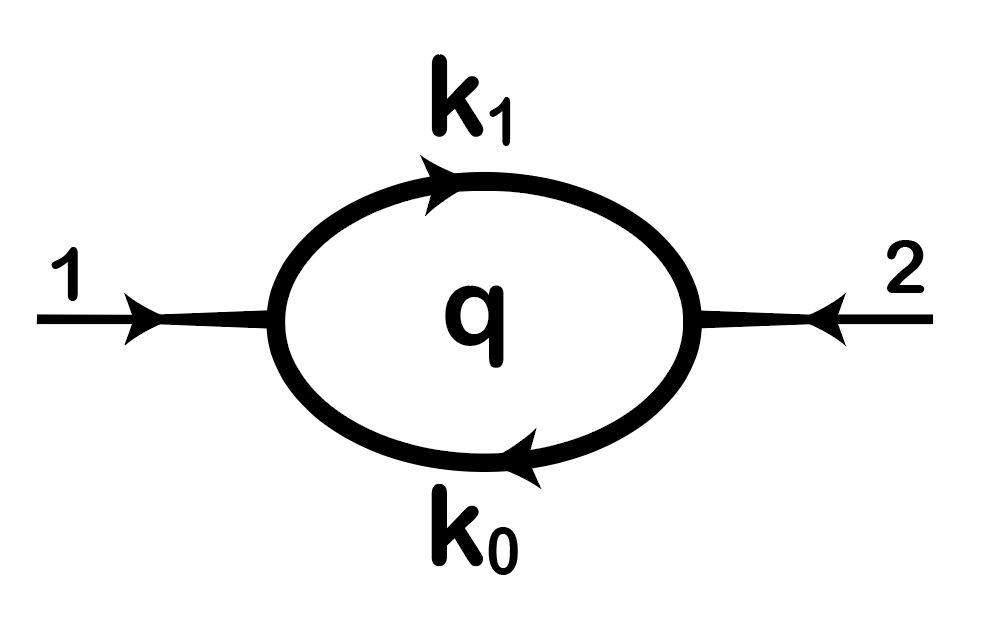}}+\parbox{3.9cm}{\includegraphics[scale=0.22]{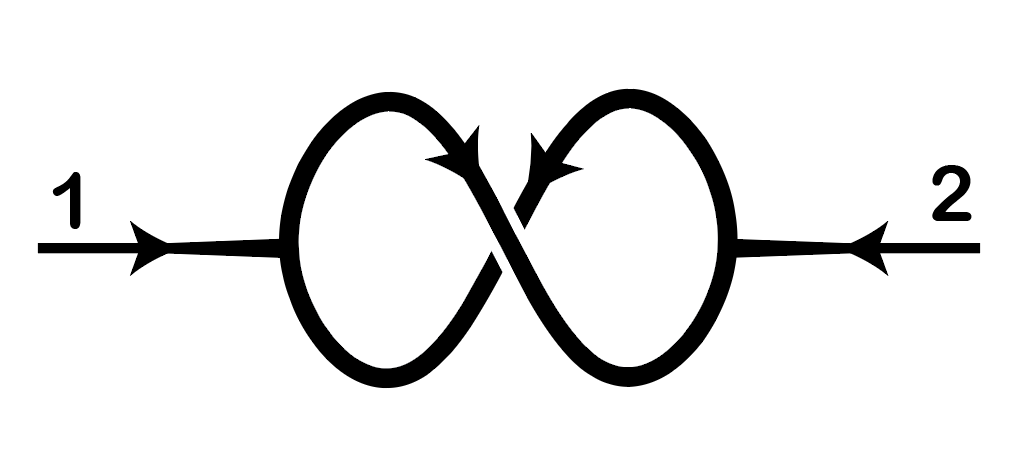}}
\end{align*}
Here, $\kvec_1,\kvec_0,\qvec$ are dual momenta\footnote{More detail about dual momenta can be found in \cite{Brandhuber:2007vm,Thorn:2004ie,Chakrabarti:2005ny,Chakrabarti:2006mb}, see also Appendix \ref{app:Thornregulator}.} 
and the external momentum is related to $\kvec$ as $\pvec_1=\kvec_1-\kvec_0$. The loop momentum is $\pvec=\qvec-\kvec_0$. 

We start our analysis by considering the simplest self-energy diagram. In order to avoid confusing and
cumbersome notation, we introduce sources $h_A^{\ B}$ that can be contracted with fields. As a result each amplitude acquires factors $\tr(hh...)$ which keeps track of the color indices. We adopt the 'world-sheet friendly' regularization \cite{Thorn:2004ie,Chakrabarti:2005ny,Chakrabarti:2006mb}, which is used in a number of theories in light-cone gauge. The one-loop self-energy reads
\begin{equation}\label{eq:selfintegrand}
\begin{split}
    \Gamma_{\text{self}}=&N\tr(h_1h_2)\sum_{\omega} \frac{(l_p)^{\Lambda_2-2}}{\beta_1^{\lambda_1}\beta_2^{\lambda_2}\Gamma(\Lambda_2-1)}\int \frac{d^4q}{(2\pi)^4} \frac{\PPb_{q-k_0,p_1}^2\delta_{\Lambda_2,2}}{(\qvec-\kvec_0)^2(\qvec-\kvec_1)^2}\\
    &-\tr(h_1)\tr(h_2)\sum_{\omega}\frac{(2l_p)^{\Lambda_2-2}}{\beta_1^{\lambda_1}\beta_2^{\lambda_2}\Gamma(\Lambda_2-1)}\int \frac{d^4q}{(2\pi)^4} \frac{\PPb_{q-k_0,p_1}^{\Lambda_2}}{(\qvec-\kvec_0)^2(\qvec-\kvec_1)^2}\,,
    \end{split}
\end{equation}
where $d^4q=dq^-d\beta d^2q_{\perp}$ and $\Lambda_2=\lambda_1+\lambda_2$. A very important feature of all loop diagrams is that the very last sum over helicities factors out, i.e. after we sum over all but one helicities running in the loop the resulting expression does not depend on the very last helicity to be summed over. Therefore, each loop diagram has an overall factor $\nu_0=\sum_\omega 1$, which we have already faced in \eqref{pdof}. Let us evaluate the leading contribution, i.e. the first term,
\begin{equation}\label{eq:selfleading}
    \Gamma_{\text{self}}^{\text{leading}}=N\tr(h_1h_2)\sum_{\omega} \frac{(l_p)^{\Lambda_2-2}}{\beta_1^{\lambda_1}\beta_2^{\lambda_2}\Gamma(\Lambda_2-1)}\int \frac{d^4q}{(2\pi)^4} \frac{\PPb_{q-k_0,p_1}^2\delta_{\Lambda_2,2}}{(\qvec-\kvec_0)^2(\qvec-\kvec_1)^2}\,.
\end{equation}
Here, we observe that the integrand is non-vanishing only when $\Lambda_2=2$. To regulate this integral, one can introduce a cut-off $\exp[-\xi q_{\perp}^2]$, where $q_{\perp}\equiv( q, \bar q)$ is the transverse part of $\qvec$. Then, using Schwinger parameterization and integrating out $q^-$ gives us $\delta\big(\beta(T_1+T_2)-T_1\beta_{k_0}-T_2\beta_{k_1}\big)$. Next, we replace\footnote{Note that whenever we write $\beta_{k_i}$, it means we consider the $k_i^+$ component of the dual 4-momentum.}
\begin{equation}
    \beta=\frac{T_1\beta_{k_0}+T_2\beta_{k_1}}{T_1+T_2}\,,
\end{equation}
and as a result the expression (\ref{eq:selfleading}) reads (omitting the prefactor)
\begin{equation}\label{eq:intermediateself1}
    \Gamma_{\text{self}}^{\text{leading}}\sim \int \PPb_{q-k_0,p_1}^2\exp\Big[-(T+\xi)\Big(q^a-\frac{T_1k_0^a+T_2k_1^a}{T+\xi}\Big)^2-\frac{T_1T_2\pvec_1^2}{T}-\frac{\xi(T_1k_0^a+T_2k_1^a)^2}{T(T+\xi)}\Big]\,,
\end{equation}
where we integrate over $q$ and over $T_i$ that are the Schwinger's parameters, $T=T_1+T_2$. It is now safe to set $\pvec_1^2$ on-shell and $\xi=0$ in the last two terms in the exponential in the expression (\ref{eq:intermediateself1}). Hence, we are left with a Gaussian integral 
\begin{equation}\label{eq:Gaussianself}
    \Gamma_{\text{self}}^{\text{leading}}\sim \int \frac{d^2q^a}{16\pi^2}\Big[(\bar{q}-\bar{k}_0)\beta_1-\bar{p}_1\big(\frac{T_1\beta_{k_0}+T_2\beta_{k_1}}{T_1+T_2}-\beta_{k_0}\big)\Big]^2e^{-(T+\xi)\Big(q^a-\frac{T_1k_0^a+T_2k_1^a}{T+\xi}\Big)^2}\,.
\end{equation}
We can evaluate (\ref{eq:Gaussianself}) noting that
\begin{equation}\label{eq:magicGuassian}
    \int d^2q_{\perp} e^{-Aq_{\perp}^2}=\frac{\pi}{A}, \qquad\qquad \int d^2 q_{\perp}\,\left(\bar{q}\right)^n e^{-Aq_{\perp}^2}=0 \quad (\text{for} \, n\geq 1)\,.
\end{equation}
As a result, we get
\begin{equation}\label{eq:selfresult}
\begin{split}
    \Gamma_{\text{self}}^{\text{leading}}&=\sum_{\omega}\frac{(l_p)^{\Lambda_2-2}\,N\tr(h_1h_2)\delta_{\Lambda_2,2}}{\beta_1^{\lambda_1-1}\beta_2^{\lambda_2-1}\Gamma[\Lambda_2-1]}\int_0^1 dx \int_0^{\infty} \frac{dT}{16\pi^2}\frac{\xi^2[x\bar{k}_0+(1-x)\bar{k}_1]^2}{(T+\xi)^3}\\
    &\xrightarrow{\xi\rightarrow 0} \nu_0\,\frac{(l_p)^{\Lambda_2-2}\,N\tr(h_1h_2)\delta_{\Lambda_2,2}}{32\pi^2\beta_1^{\lambda_1-1}\beta_2^{\lambda_2-1}\Gamma[\Lambda_2-1]}\int_0^1 dx[x\bar{k}_0+(1-x)\bar{k}_1]^2\\
    &=\nu_0\,\delta_{\Lambda_2,2}\frac{(l_p)^{\Lambda_2-2}\,N\tr(h_1h_2)\, (\bar{k}_0^2+\bar{k}_0\bar{k}_1+\bar{k}_1^2)}{96\pi^2\beta_1^{\lambda_1-1}\beta_2^{\lambda_2-1}\Gamma[\Lambda_2-1]}\,,
    \end{split}
\end{equation}
where we made a change of variables $x=T_1/T$. Here, the $x$-integral in (\ref{eq:selfresult}) is perfectly finite and $\Gamma_{\text{self}}^{\text{leading}}$ is reminiscent of $\Pi^{++}$ amplitude in \cite{Thorn:2005ak,Chakrabarti:2005ny,Chakrabarti:2006mb}. The important feature of the computation above is that the loop diagrams have the number of physical degrees of freedom $\nu_0$ as an overall factor, which guarantees that the contribution above vanishes and does not require a counterterm. We note that the Lorentz invariance forbids helicity flips for an isolated spinning particle. Therefore, if we were to find a non-vanishing contribution to $\Gamma_{\text{self}}^{\text{leading}}$ we would have to introduce local counterterms to cancel it. 

Let us also consider the sub-leading term for the self-energy correction by repeating the procedure given above. The sub-leading contribution before taking the $T$-integral is
\begin{align}\label{eq:selfsub1}
    \Gamma^{\subleading}_{\self}&=\nu_0\,\frac{(2l_p)^{\Lambda_2}(-)^{\lambda_1}\tr(h_1)\tr(h_2)}{16\pi^2\Gamma[\Lambda_2-1]}\int_0^1dx \int_0^{\infty}dT \frac{\xi^{\Lambda_2}[x\bar{k}_0+(1-x)\bar{k}_1]^{\Lambda_2}}{(T+\xi)^{\Lambda_2+1}}\,,
\end{align}
which can be obtained using the holomorphic integral (\ref{eq:magicGuassian}). Eq. \eqref{eq:selfsub1} assumes that $\Lambda_2>1$. We now have a convergent integral and the result is
\begin{equation}\label{eq:selfsub2}
\begin{split}
    \Gamma^{\subleading}_{\self}&=\nu_0\,\frac{(-)^{\lambda_1}(2l_p)^{\Lambda_2}\,\tr(h_1)\tr(h_2)(\Lambda_2-1)}{16\pi^2\Gamma[\Lambda_2+1]} \int_0^1 dx[x\bar{k}_0+(1-x)\bar{k}_1]^{\Lambda_2}\\
    &=\nu_0\,\frac{(-)^{\lambda_1}(2l_p)^{\Lambda_2}\,\tr(h_1)\tr(h_2)(\Lambda_2-1)}{16\pi^2\Gamma[\Lambda_2+2]}\times\frac{\bar{k}_0^{\Lambda_2+1}-\bar{k}_1^{\Lambda_2+1}}{\bar{k}_0-\bar{k}_1}\,, \quad (\Lambda_2\geq 0)\,.
    \end{split}
\end{equation} 
Since $\Lambda_2>1$ the potentially dangerous non-local contribution is zero. The kinematic part of $\Gamma_{\self}^{\subleading}$ is finite and, hence, $\Gamma_{\self}^{\subleading}$ vanishes again
due to the factorization of $\nu_0$, which takes place
regardless of the value of $\Lambda_2$. This implies that the self-energy correction of Chiral Theory does not break Lorentz invariance. 

Finally let us mention that, alternatively, one can use the original momentum $\pvec_i$ and the loop momentum $\pvec$ together with the cut-off $\exp[-\xi p_{\perp}^2]$ for the loop computations. In the case of the self-energy the corresponding integral reads
\begin{equation}
    \begin{split}
    \Gamma^{\subleading}_{\self}&=\nu_0\,\frac{(2l_p)^{\Lambda_2}}{\beta_1^{\lambda_1}\beta_2^{\lambda_2}\Gamma[\Lambda_2-1]}\int \frac{d^4p}{(2\pi)^4}\frac{\PPb_{p1}^{\Lambda_2}}{\pvec^2(\pvec+\pvec_1)^2}\\
    &=\nu_0\,\frac{(2l_p)^{\Lambda_2}(-)^{\lambda_1}\tr(h_1)\tr(h_2)}{16\pi^2\Gamma[\Lambda_2-1]}\int_0^1dx \int_0^{\infty}dT \frac{\xi^{\Lambda_2}[x\bar{p}_1]^{\Lambda_2}}{(T+\xi)^{\Lambda_2+1}}\\
    &=\nu_0\,\frac{(2l_p)^{\Lambda_2}(-)^{\lambda_1}\tr(h_1)\tr(h_2)(\Lambda_2-1)}{16\pi^2\Gamma[\Lambda_2+1]}\int_0^1dx  [x\bar{p}_1]^{\Lambda_2}\\
    &=\nu_0\,\frac{(2l_p)^{\Lambda_2}(-)^{\lambda_1}\tr(h_1)\tr(h_2)(\Lambda_2-1)}{16\pi^2\Gamma[\Lambda_2+2]}\bar{p}_1^{\Lambda_2}, \qquad (\Lambda_2\geq 0)\,.
\end{split}
\end{equation}
%%%%%%%%%%%%%%%%%%%%%%%%%%%%%
\subsection{Vertex correction}
\label{sec:vertexcorrection}
The next case is to consider the vertex correction diagrams 
\begin{align*}
    \parbox{3.3cm}{\includegraphics[scale=0.19]{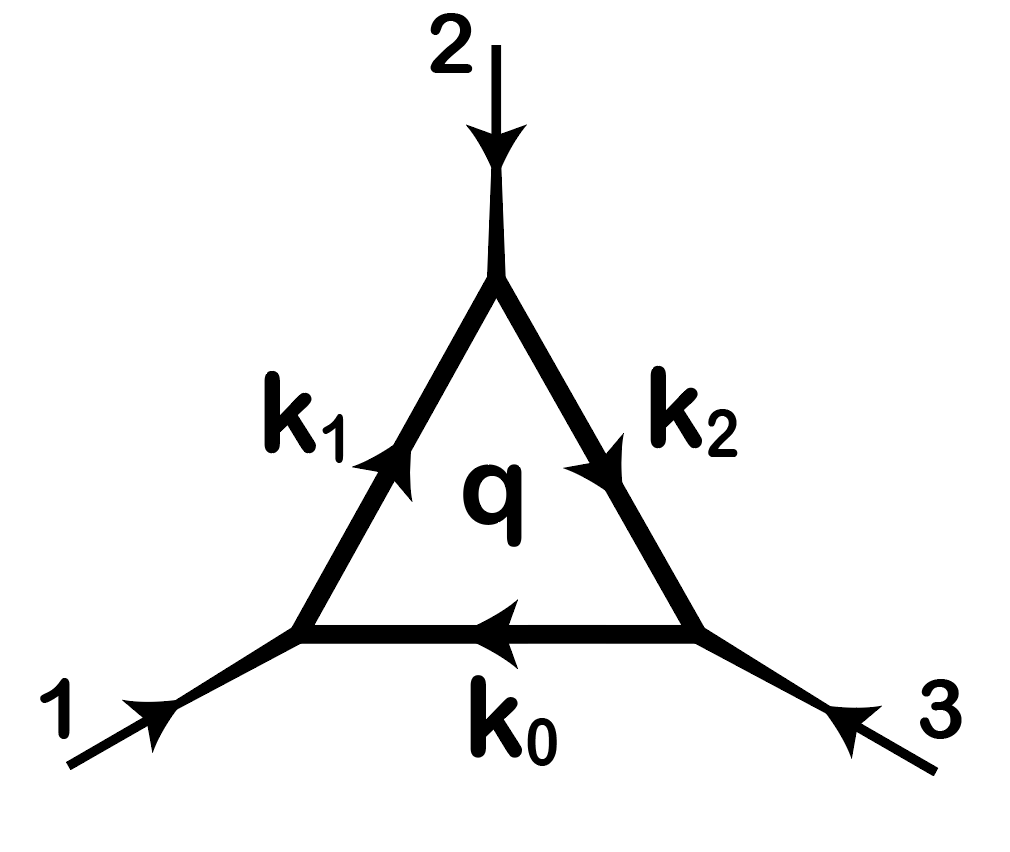}}+\parbox{2.9cm}{\includegraphics[scale=0.24]{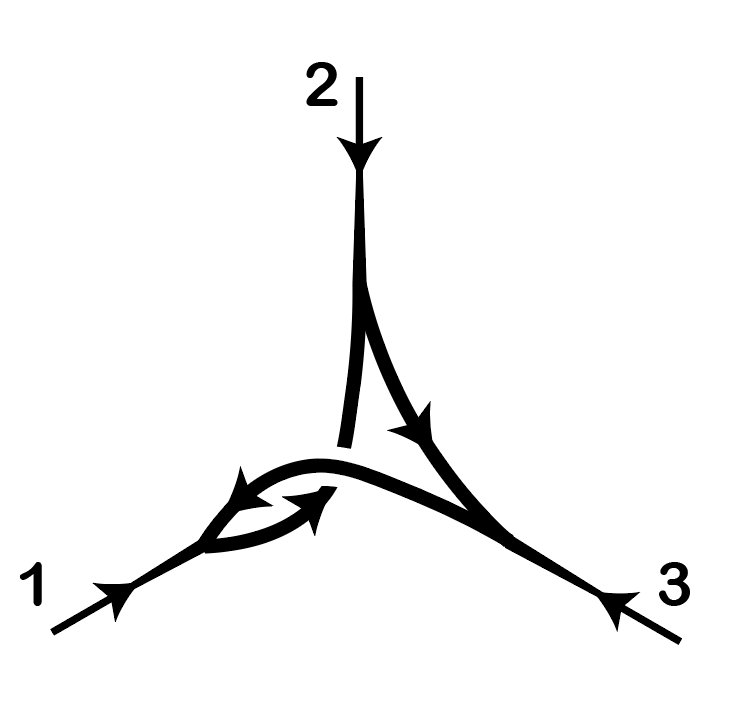}}+\parbox{3.1cm}{\includegraphics[scale=0.24]{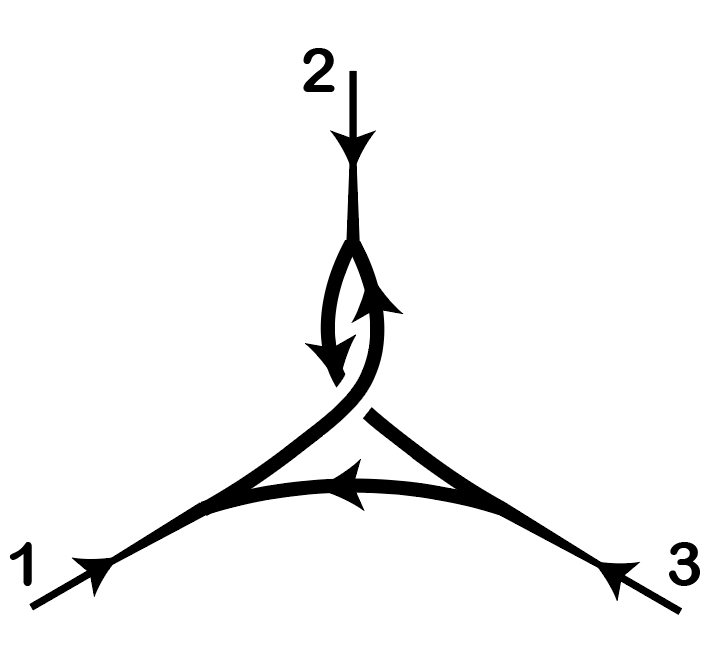}}+\parbox{3.3cm}{\includegraphics[scale=0.24]{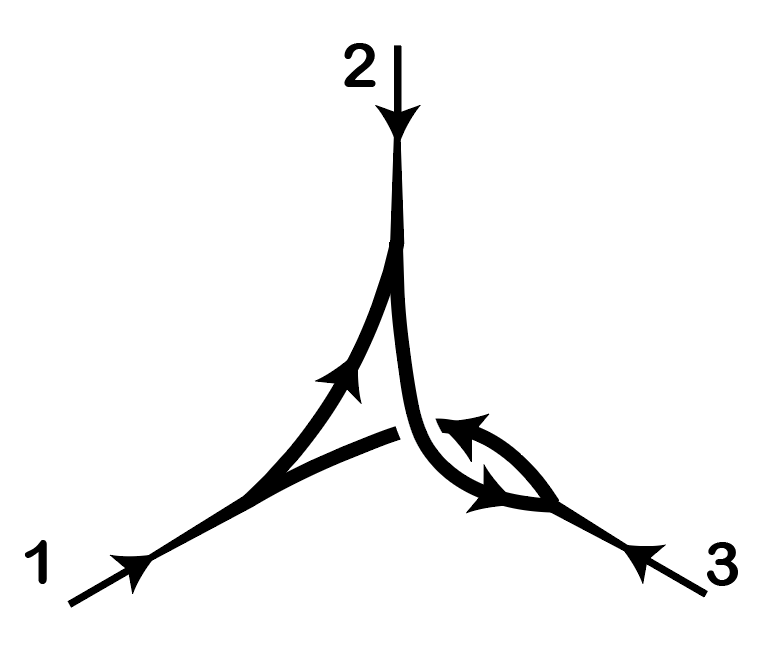}}
\end{align*}
The dual momenta in this case are $\qvec,\kvec_i$ with $i=0,1,2$. The loop momentum can be chosen to be $\pvec=\qvec-\kvec_0$ and the relation between the external momenta and dual regional momenta are $\pvec_i=\kvec_i-\kvec_{i-1}$ with $\kvec_3\equiv \kvec_0$. In other words, with clockwise order $\pvec_i$ is the difference between the outgoing dual momenta and the ingoing dual momenta as depicted in the above figures. We keep the third leg off-shell, i.e $\pvec_3^2\neq 0$, and find the leading contribution to be
\small
\begin{equation}\label{eq:vertexcorrection}
\begin{split}
    \Gamma_{\tri}^{\text{lead}}=\nu_0\,\frac{\Omega_3^{\leading}(l_p \PPb_{12})^{\Lambda_3-3}}{\prod_{i=1}^3\beta_i^{\lambda_i}\Gamma[\Lambda_3-2]}\int \frac{d^4q}{(2\pi)^4} \frac{\PPb_{q-k_0,p_1}(\PPb_{q-k_1,p_2}+\PPb_{12})\PPb_{q-k_2,p_3}}{(\qvec-\kvec_0)^2(\qvec-\kvec_1)^2(\qvec-\kvec_2)^2}\,.
    \end{split}
\end{equation}
\normalsize
The sub-leading terms come with a twist at one of the three vertices and they read
\small
\begin{equation}
\begin{split}
    \Gamma_{\tri}^{\subleading}=&-\mathcal{N}_{\tri}\tr(h_1)\tr(h_2h_3)\int\frac{d^4q}{(2\pi)^4}\frac{\PPb_{q-k_0,p_1}(\PPb_{q-k_1,p_2}+\PPb_{12})\PPb_{q-k_2,p_3}(\PPb_{12}-2\PPb_{q-k_0,p_1})^{\Lambda_3-3}}{(\qvec-\kvec_0)^2(\qvec-\kvec_1)^2(\qvec-\kvec_2)^2} \\
    &-\mathcal{N}_{\tri}\tr(h_2)\tr(h_3h_1)\int\frac{d^4q}{(2\pi)^4}\frac{\PPb_{q-k_0,p_1}(\PPb_{q-k_1,p_2}+\PPb_{12})\PPb_{q-k_2,p_3}(-2\PPb_{q-k_1,p_2}-\PPb_{12})^{\Lambda_3-3}}{(\qvec-\kvec_0)^2(\qvec-\kvec_1)^2(\qvec-\kvec_2)^2}\\
    &-\mathcal{N}_{\tri}\tr(h_3)\tr(h_1h_2)\int\frac{d^4q}{(2\pi)^4}\frac{\PPb_{q-k_0,p_1}(\PPb_{q-k_1,p_2}+\PPb_{12})\PPb_{q-k_2,p_3}(\PPb_{12}-2\PPb_{q-k_2,p_3})^{\Lambda_3-3}}{(\qvec-\kvec_0)^2(\qvec-\kvec_1)^2(\qvec-\kvec_2)^2} 
    \end{split}\notag
\end{equation}
\normalsize
where $\mathcal{N}_{\tri}=\nu_0\,\frac{(l_p)^{\Lambda_3-3}}{\prod_{i=1}^3\beta_i^{\lambda_i}\Gamma[\Lambda_3-2]}$. Next, let us show how to evaluate the integral for the leading contribution. Proceeding as in Section \ref{sec:selfenergy1loop} and Appendix \ref{app:Thornregulator}, we arrive at
\small
\begin{equation}\label{eq:vertexintermediate1}
\begin{split}
   \Gamma_{\tri}^{\leading}=\nu_0\,\frac{\Omega_3^{\leading}\, (l_p \PPb_{12})^{\Lambda_3-3}}{16\pi^2\prod_{i=1}^3\beta_i^{\lambda_i}\Gamma[\Lambda_3-2]}\int \frac{\prod_{i=1}^3dT_i}{T(T+\xi)}e^{-\frac{T_1T_3\pvec_3^2}{T}}\prod_{i=1}^3\Bigg[\frac{T_{i+2}\overline{\mathbb{K}}}{T}-\xi\frac{\beta_i(\sum_{i=1}^3 T_i\bar{k}_{i-1})}{T(T+\xi)}\Bigg]\,,
   \end{split}
\end{equation}
\normalsize
where $\Omega_3^{\leading}=N\tr(h_1h_2h_3)$ and 
\begin{equation}
    \overline{\mathbb{K}}\equiv(\bar{k}_1-\bar{k}_0)\beta_2-(\bar{k}_2-\bar{k}_1)\beta_1=\PPb_{12}\,.
\end{equation}
It is important to note that the integral in (\ref{eq:vertexcorrection}) is finite without the need for the cut-off $\exp[-\xi q^2_{\perp}]$. In (\ref{eq:vertexintermediate1}), we identify $T_4=T_1$ and $T_5=T_2$. Now, it is safe to take $\xi\rightarrow 0$, and we obtain
\begin{equation}\label{eq:vertexfinalresult}
\begin{split}
    \Gamma_{\tri}^{\leading}&=\nu_0\,\frac{\Omega_3^{\leading}\,(l_p)^{\Lambda_3-3}\PPb_{12}^{\Lambda_3}}{16\pi^2\Gamma(\Lambda_3-2)}\int \frac{dT_1dT_2dT_3}{\prod_{i=1}^3\beta_i^{\lambda_i}}\frac{T_1T_2T_3}{T^5}e^{-\frac{T_1T_3\pvec_3^2}{T}}\\
    &=\nu_0\,\frac{\Omega_3^{\leading}\,(l_p)^{\Lambda_3-3}\PPb_{12}^{\Lambda_3}}{16\pi^2\Gamma(\Lambda_3-2)\prod_{i=1}^3\beta_i^{\lambda_i}}\int_{x+y<1}dxdy\int_0^{\infty}dT\, xy(1-x-y)e^{-Tx(1-x-y)\pvec_3^2}\\
    &=\nu_0\,\frac{\Omega_3^{\leading}\,(l_p)^{\Lambda_3-3}\PPb_{12}^{\Lambda_3}}{96\pi^2\prod_{i=1}^{3}\beta_i^{\lambda_i}\Gamma(\Lambda_3-2)\pvec_3^2}\,.
    \end{split}
\end{equation}
To obtain the above result, instead of using dual momenta, one can also start with the original momenta $\pvec_i$. In terms of these variables   the vertex correction reads
\begin{equation}\label{eq:vertexalternative}
    \Gamma_{\tri}^{\leading}=\sum_{\omega}\frac{\Omega_3^{\leading}\,(l_p)^{\Lambda_3-3}\PPb_{12}^{\Lambda_3-3}}{\prod_{i=1}^3\beta_i^{\lambda_i}\Gamma(\Lambda_3-2)}\int \frac{d^4p}{(2\pi)^4}\frac{\PPb_{p1}(\PPb_{p2}+\PPb_{12})\PPb_{p3}}{\pvec^2(\pvec+\pvec_1)^2(\pvec+\pvec_1+\pvec_2)^2}\,.
\end{equation}
Omitting the prefactor and proceeding as before, we find the integral in (\ref{eq:vertexalternative}) to be 
\begin{equation}
\begin{split}
   \frac{\pi}{T(T+\xi)}\prod_{i=1}^3\Bigg[\frac{T_{i+2}\PPb_{12}}{T}-\xi\frac{\beta_i\big[(T_2+T_3)\bar{p}_1+T_3\bar{p}_2\big]}{T(T+\xi)}\Bigg] \xrightarrow{\xi\rightarrow 0}\frac{\pi T_1T_2T_3\PPb_{12}^3}{T^5} \,,
   \end{split}
\end{equation}
which is the same as (\ref{eq:vertexfinalresult}). One can immediately recognize that the final result is reminiscent of the $\Gamma^{+++}$ amplitude for QCD \cite{Thorn:2005ak,Chakrabarti:2005ny,Chakrabarti:2006mb} in the large-$N$ limit. It contains the part of self-dual Yang-Mills theory dressed with the Chiral Theory factor.\footnote{It would be interesting to see if one can apply the hidden self-duality of Chiral Theory \cite{Ponomarev:2017nrr} to simplify the computations in this section.} The overall factor $\nu_0$ makes the vertex correction vanish. 

Although we do not compute the integral for the sub-leading terms of the vertex correction, the following arguments show that these terms are finite. 
Indeed, higher power of $\bar{q}$ entering the Gaussian integral of type (\ref{eq:magicGuassian}) will give zero and improve the UV-behaviour of the integral. The only place where one can potentially get a divergence is the $T$-integral. The $T$-integral will have the form
\begin{equation}
    \int_0^{\infty}dT \frac{\xi^a}{(T+\xi)^b}\,.
\end{equation}
It will pick up poles of the form $1/\xi^{b-a-1}$ whenever $b\geq a+2$. However, due to simple power counting and the magic of the holomorphic integral (\ref{eq:magicGuassian}), we shall find convergent integrals. The $\nu_0$-factor will make all of the sub-leading terms vanish due to the zeta-function regularization. 

%%%%%%%%%%%%%%%%%%%%%%%%%%%%%%%%%%%%%%%%%%%%%%%%%%%%%%%%%%
\subsection{Four-point amplitude}
\label{sec:boxandtri}
%%%%%%%%%%%%%%%%%%%%%%%%%%%%%%%%%%%%%%%%%%%%%%%%%%%%%%%%%%
Next, we consider the one-loop diagram with four external legs in the large-$N$ limit. The large-$N$ limit simplifies computations as we do not need to consider contributions coming from non-planar diagrams. Let us take a look at the relevant one-loop diagrams and prove that they are UV-finite. 
\paragraph{Box and triangle-like diagrams.} We take first the vertex insertions into the four-point function and choose (1234) color order as an example:
\small
\begin{equation}\notag
    \begin{split}
        \parbox{1.6cm}{\includegraphics[scale=0.21]{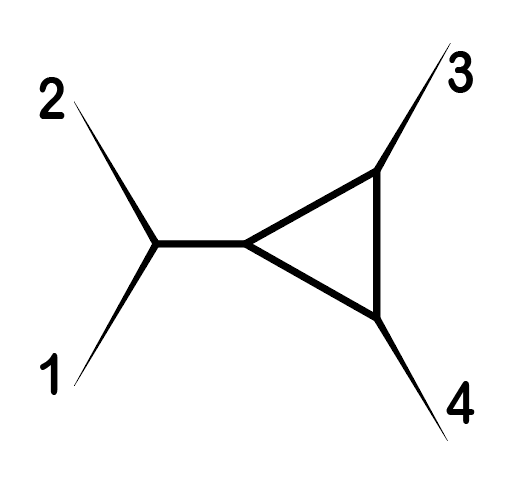}}=\Gamma_{\Delta}(1234)&=\frac{\nu_0\, (l_p)^{\Lambda_4-4}\alpha_4^{\Lambda_4-4}}{\Gamma(\Lambda_4-3)\prod_{i=1}^4\beta_i^{\lambda_i}}\frac{\PPb_{12}}{s_{12}}\int \frac{d^4p}{(2\pi)^4}\frac{(\PPb_{p1}+\PPb_{p2})(\PPb_{p3}+\PPb_{13}+\PPb_{23})\PPb_{p4}}{\pvec^2(\pvec+\pvec_1+\pvec_2)^2(\pvec+\pvec_1+\pvec_2+\pvec_3)^2}\\
        &=\frac{\nu_0\,(l_p)^{\Lambda_4-4}\alpha_4^{\Lambda_4-4}}{\Gamma(\Lambda_4-3)\prod_{i=1}^4\beta_i^{\lambda_i}}\frac{\PPb_{12}\PPb_{34}^3}{96\pi^2 s_{34}^2}\\
        &=-\frac{\nu_0\, (l_p)^{\Lambda_4-4}\alpha_4^{\Lambda_4-4}}{\Gamma(\Lambda_4-3)\prod_{i=1}^4\beta_i^{\lambda_i}}\frac{\PPb_{34}^2\PPb_{41}\PPb_{23}}{96\pi^2 s_{12}s_{23}}\,.
    \end{split}
\end{equation}
\normalsize
Similarly,
\besubeqs
\begin{align}
    \Gamma_{\Delta}(2341)&=\frac{\nu_0\,(l_p)^{\Lambda_4-4}\alpha_4^{\Lambda_4-4}}{\Gamma(\Lambda_4-3)\prod_{i=1}^4\beta_i^{\lambda_i}}\frac{\PPb_{23}\PPb_{41}^3}{96\pi^2 s_{23}^2}=-\frac{\nu_0\,(l_p)^{\Lambda_4-4}\alpha_4^{\Lambda_4-4}}{\Gamma(\Lambda_4-3)\prod_{i=1}^4\beta_i^{\lambda_i}}\frac{\PPb_{41}^2\PPb_{12}\PPb_{34}}{96\pi^2 s_{12}s_{23}}\,,\\
    \Gamma_{\Delta}(3412)&=\frac{\nu_0\,(l_p)^{\Lambda_4-4}\alpha_4^{\Lambda_4-4}}{\Gamma(\Lambda_4-3)\prod_{i=1}^4\beta_i^{\lambda_i}}\frac{\PPb_{34}\PPb_{12}^3}{96\pi^2 s_{34}^2}=-\frac{\nu_0\,(l_p)^{\Lambda_4-4}\alpha_4^{\Lambda_4-4}}{\Gamma(\Lambda_4-3)\prod_{i=1}^4\beta_i^{\lambda_i}}\frac{\PPb_{12}^2\PPb_{23}\PPb_{41}}{96\pi^2 s_{12}s_{23}}\,,\\
    \Gamma_{\Delta}(4123)&=\frac{\nu_0\,(l_p)^{\Lambda_4-4}\alpha_4^{\Lambda_4-4}}{\Gamma(\Lambda_4-3)\prod_{i=1}^4\beta_i^{\lambda_i}}\frac{\PPb_{41}\PPb_{23}^3}{96\pi^2 s_{41}^2}=-\frac{\nu_0\,(l_p)^{\Lambda_4-4}\alpha_4^{\Lambda_4-4}}{\Gamma(\Lambda_4-3)\prod_{i=1}^4\beta_i^{\lambda_i}}\frac{\PPb_{23}^2\PPb_{34}\PPb_{12}}{96\pi^2 s_{12}s_{23}}\,.
\end{align}
\esubeqs
As it was discussed in \cite{Chakrabarti:2005ny,Chakrabarti:2006mb}
for the QCD case, 
one can reduce the more complicated box integral to the triangle-like integral. We would like to see if this can be done for the higher spin case. The box contribution reads
\begin{equation}
\begin{split}
    \Gamma_{\square}=\sum_{\omega}\frac{\nu_0 \,(l_p)^{\Lambda_4-4}\alpha_4^{\Lambda_4-4}}{\prod_{i=1}^4\beta_i^{\lambda_i}\Gamma(\Lambda_4-3)}\int \frac{d^4p}{(2\pi)^4}\frac{\PPb_{p1}(\PPb_{p2}+\PPb_{12})(\PPb_{p3}+\PPb_{34})\PPb_{p4}}{\pvec^2(\pvec+\pvec_1)^2(\pvec+\pvec_1+\pvec_2)^2(\pvec-\pvec_4)^2}\,.
\end{split}
\end{equation}
Since $\pvec$ is off-shell, we can use the following identity:
\begin{equation}\label{eq:cuttrick}
    \PPb_{pi}\PP_{pi}=-\frac{\beta_i\beta}{2}(\pvec+\pvec_i)^2+\frac{\beta_i(\beta_i+\beta)\pvec^2}{2}
\end{equation} 
to arrive at 
\begin{equation}
    \frac{\PPb_{p1}}{(\pvec+\pvec_1)^2}=-\frac{\beta\beta_1}{2\PP_{p1}}+\frac{\beta_1(\beta_1+\beta)\pvec^2}{2\PP_{p1}(\pvec+\pvec_1)^2}, \qquad \frac{\PPb_{p4}}{(\pvec-\pvec_4)^2}=\frac{\beta\beta_4}{2\PP_{p4}}-\frac{\beta_4(\beta-\beta_4)\pvec^2}{2\PP_{p4}(\pvec-\pvec_4)^2}\,.
\end{equation}
We can reduce the box integral to a triangle-like integral by canceling out one propagator in the denominator using (\ref{eq:cuttrick}). Next, we multiply $\Gamma_{\square}$ by two for a moment, then
\small
\begin{equation}
\begin{split}
 2\Gamma_{\square}=&\frac{\nu_0\,(l_p)^{\Lambda_4-4}\alpha_4^{\Lambda_4-4}}{\prod_{i=1}^4\beta_i^{\lambda_i}\Gamma(\Lambda_4-3)}\int\frac{d^4p}{(2\pi)^4}\Bigg[\frac{(\PPb_{p2}+\PPb_{12})(\PPb_{p3}+\PPb_{34})}{\pvec^2(\pvec+\pvec_1+\pvec_2)^2}\Bigg( \frac{\beta\beta_4\PPb_{p1}}{2\PP_{p4}(\pvec+\pvec_1)^2} -\frac{\beta\beta_1\PPb_{p4}}{2\PP_{p1}(\pvec-\pvec_4)^2}\Bigg)\\
&+\frac{(\PPb_{p2}+\PPb_{12})(\PPb_{p3}+\PP_{34})}{(\pvec+\pvec_1)^2(\pvec+\pvec_1+\pvec_2)^2(\pvec-\pvec_4)^2}\Bigg(\frac{\beta_1(\beta+\beta_1)\PPb_{p4}}{2\PP_{p1}}-\frac{\beta_4(\beta-\beta_4)\PPb_{p1}}{2\PP_{p4}}\Bigg)\Bigg]\,.
 \end{split}\notag
\end{equation}
\normalsize
Using Bianchi-like identity $\beta_{[i}\PPb_{jk]}=0$, we find 
\begin{equation}
    \frac{\beta\beta_4}{2\PP_{p4}}=\frac{\PPb_{41}}{s_{41}}+\frac{\beta_4^2\PP_{p1}}{2\PP_{p4}\PP_{41}}, \qquad\quad -\frac{\beta\beta_1}{2\PP_{p1}}=\frac{\PPb_{41}}{s_{41}}+\frac{\beta_1^2\PP_{p4}}{2\PP_{p1}\PP_{41}}\,.
\end{equation}
Then, after some straightforward algebra, the box integral becomes
\small
\begin{equation}\notag
    \begin{split}
    2\Gamma_{\square}=\nu_0\,\mathcal{N}_{\square}\frac{\PPb_{41}}{s_{41}}\int \frac{d^4p}{(2\pi)^4}\frac{(\PPb_{p2}+\PPb_{12})(\PPb_{p3}+\PPb_{34})}{\pvec^2(\pvec+\pvec_1+\pvec_2)^2}\Bigg[\frac{\PPb_{p1}}{(\pvec+\pvec_1)^2}+\frac{\PPb_{p4}}{(\pvec-\pvec_4)^2}-\frac{(\PPb_{p4}+\PPb_{p1}-\PPb_{41})\pvec^2}{(\pvec+\pvec_1)^2(\pvec-\pvec_4)^2}\Bigg]+\Gamma_{\square}\,,
\end{split}
\end{equation}
\normalsize
where $\mathcal{N}_{\square}=\frac{(l_p)^{\Lambda_4-4}\alpha_4^{\Lambda_4-4}}{\prod_{i=1}^4\beta_i^{\lambda_i}\Gamma(\Lambda_4-3)}$. Hence, 
\small
\begin{equation}\label{eq:box expression}
    \begin{split}
    \Gamma_{\square}&=\nu_0\,\mathcal{N}_{\square}\frac{\PPb_{41}}{s_{41}}\int \frac{d^4p}{(2\pi)^4}\frac{(\PPb_{p2}+\PPb_{12})(\PPb_{p3}+\PPb_{34})}{\pvec^2(\pvec+\pvec_1+\pvec_2)^2}\Bigg[\frac{\PPb_{p1}}{(\pvec+\pvec_1)^2}+\frac{\PPb_{p4}}{(\pvec-\pvec_4)^2}-\frac{(\PPb_{p4}+\PPb_{p1}-\PPb_{41})\pvec^2}{(\pvec+\pvec_1)^2(\pvec-\pvec_4)^2}\Bigg]\\
    &=\nu_0\,\mathcal{N}_{\square}\left[\frac{\PPb_{41}\big[\,\PPb_{12}^2(\PPb_{23}+\PPb_{34})+(\PPb_{12}+\PPb_{23})\PPb_{34}^2\big]}{96\pi^2 s_{12}s_{23}}+\frac{\PPb_{41}\PPb_{23}^3}{96\pi^2s_{41}^2}\right]\,.
    \end{split}
\end{equation}
\normalsize
The last term in (\ref{eq:box expression}) cancels against the triangle $\Gamma_{\Delta}(4123)$ diagram. Finally, we obtain 
\begin{equation}
\begin{split}
   \parbox{2cm}{\includegraphics[scale=0.13]{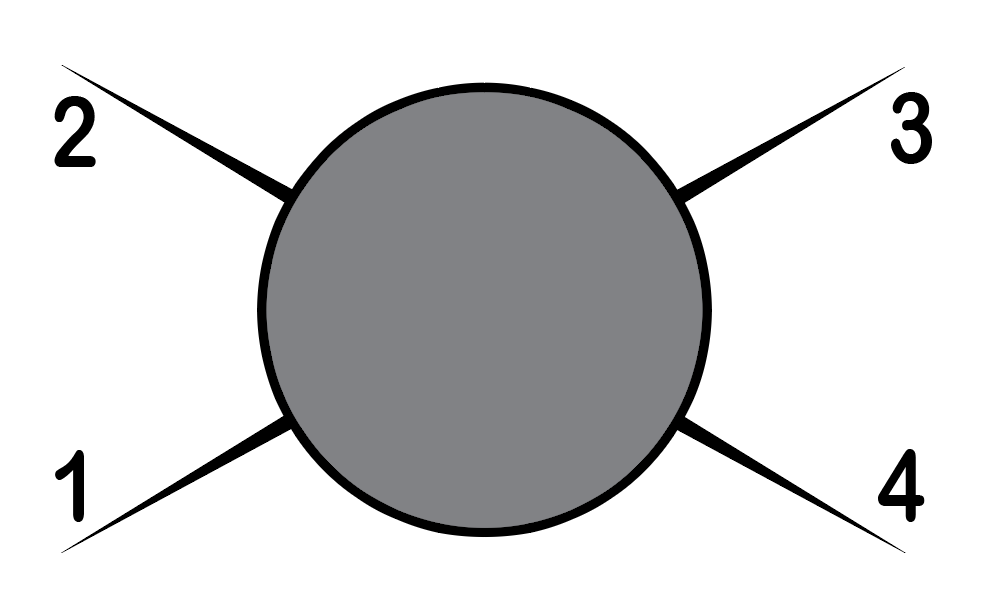}}= \Gamma_4&=\Gamma_{\square}+\big[\Gamma_{\Delta}(1234)+\text{cycl.}\big]\\
   &=\nu_0\,\frac{\mathcal{N}_{\square}}{96\pi^2}\frac{\PPb_{12}\PPb_{34}\PPb_{41}(\PPb_{12}+\PPb_{34}-\PPb_{41})}{s_{12}s_{23}}\\
   &=\nu_0\,\frac{\mathcal{N}_{\square}}{96\pi^2}\frac{\PPb_{12}\PPb_{23}\PPb_{34}\PPb_{41}}{s_{12}s_{23}}\,,
   \end{split}
\end{equation}
which is similar to the QCD result for $\Gamma_4^{++++}$ amplitude \cite{Chakrabarti:2005ny}, see also \cite{Brandhuber:2007vm}.

%%%%%%%%%%%%%%%%%%%%%%%%%%%%%%%%%%%%%%%%%%%%%%%%%%%%%%%%%%
\paragraph{The bubbles.}
%\label{sec:leggedbubbles}
%%%%%%%%%%%%%%%%%%%%%%%%%%%%%%%%%%%%%%%%%%%%%%%%%%%%%%%%%%
As discussed in \cite{Chakrabarti:2005ny}, 
%Zvi Bern observed that
the sum over bubbles, triangle like and box diagrams should add up to zero in the case of all-plus 4--point one-loop amplitude for QCD.
We would like to see whether Chiral Theory 
has a similar property. First, let us look at the bubble insertions into the internal propagator, which come in two channels, $s$ and $t$, for $U(N)$ factors:
\begin{figure}[h]
    \centering
    \includegraphics[scale=0.35]{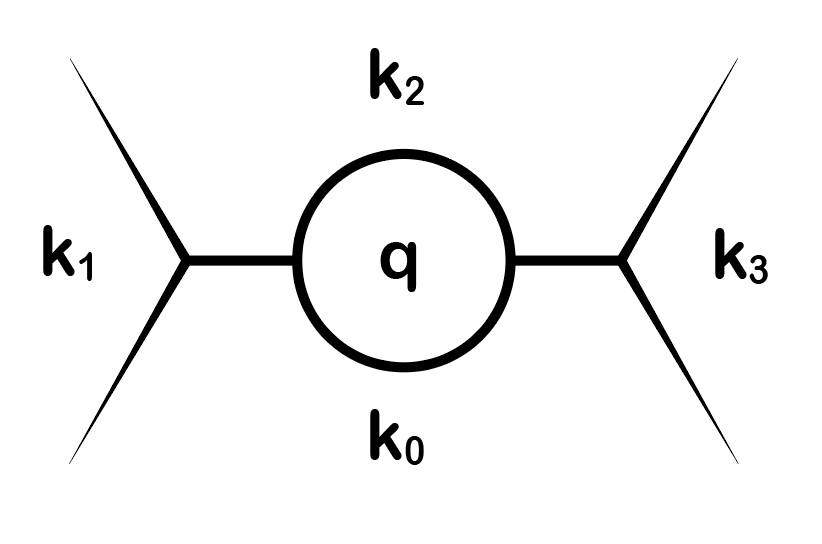}
\end{figure}\\
Here, we divided the space
of dual momenta $\kvec_i$
into four regions. The external momenta $\pvec_i$ can be read off by using two adjacent regional dual momenta. For example, $\pvec_1=\kvec_1-\kvec_0$, $\pvec_2=\kvec_2-\kvec_1$, etc. In general, whenever we have a closed loop, we can 'put' the dual momentum $\qvec$ inside it and the loop momentum can be obtained as the difference between $\qvec$ and the nearest dual regional momentum. In the above figure, $\pvec=\qvec-\kvec_0$. Now, it is a matter of a direct calculation to show the 'internal' self-energy diagram with the four external legs labeled in clockwise order to be 
\begin{equation}
\begin{split}
    \Gamma^{\text{in}}_{\bigcirc}(1234)&=\sum_{\omega}\frac{(l_p)^{\Lambda_4-4}\alpha_4^{\Lambda_4-4}}{\prod_{i=1}^4\beta_i^{\lambda_i}\Gamma(\Lambda_4-3)}\frac{\PPb_{12}\PPb_{34}(\beta_1+\beta_2)(\beta_3+\beta_4)(\bar{k}_0^2+\bar{k}_0\bar{k}_2+\bar{k}_2^2)}{96\pi^2 s_{12}^2}\\
    &=-\sum_{\omega}\frac{(l_p)^{\Lambda_4-4}\alpha_4^{\Lambda_4-4}}{\prod_{i=1}^4\beta_i^{\lambda_i}\Gamma(\Lambda_4-3)}\frac{\PPb_{41}\PPb_{23}(\beta_1+\beta_2)(\beta_3+\beta_4)
    (\bar{k}_0^2+\bar{k}_0\bar{k}_2+\bar{k}_2^2)}{96\pi^2 s_{12}s_{23}}\,.
    \end{split}
\end{equation}
Similarly,
\begin{equation}
\begin{split}
    \Gamma^{\text{in}}_{\bigcirc}(2341)&=\sum_{\omega}\frac{(l_p)^{\Lambda_4-4}\alpha_4^{\Lambda_4-4}}{\prod_{i=1}^4\beta_i^{\lambda_i}\Gamma(\Lambda_4-3)}\frac{\PPb_{23}\PPb_{41}(\beta_2+\beta_3)(\beta_4+\beta_1)(\bar{k}_1^2+\bar{k}_1\bar{k}_3+\bar{k}_3^2)}{96\pi^2 s_{23}^2} \\
    &=- \sum_{\omega}\frac{(l_p)^{\Lambda_4-4}\alpha_4^{\Lambda_4-4}}{\prod_{i=1}^4\beta_i^{\lambda_i}\Gamma(\Lambda_4-3)}\frac{\PPb_{12}\PPb_{34}(\beta_2+\beta_3)(\beta_4+\beta_1)
    (\bar{k}_1^2+\bar{k}_1\bar{k}_3+\bar{k}_3^2)}{96\pi^2 s_{12}s_{23}}\,.
    \end{split}
\end{equation}
Next, we move to the graphs where we have vacuum bubbles on the external legs. In this case, we have in total eight diagrams. Take the following diagram as an example \\
\begin{figure}[h]
    \centering
    \includegraphics[scale=0.35]{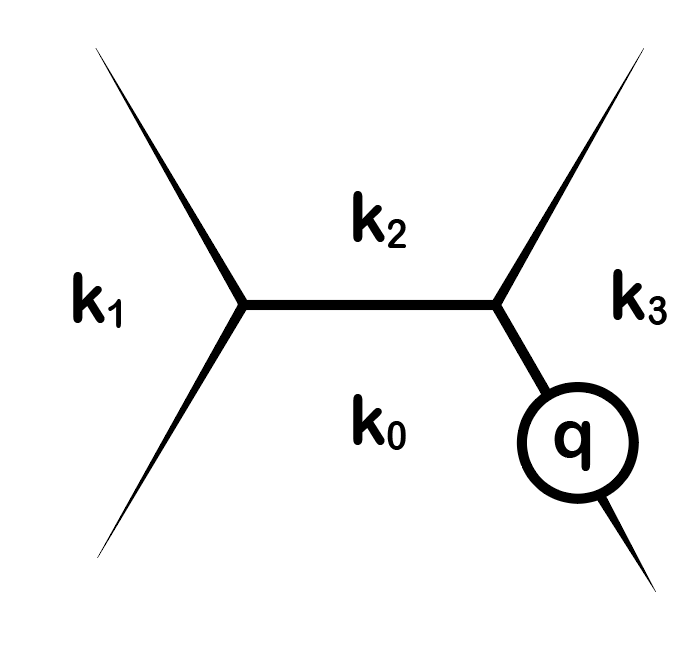}
\end{figure}\\
Here, the loop momentum is $\pvec=\qvec-\kvec_0$ and external momenta remain to be the same as $\pvec_i=\kvec_i-\kvec_{i-1}$. We denote the result of the bubble insertion into the $i$-th leg as $\Gamma_{\bigcirc}^{i}$. It reads (remember that we have two different channels for each diagram due to the color ordering)
{\allowdisplaybreaks
\besubeqs
\begin{align}
    \Gamma_{\bigcirc}^1&=-\sum_{\omega}\frac{(l_p)^{\Lambda_4-4}\alpha_4^{\Lambda_4-4}}{\prod_{i=1}^4\beta_i^{\lambda_i}\Gamma(\Lambda_4-3)}\frac{\PPb_{23}\PPb_{34}\beta_1^2(\bar{k}_0^2+\bar{k}_0\bar{k}_1+\bar{k}_1^2)}{96\pi^2 s_{12}s_{23}}\,,\\
    \Gamma_{\bigcirc}^2&=-\sum_{\omega}\frac{(l_p)^{\Lambda_4-4}\alpha_4^{\Lambda_4-4}}{\prod_{i=1}^4\beta_i^{\lambda_i}\Gamma(\Lambda_4-3)}\frac{\PPb_{34}\PPb_{41}\beta_2^2(\bar{k}_1^2+\bar{k}_1\bar{k}_2+\bar{k}_2^2)}{96\pi^2 s_{12}s_{23}}\,,\\
    \Gamma_{\bigcirc}^3&=-\sum_{\omega}\frac{(l_p)^{\Lambda_4-4}\alpha_4^{\Lambda_4-4}}{\prod_{i=1}^4\beta_i^{\lambda_i}\Gamma(\Lambda_4-3)}\frac{\PPb_{41}\PPb_{12}\beta_3^2(\bar{k}_2^2+\bar{k}_2\bar{k}_3+\bar{k}_3^2)}{96\pi^2 s_{12}s_{23}}\,,\\
    \Gamma_{\bigcirc}^4&=-\sum_{\omega}\frac{(l_p)^{\Lambda_4-4}\alpha_4^{\Lambda_4-4}}{\prod_{i=1}^4\beta_i^{\lambda_i}\Gamma(\Lambda_4-3)}\frac{\PPb_{12}\PPb_{23}\beta_4^2(\bar{k}_3^2+\bar{k}_3\bar{k}_0+\bar{k}_0^2)}{96\pi^2 s_{12}s_{23}}\,.
\end{align}
\esubeqs}%
\noindent Equivalently, we can write them as
\small
\begin{align*}
        \Gamma^1_{\bigcirc}&=-\nu_0\mathcal{N}_{\square}\frac{\PPb_{23}\PPb_{34}(\beta_1\beta_3\PPb_{41}\PPb_{12}+\beta_1(\beta_1+\beta_4)\PPb_{12}\PPb_{34}+\beta_1(\beta_1+\beta_2)\PPb_{23}\PPb_{41})(\bar{k}_0^2+\bar{k}_0\bar{k}_1+\bar{k}_1^2)}{96\pi^2 s_{12}s_{23}}\,,\\
        \Gamma_{\bigcirc}^2&=-\nu_0\mathcal{N}_{\square}\frac{(\beta_2\beta_3\PPb_{41}\PPb_{12}+\beta_2(\beta_1+\beta_2)\PPb_{23}\PPb_{41})(\bar{k}_1^2+\bar{k}_1\bar{k}_2+\bar{k}_2^2)}{96\pi^2 s_{12}s_{23}}\,,\\
        \Gamma_{\bigcirc}^3&=-\nu_0\mathcal{N}_{\square}\frac{\PPb_{41}\PPb_{12}\beta_3^2(\bar{k}_2^2+\bar{k}_2\bar{k}_3+\bar{k}_3^2)}{96\pi^2 s_{12}s_{23}}\,,\\
        \Gamma_{\bigcirc}^4&=-\nu_0\mathcal{N}_{\square}\frac{(\beta_3\beta_4\PPb_{41}\PPb_{12}+\beta_4(\beta_1+\beta_4)\PPb_{12}\PPb_{34})(\bar{k}_3^2+\bar{k}_3\bar{k}_0+\bar{k}_0^2)}{96\pi^2 s_{12}s_{23}}\,.
\end{align*}
\normalsize
Collecting the results and remembering that $\pvec_i=\kvec_i-\kvec_{i-1}$, we obtain 
\begin{equation}
    \Gamma_{\text{bubbles}}=\sum_{i=1}^4\Gamma^i_{\bigcirc}+2\Gamma^{\text{in}}_{\bigcirc}=-\nu_0\,\frac{(l_p)^{\Lambda_4-4}\alpha_4^{\Lambda_4-4}}{96\pi^2\Gamma(\Lambda_4-3)\prod_{i=1}^{\lambda_i}\beta_i^{\lambda_i}}\frac{\PPb_{12}\PPb_{23}\PPb_{34}\PPb_{41}}{s_{12}s_{23}}\,.
\end{equation}
Finally, we proved the higher spin analog of the QCD relation:
\begin{equation}
    \Gamma_4=\parbox{2.5cm}{\includegraphics[scale=0.13]{newpics/4ptfull.png}}+2\times \parbox{2.4cm}{\includegraphics[scale=0.2]{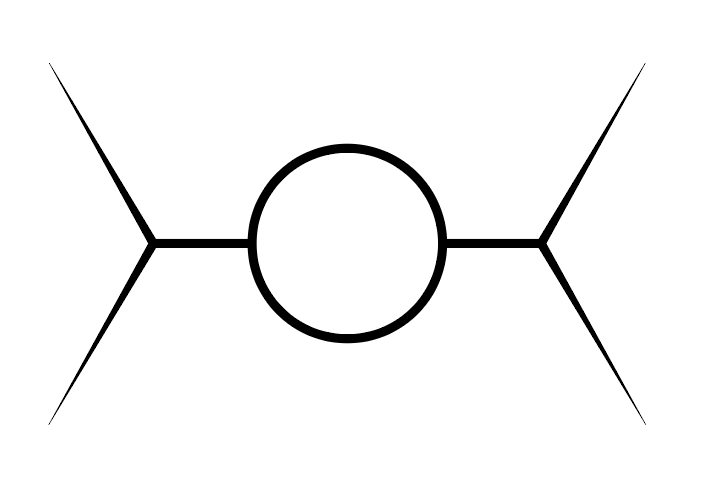}}+8\times \parbox{2cm}{\includegraphics[scale=0.18]{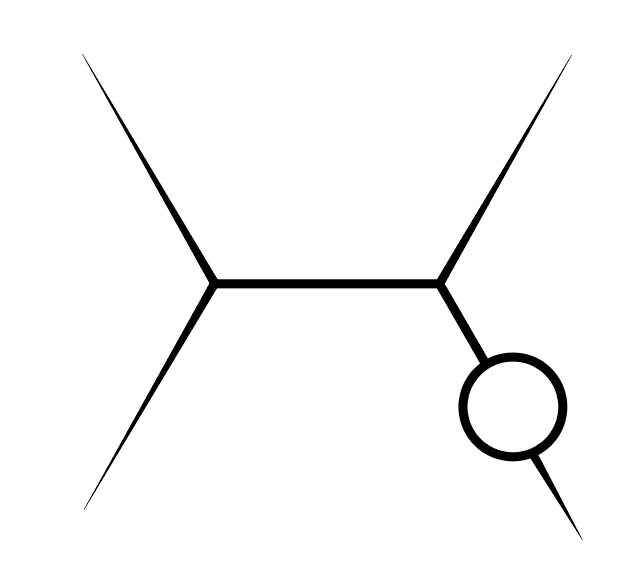}}=0\,.
\end{equation}
Therefore, the $4$-point function at one loop does not have any UV-divergences since it can be reduced to UV-convergent integrals we have already analyzed. The complete $4$-point amplitude vanishes due to the same $\nu_0$-factor. 

%%%%%%%%%%%%%%%%%%%%%%%%%%%%%%%%%%%%%%%%%%%%%%%%%%%%%%%%%%
\subsection{Sun Diagrams and Multiloop Amplitudes}
\label{sec:multiloop}
%%%%%%%%%%%%%%%%%%%%%%%%%%%%%%%%%%%%%%%%%%%%%%%%%%%%%%%%%%
For multiloop amplitudes in the large-$N$ limit, one can start with the sun-like diagrams that have some of the legs off-shell and then glue them together.
\begin{figure}[h]
    \centering
    \includegraphics[scale=0.3]{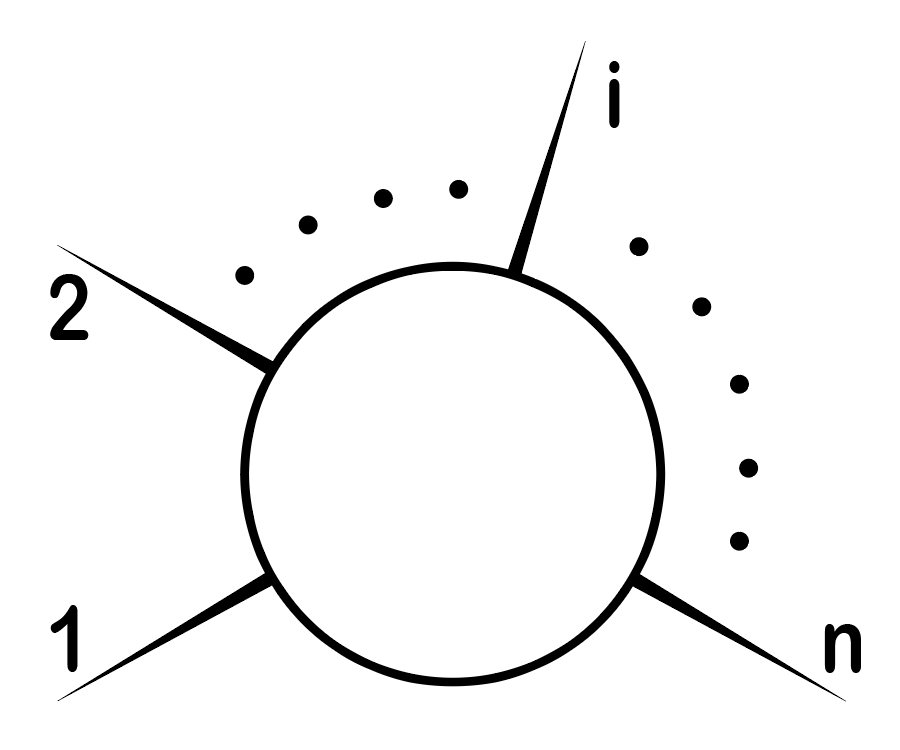}
\end{figure}
The kinematic part of the sun-like diagrams can be simply written as (for the moment we omit the overall $\beta_i^{\lambda_i}$-factors)
\small
\begin{equation}
    \sum_{\{\omega_i\}}\frac{\PPb_{\pvec_1,\pvec,-\pvec-\pvec_1}^{\lambda_1+\omega_1-\omega_n}}{\Gamma(\lambda_1+\omega_1-\omega_n)}\frac{\PPb_{\pvec_2,\pvec+\pvec_1,-\pvec-\pvec_1-\pvec_2}^{\lambda_2-\omega_1+\omega_2}}{\Gamma(\lambda_2-\omega_1+\omega_2)}\cdots \frac{\PPb_{\pvec_n,\pvec-\pvec_n,-\pvec}^{\lambda_n-\omega_{n-1}+\omega_n}}{\Gamma(\lambda_n-\omega_{n-1}+\omega_n)}=\sum_{\omega_n}\frac{\alpha_n^{\Lambda_n-n}\mathcal{K}_n}{\Gamma(\Lambda_n-(n-1))}\,,
\end{equation}
\normalsize
where $i=1,...n$ and $\mathcal{K}_n$ is the kinematic part that contains $\PPb_{pi},\PPb_{ij}$. Putting the propagator and coupling constant together, one gets the following general form for the one-loop diagram with $n$-external legs, some of which can be off-shell,
\begin{equation}
    \Gamma_{n}=\nu_0\,\frac{(l_p)^{\Lambda_n-n}\alpha_n^{\Lambda_n-n}}{\Gamma(\Lambda_n-(n-1))\prod_{i=1}^n\beta_i^{\lambda_i}}\int\frac{d^4p}{(2\pi)^4}\frac{\mathcal{K}_n(\PPb)}{\pvec^2(\pvec+\pvec_1)^2 ... (\pvec-\pvec_n)^2}\,,
\end{equation}
with certain numerator $\mathcal{K}_n(\PPb)$. The
factorization of the sum over helicities $\nu_0$ is crucial to make the contribution vanish even though we do not evaluate the integral explicitly. It should not be hard to show that it is UV-finite. Consequently, all multiloop amplitudes should vanish confirming that $S=1$.

%%%%%%%%%%%%%%%%%%%%%%%%%%%%%%%%%%%%%%%%%%%%%%%%%%%%%%%%%%
\section{Conclusions and Discussion}
\label{sec:conclusions}
%%%%%%%%%%%%%%%%%%%%%%%%%%%%%%%%%%%%%%%%%%%%%%%%%%%%%%%%%%
The results of the present paper strengthen those of \cite{Skvortsov:2018jea} and provide further details. Chiral Theory reveals a remarkable cancellation of UV-divergences and should be an example of a quantum consistent Higher Spin Gravity, which is the very first and the only higher spin model with propagating massless higher spin fields at present where quantum corrections can be computed. 

The tree level amplitudes can be shown to vanish on-shell, which is a result of highly nontrivial cancellations in the total sum over Feynman diagrams. This is required by the Weinberg low energy theorem. The chirality of interactions restricts all spin sums on the internal lines in such a way that they are always over a finite range (assuming the external helicities are fixed). In generic higher spin theories we would expect an infinite sum over all spins already for tree level diagrams. This does not happen for Chiral Theory and infinite spin sums show up only at the loop level.

The loop diagrams that we have analyzed turn out to consist of two factors: the UV-convergent integral and a purely numerical factor $\nu_0=\sum_\lambda 1$. The UV-convergence is a very important property that again relies on the presence of higher spin fields. This effect is reminiscent of $\mathcal{N}=4$ Yang-Mills Theory \cite{Mandelstam:1982cb,Brink:1982wv}, in which the supersymmetry forces one momentum to eventually factor out and makes the integrals convergent. Higher spin symmetry amplifies this effect. Chiral Theory has infinitely many non-renormalizable interactions, which include the two-derivative graviton self-coupling. Higher spin symmetry forces enough momenta to factor out in every loop integral and makes all loop integrals free of UV-divergences. Overall factor $\nu_0$ is to be expected in any theory with infinitely many fields and some value needs to be assigned to the sum. It is natural to set $\nu_0=0$, which is achieved via the zeta-function regularization. Such an assignment is consistent both with the Weinberg theorem and with the large web of results on one-loop determinants in holographic higher spin theories. 

As a result, we see that $S=1$ for Minkowski Chiral Theory, as expected. However, once the cosmological constant is turned on the holographic $S$-matrix turns out to be nontrivial \cite{Skvortsov:2018uru}. Therefore, we consider Minkowski Chiral Theory as a useful toy model to check the cancellation of UV-divergences thanks to higher spin symmetry. It is exactly the effect that Higher Spin Gravities have long been expected to have. 

The class of Chiral Higher Spin Gravities has been extended to incorporate Yang-Mills gaugings. Even though we do not see any immediate relation to string theory, it is quite surprising that higher spin fields can be made charged with respect to the spin-one field via the method that is very similar to the Chan-Paton approach. 

Higher spin symmetry seems to be powerful enough as to make graviton be part of a quantum consistent theory. Nevertheless, it should be possible to combine higher spin symmetry with supersymmetry and construct supersymmetric Chiral Theories \cite{Metsaev:2019dqt,Metsaev:2019aig}.\footnote{See also e.g \cite{Hutomo:2017nce,Buchbinder:2017nuc,Buchbinder:2018nkp} for a recent progress for interacting supersymmetric higher spin theories on flat and AdS spaces.}

Chiral Theory is the only class at present with propagating massless higher spin fields and an action. Nevertheless, there is a handful of other higher spin models with action that are of great interest.\footnote{We do not discuss Formal Higher Spin Gravities, i.e. formally consistent equations of motion, which are available for many cases 
\cite{Vasiliev:1990en,Vasiliev:2003ev,Sagnotti:2005ns,Bekaert:2013zya,Bonezzi:2016ttk,Bekaert:2017bpy,Grigoriev:2018wrx,Sharapov:2019vyd}. The general solution of this problem is given in \cite{Sharapov:2019vyd}. Other interesting proposals include \cite{deMelloKoch:2018ivk,Sperling:2017dts}. } There are topological theories in three dimension: purely massless \cite{Blencowe:1988gj,Bergshoeff:1989ns,Campoleoni:2010zq,Henneaux:2010xg} and conformal \cite{Pope:1989vj,Fradkin:1989xt,Grigoriev:2019xmp}. Another class is $4d$ conformal higher spin gravity \cite{Segal:2002gd,Tseytlin:2002gz,Bekaert:2010ky}, which is an extension of conformal gravity. There also has been some progress in two dimensions \cite{Alkalaev:2020kut}. Topological models are, of course, free of UV-divergences. There are encouraging results on quantum checks for conformal higher spin gravity \cite{Beccaria:2016syk,Joung:2015eny} that indicate that the conformal higher spin symmetry also makes $S$-matrix trivial in flat space. The $2d$-models of \cite{Alkalaev:2020kut} involve propagating matter fields with interactions mediated via topological higher spin fields, thereby providing interesting toy models for quantum checks. Lastly, it would be very important to directly verify that $AdS_4$ Chiral Theory is free of UV-divergences.

%%%%%%%%%%%%%%%%%%%%%%%%%%%%%%%%%%%%%%%%%%%%%%%%%%%%%%%%%%
\section*{Acknowledgments}
\label{sec:Aknowledgements}
%%%%%%%%%%%%%%%%%%%%%%%%%%%%%%%%%%%%%%%%%%%%%%%%%%%%%%%%%%
We would like to thank Sudarshan Ananth, Nicolas Boulanger, Andrea Compoleoni, Dario Francia, Gregory Korchemsky, Kirill Krasnov, Ruslan Metsaev, Julian Miczajka, Yasha Neiman, Jan Plefka, Dmitry Ponomarev, Radu Roiban, Adam Schwimmer, Stefan Theisen, Arkady Tseytlin and Edward Witten for useful discussions and comments.  The work of E.S. was supported by the Russian Science Foundation grant 18-72-10123 in association with the Lebedev Physical Institute. The work of T.T. is supported by the International Max Planck Research School for Mathematical and Physical Aspects of Gravitation, Cosmology and Quantum Field Theory. The work of M.T. was supported by the Quantum Gravity Unit of the Okinawa Institute of Science and Technology Graduate University (OIST). M.T.  would like to thank the Department of Mathematics, the University of Auckland for their kind hospitality during the last stage of the project.

%%%%%%%%%%%%%%%%%%%%%%%%%%%%%%%%%%%%%%%%%%%%%%%%%%%%%%%%%%
\begin{appendix}
\renewcommand{\thesection}{\Alph{section}}
\renewcommand{\theequation}{\Alph{section}.\arabic{equation}}
\setcounter{equation}{0}\setcounter{section}{0}
%%%%%%%%%%%%%%%%%%%%%%%%%%%%%%%%%%%%%%%%%%%%%%%%%%%%%%%%%%

%%%%%%%%%%%%%%%%%%%%%%%%%%%%%%%%%%%%%%%%%%%%%%%%%%%%%%%%%%
\section{Crash Course on Light Front Approach}
\label{app:LC}
\setcounter{equation}{0}
%%%%%%%%%%%%%%%%%%%%%%%%%%%%%%%%%%%%%%%%%%%%%%%%%%%%%%%%%%
The main idea of the light cone approach is that any classical or quantum field theory in flat space should provide a realization of the Poincare algebra 
\besubeqs
\begin{align}
[P^A,P^B]&=0\,,\\
[J^{AB},P^C]&=P^A\eta^{BC}-P^B\eta^{AC}\,,\\
[J^{AB},J^{CD}]&=J^{AD}\eta^{BC}-J^{BD}\eta^{AC}-J^{AC}\eta^{BD}+J^{BC}\eta^{AD}\,,
\end{align}
\esubeqs
where generators of Lorentz transformations are $J^{AB}$ and generators of translations are $P^A$. We refer to \cite{Bengtsson:1983pg,Bengtsson:1983pd,Brink:2005wh,Metsaev:2005ar,Ponomarev:2016cwi,Ponomarev:2016lrm} for more detail. In free field theory the generators are known to be bilinear in the fields. Interactions, whether classical or quantum, append some of the generators with non-linear corrections. The generators that get deformed by the interactions are called dynamical, the rest of the generators are called kinematical. An important observation is that the number of dynamical generators depends on how we quantize the fields: those generators need to be deformed that do not preserve the Cauchy surface. The usual choice is $x^0$ for time and the quantization surface is taken to be a spacial slice. The least number of dynamical generators is achieved for the light-like surface, e.g. $x^+=0$. Then, $x^+$ is treated as time and $H=P^-$ is the Hamiltonian. The ten generators of the Poincare algebra split as
\begin{align}
\text{kinematical}&: && P^{+}, P^{a}, J^{a+}, J^{+-}, J^{ab} &&: 7\\
\text{dynamical}&: && P^{-}, J^{a-} &&:3
\end{align}
It is sufficient to construct the Poincare algebra at $x^+=0$ and then evolve all the generators according to $\dot{G}=i[H,G]$. Therefore, the equations to be solved are
\begin{align}
[J^{a-},J^{c-}]&=0\,, &
[J^{a-},P^{-}]&=0\,.\label{hardequations}
\end{align}
As a historical note, it is these equations 
from which the critical dimension and the intercept of
string theory where first obtained \cite{Goddard:1973qh}. It is convenient to work with partial Fourier transforms
\begin{align}
    \Phi(x,x^+)&=(2\pi)^{-\tfrac{d-1}2} \int e^{+i(x^-p^++x\cdot p)} \Phi(p,x^+)\, d^{d-1}p\,,\\
    \Phi(p,x^+)&=(2\pi)^{-\tfrac{d-1}2}  \int e^{-i(x^-p^++p\cdot x)} \Phi(x,x^+)\, d^{d-1}x\,.
\end{align}
In four dimensions a massless spin-$s$ particle leads to two helicity $(\pm s)$ states $\Phi^{\pm s}(p,x^+)$. The classical Poisson brackets are 
\begin{align}\label{equaltime}
    [\Phi^{\mu}(p,x^+),\Phi^{\lambda}(q,x^+)]&=\delta^{\mu,-\lambda}\frac{\delta^{3}(p+q)}{2p^+}\,.
\end{align}
Here $\mu,\lambda,...$ are helicity labels, $\mu=\pm s$ and $s=0,1,2,...$. The kinematical generators
that will not be affected by interactions are\footnote{Note that $\beta$ is used instead of $p^+$ 
in order to simplify notation. The spacial momenta are complexified to $p$ and $\pb$. Also, $x^+=0$ from now on.}
{\allowdisplaybreaks
\besubeqs\begin{align}
\hat{P}^+&=\beta\,, & \hat{P}&=p\,, & \hat{\PPb}&=\pb\,,\\
\hat{J}^{z+}&=-\beta\pfrac{\pb}\,, & \hat{J}^{\zb+}&=-\beta\pfrac{p}\,, & \hat{J}^{-+}&=-N_\beta-1=-\frac{\pl}{\pl \beta} \beta\,,\\
\hat{J}^{z\zb}&= N_p-N_\pb -\lambda\,,
\end{align}\esubeqs}%
\noindent where $N_p=p\pl_p$ is the Euler operator. The dynamical generators at the free level are
\begin{align}
    H_2&=-\frac{p\pb}{\beta}\,, &&
    \begin{aligned}
        \hat{J}^{z-}_2&= \pfrac{\pb} \frac{ p\pb}{\beta} +p \pfrac{\beta} +\lambda\frac{p}{\beta}\,,\\
         \hat{J}^{\zb-}_2&= \pfrac{p} \frac{ p\pb}{\beta} +\pb \pfrac{\beta} -\lambda\frac{\pb}{\beta}  \,.     
    \end{aligned}
\end{align}
The Poincare algebra is then realized by charges
\begin{align}
Q_\xi&= \int p^+\, d^{3}p\,\Phi^{-\mu}_{-p} O_\xi(p,\pl_p)\Phi^{\mu}_p\,,
\end{align}
that act via commutators
\begin{align}
    \delta_\xi \Phi^\mu(p,x^+)&= [\Phi^\mu(p,x^+),Q_\xi]\,.
\end{align}
At the interaction level one assumes the following expansion for the dynamical generators
{\allowdisplaybreaks\besubeqs\begin{align}
    H&=H_2+\sum_n\int d^{3n}q\,\delta\left(\sum q_i\right) h_{\lambda_1...\lambda_n}^{q_1,...,q_n}\, \Phi^{\lambda_1}_{q_1}...\Phi^{\lambda_n}_{q_n}\,,\\
    J^{z-}&=J^{z-}_2+\sum_n\int d^{3n}q\, \deltas{\sum q_i}\left[ j_{\lambda_1...\lambda_n}^{q_1,...,q_n}-\frac{1}{n}h_{\lambda_1...\lambda_n}^{q_1,...,q_n}\left(\sum_k \pfrac{\bar{q}_k}\right)\right]\, \Phi^{\lambda_1}_{q_1}...\Phi^{\lambda_n}_{q_n}   \,,\\ 
    J^{\zb-}&=J^{\zb-}_2+\sum_n\int d^{3n}q\, \deltas{\sum q_i}\left[ \jb_{\lambda_1...\lambda_n}^{q_1,...,q_n}-\frac{1}{n} h_{\lambda_1...\lambda_n}^{q_1,...,q_n}\left(\sum_k \pfrac{q_k}\right)\right]\, \Phi^{\lambda_1}_{q_1}...\Phi^{\lambda_n}_{q_n}   \,,  
\end{align}\esubeqs}\noindent
The Poincare algebra is maintained up to the cubic order \cite{Metsaev:1991nb,Metsaev:1991mt} provided that
\besubeqs\label{famouscubic}\begin{empheq}{align}
    h_{\lambda_1,\lambda_2,\lambda_3}&= C^{\lambda_1,\lambda_2,\lambda_3} \frac{\PPb^{\lambda_1+\lambda_2+\lambda_3}}{\beta_1^{\lambda_1}\beta_2^{\lambda_2}\beta_3^{\lambda_3}}+\bar{C}^{-\lambda_1,-\lambda_2,-\lambda_3} \frac{\PP^{-\lambda_1-\lambda_2-\lambda_3}}{\beta_1^{-\lambda_1}\beta_2^{-\lambda_2}\beta_3^{-\lambda_3}}\,,\\
    j_{\lambda_1,\lambda_2,\lambda_3}&=+\frac23 C^{+\lambda_1,+\lambda_2,+\lambda_3} \frac{\PPb^{+\lambda_1+\lambda_2+\lambda_3-1}}{\beta_1^{+\lambda_1}\beta_2^{+\lambda_2}\beta_3^{+\lambda_3}}\chi^{\lambda_1,\lambda_2,\lambda_3}\,,\\
    \jb_{\lambda_1,\lambda_2,\lambda_3}&=-\frac23\bar{C}^{-\lambda_1,-\lambda_2,-\lambda_3} \frac{\PP^{-\lambda_1-\lambda_2-\lambda_3-1}}{\beta_1^{-\lambda_1}\beta_2^{-\lambda_2}\beta_3^{-\lambda_3}}\chi^{\lambda_1,\lambda_2,\lambda_3}\,,
\end{empheq}\esubeqs
where
\begin{align}
    \chi=\beta_1(\lambda_2-\lambda_3)+\beta_2(\lambda_3-\lambda_1)+\beta_3(\lambda_1-\lambda_2)\,.
\end{align}
Here $C^{\lambda_1,\lambda_2,\lambda_3}$ and $\bar{C}^{-\lambda_1,-\lambda_2,-\lambda_3}$ are a priori independent coupling constants that, as usual, are not fixed by the cubic analysis. 

Chiral Theory results from the nontrivial fact that the following Hamiltonian makes the Poincare algebra valid to all orders \cite{Ponomarev:2016lrm}:
\begin{align}
H&=\int \Phi^{-\lambda}_{-p} \,({p\pb})\,\Phi^\lambda_p + \int \frac{(l_p)^{\lambda_1+\lambda_2+\lambda_3-1}}{\Gamma(\lambda_1+\lambda_2+\lambda_3)} \frac{\PPb^{\lambda_1+\lambda_2+\lambda_3}}{\beta_1^{\lambda_1}\beta_2^{\lambda_2}\beta_3^{\lambda_3}}\Phi^{\lambda_1}_{p_1}\Phi^{\lambda_2}_{p_2}\Phi^{\lambda_3}_{p_3}\delta^{3}(p_1+p_2+p_3)\,.
\end{align}
The essential part here is that the Poincare algebra at the quartic order is violated by three types of terms \cite{Metsaev:1991nb,Metsaev:1991mt}: $CC$, $\bar{C}\bar{C}$ and $C\bar{C}$. The $CC$-terms can be made zero by fine-tuning the coupling constants to be the $\Gamma$-function \cite{Metsaev:1991nb,Metsaev:1991mt} and a detailed analysis can be found in \cite{Ponomarev:2016lrm}. All the other terms vanish for Chiral Theory where $\bar{C}=0$. The action (\ref{eq:chiralaction}) given 
in the main text we give the action corresponds to the Hamiltonian above.

%%%%%%%%%%%%%%%%%%%%%%%%%%%%%%%%%%%%%%%%%%%%%%%%%%%%%%%%%%
\section{Kinematics}
\label{app:kinematics}
\setcounter{equation}{0}
%%%%%%%%%%%%%%%%%%%%%%%%%%%%%%%%%%%%%%%%%%%%%%%%%%%%%%%%%%
We collect below some identities that are used for the calculations in the main text. Suppose we are interested in some quantities in four dimensions,
e.g. Hamiltonian or off-shell amplitudes, that can depend on $N$ external momenta. The Poincare algebra implies that transverse momenta should appear only in the following two combinations
\begin{align}
   \PP_{km}&=p_k\beta_m-p_m\beta_k\,, & \PPb_{km}&=\pb_k\beta_m-\pb_m\beta_k \,.
\end{align}
Also, it can be shown that only $N-2$ out of $N(N-1)/2$
combinations $\PP_{ij}$ are independent and likewise for $\PPb$. In particular, for the three-point case there is just one independent transverse momenta (and its conjugate).
\begin{align}
    \PP^a_{12}&=...=\PP^a=\frac13\left[ (\beta_1-\beta_2)p_3+(\beta_2-\beta_3)p_1+(\beta_3-\beta_1)p_2\right]\,.
\end{align}
All $\PP_{ij}$ are anti-symmetric under permutations:
\begin{align}
    \sigma_{123}\PP&=\PP\,, \qquad\qquad \sigma_{12}\PP=\sigma_{23}\PP=\sigma_{13}\PP=-\PP\,,
\end{align}
where the conservation of the total momenta has been used. Also, for three points we have
 \begin{align}
      -\sum_i \frac{p_i \pb_i}{\beta_i}&=\frac{\PP \PPb}{\beta_1\beta_2\beta_3}= \frac{\PP \cdot \PP}{2\beta_1\beta_2\beta_3}\,.
\end{align}
We have a number of useful identities, such as 
%(we use $d$-dimensional notation sometimes, $a=z,\bar{z}$ in $4d$). 
the Bianchi-like identities:
\begin{align}\label{eq:Bianchilike}
    \sum_i \PP^a_i&=0\,,&
    \beta_{[i} \PP^a_{jk]}&\equiv0\,,&
    \PP^a_{i[j}\PP^a_{kl]}&\equiv0\,.
\end{align}
Other kinematic identities include
\begin{align} \label{C1}
    \sum_j \frac{\PP_{ij} \PPb_{jk}}{\beta_j}&= -\frac12\beta_i\beta_k \sum_j \frac{\pvec_j^2}{\beta_j}\,,\\
    \sum_j \frac{\PP_{ij} \PPb_{jk}}{\beta_j}&= -\beta_i\beta_k \sum_j {p_j \pb_j}\,,\\ \label{C3}
    \PP_{ij}\PPb_{ij} &=-\frac12\beta_i\beta_j (\pvec_i+\pvec_j)^2\qquad \text{for}\quad\pvec_i^2,\pvec_j^2=0 \
\end{align}
and one of the most important for dealing with one off-shell leg is ($s_{ik}=(\pvec_i+\pvec_k)^2$):
\begin{equation}\label{eq:magicidentity}
    \PPb_{ik}\PP_{ik}=-\frac{\beta_i\beta_k}{2}s_{ik}+\frac{1}{2}\beta_i(\beta_k+\beta_i)\pvec_k^2, \qquad \pvec_i^2=0,\quad \pvec_k^2\neq 0
\end{equation}

%%%%%%%%%%%%%%%%%%%%%%%%%%%%%%%%%%%%%%%%%%%%%%%%%%%%%%%%%%
\section{Yang-Mills/Chan-Paton Gauging}
\label{app:gauging}
%%%%%%%%%%%%%%%%%%%%%%%%%%%%%%%%%%%%%%%%%%%%%%%%%%%%%%%%%%
As is explained in Appendix \ref{app:LC}, see also \cite{Metsaev:1991nb,Metsaev:1991mt,Ponomarev:2016lrm}, the main equation to be solved within the light-cone approach reads
\begin{equation}\label{eq:Dconstraint}
    [H_3(\PPb),J_3^{a-}]=0\,.
\end{equation}
Assuming that the fields take values in some matrix algebra with generators $T$ that may depend on helicity $\lambda$
\begin{align}
    \Phi^{\lambda}(\pvec)\equiv \Phi_a^{\lambda}(\pvec)T^{a, \lambda} \equiv (\Phi^{\lambda}_{\pvec})^A_{\ B}\,,
\end{align}
we would like to see what are the restrictions on $T$ from \eqref{eq:Dconstraint}. The explicit form of the cubic Hamiltonian $H_3$ is
\small
\begin{equation}
    H_3=\sum_{\lambda_i}\int \D p\,  \delta^3\left(\sum_{i}p_i\right)h_3^{\lambda_i}(p_i)\mathrm{Tr}\left[\Phi^{\lambda_1}_{p_1}\Phi^{\lambda_2}_{p_2}\Phi^{\lambda_3}_{p_3}\right], \quad h_3^{\lambda_i}=C^{\lambda_1,\lambda_2,\lambda_3}\frac{\PPb^{\lambda_1+\lambda_2+\lambda_3}}{\beta_1^{\lambda_1}\beta_2^{\lambda_2}\beta_3^{\lambda_3}}\,,
\end{equation}
\normalsize
where the measure is $\D p=\prod_{i=1}^3d^3p_i$. Similarly, the dynamical boost generator $J_3$ 
\small
\begin{equation}
    J_3=\sum_{\lambda_i}\int\prod_{i=1}^3 d^3p_i\delta^3\left(\sum_{i}p_i\right)
    \Bigg[j_3^{\lambda_i}(p_i)-\frac{h_3^{\lambda_i}(p_i)}{3}\Big(\sum_k\frac{\partial}{\partial p_k}\Big)\Bigg]\mathrm{Tr}\prod_{i=1}^3\Phi^{\lambda_i}_{p_i}\,,
\end{equation}
\normalsize
where
\small
\begin{equation}
    j_3^{\lambda_i}=\frac{2}{3} C^{\lambda_1,\lambda_2,\lambda_3} \frac{\PPb^{\lambda_1+\lambda_2+\lambda_3-1}}{\beta_1^{\lambda_1}\beta_2^{\lambda_2}\beta_3^{\lambda_3}}\chi^{\lambda_1,\lambda_2,\lambda_3}\quad \text{and} \quad \chi=(\lambda_1-\lambda_2)\beta_3+(\lambda_2-\lambda_3)\beta_1+(\lambda_3-\lambda_1)\beta_2\,.
\end{equation}
\normalsize
Then, the constraint (\ref{eq:Dconstraint}) gives
\small
\begin{equation}\label{eq:constraintdensity}
    \begin{split}
        [H_3,J_3]&=\sum_{\lambda_i,\mu_j}\int \D p\, \D q\, \delta^3\left(\sum_j q_j\right)\Bigg[j_3^{\mu_j}(q_j)-\frac{h_3^{\mu_j}(q_j)}{3}\Big(\sum_k \frac{\partial}{\partial q_k}\Big)\Bigg]\\
        &\times \delta^3\left(\sum_i p_i\right)h_3^{\lambda_i}(p_i)\Big[\prod_{i=1}^3\Phi^{\lambda_i}_{p_i}, \prod_{j=1}^3\Phi^{\mu_j}_{q_j}\Big].
    \end{split}
\end{equation}
\normalsize
Since both $h_3$ and $j_3$ are cyclic invariant, the  fields can be put back to the same color order. Hence, the Poisson bracket in (\ref{eq:constraintdensity}) can be written as
\begin{equation}\label{eq:fieldsPoisson}
    \Big[\prod_{i=1}^3\Phi^{\lambda_i}_{p_i}, \prod_{j=1}^3\Phi^{\mu_j}_{q_j}\Big]=\prod_{i,j=1}^2\Phi^{\lambda_i}_{p_i}\Phi^{\mu_j}_{q_j}\Big[\Phi^{\lambda_3}_{p_3},\Phi^{\mu_3}_{q_3}\Big]\,,
\end{equation}
where we choose to contract fields $\Phi^{\lambda_3}_{p_3}$ and $\Phi^{\mu_3}_{q_3}$ in $H_3$ and $J_3$. Now we are ready to analyze the equation (\ref{eq:Dconstraint}) for the case of various gauge groups.  

\paragraph{$\boldsymbol{U(N)}$-gauging.} We first look at the case where fields are $u(N)$-valued
\begin{align}
    \Phi^{\lambda}(\pvec)\equiv \Phi_a^{\lambda}(\pvec)T^a \equiv (\Phi^{\lambda}_{\pvec})^A_{\ B}\,,
\end{align}
so that the trace in (\ref{eq:generalvertex}) is over $u(N)$ indices.  The Poisson bracket can be defined as
\begin{equation}\label{eq:PoissonU(N)}
    [(\Phi^{\lambda}_{p})^A_{\ B},(\Phi^{\mu}_{q})^C_{\ D}]=\frac{\delta^{\lambda,-\mu}\delta^3(p+q)}{2q^+}\times [\theta_{\lambda}\delta^C_{ \ B}\delta^A_{\ D}]\,,
\end{equation}
where $\theta_{\lambda}$ is some phase factor to be determined later. \eqref{eq:Dconstraint} leads to 
\small
\begin{equation}\label{eq:dynamicU(N)}
\begin{split}
   0= &\sum_{\omega}\text{Sym}\,(-)^{\omega}\theta_{\omega}\tr(\Phi_1\Phi_2\Phi_3\Phi_4)\\
    &\times\Big[\frac{(\lambda_1+\omega-\lambda_2)\beta_1-(\lambda_2+\omega-\lambda_1)\beta_2}{\beta_1+\beta_2}C^{\lambda_1,\lambda_2,\omega}C^{\lambda_3,\lambda_4,-\omega}\PPb_{12}^{\lambda_1+\lambda_2+\omega-1}\PPb_{34}^{\lambda_3+\lambda_4-\omega}\Big]\,,
    \end{split}
\end{equation}
\normalsize
where $\Phi_i\equiv \Phi^{\lambda_i}(p_i)$. 
Next, we let $\theta_{\omega}=e^{ix\omega}$ to be an arbitrary phase factor and determine the value of $x$ so that the coupling constant (\ref{eq:magicalcoupling}) is a solution of (\ref{eq:dynamicU(N)}). Note that the symmetrized sum in (\ref{eq:dynamicU(N)}) appears from the contraction between fields \cite{Ponomarev:2016lrm} that preserve all possible color-orderings. If we denote $\tr(\Phi_i\Phi_j\Phi_k\Phi_l)E(i,j,k,l)$ as $[i,j,k,l]$,  where $E$ is the kinematic part of \eqref{eq:dynamicU(N)}, then we have in total six \textit{partial color-ordered contributions} (or $partial$-$contributions$ for short) appearing in (\ref{eq:Dconstraint}). In terms of $[i,j,k,l]$
these contributions are:
\small
\begin{equation}\label{eq:colormodulo}
    0=[1,2,3,4]+[1,3,4,2]+[1,4,2,3]+[1,3,2,4]+[1,2,4,3]+[1,4,3,2]\,.
\end{equation}
\normalsize
In order to satisfy (\ref{eq:dynamicU(N)})
each of the terms in (\ref{eq:colormodulo}) has to vanish separately 
 since it is impossible for different partial contributions to
 cancel each others. Let us  take $[1,2,3,4]$ as an example. It is a combination of the following permutations that preserve the color-ordering of $\tr(\Phi_1\Phi_2\Phi_3\Phi_4)$ 
\begin{equation}
    [1,2,3,4]=\{1,2,3,4\}+\{2,3,4,1\}+\{3,4,1,2\}+\{4,1,2,3\}\,,
\end{equation}
where the curly brackets $\{i,j,k,l\}$ notation is for permutations with $i,j,k,l$ are indices of left-over external sources. First of all, when we consider the permutation $\{1,2,3,4\}\rightarrow \{3,4,1,2\}$ with $\omega\rightarrow-\omega$, the two terms combine as
\begin{equation}\label{eq:dynamic1234}
\begin{split}
    &\sum_{\omega}e^{ix\omega}\frac{\PPb_{12}^{\lambda_1+\lambda_2+\omega-1}}{\Gamma(\lambda_1+\lambda_2+\omega)}\frac{\PPb_{34}^{\lambda_3+\lambda_4-\omega-1}}{\Gamma(\lambda_3+\lambda_4-\omega)}\tr(\Phi_1\Phi_2\Phi_3\Phi_4) \\
    &\times\Bigg[e^{ix\omega}(\lambda_1-\lambda_2)\PPb_{34}+e^{-ix\omega}(\lambda_3-\lambda_4)\PPb_{12}+\omega\Big(e^{ix\omega}\frac{\beta_1-\beta_2}{\beta_1+\beta_2}\PPb_{34}+e^{-ix\omega}\frac{\beta_3-\beta_4}{\beta_1+\beta_2}\PPb_{12}\Big)\Bigg]\,.
    \end{split}
\end{equation}
Secondly, for the combination of $\{2,3,4,1\}\xrightarrow{\omega\rightarrow-\omega}\{4,1,2,3\}$, we get
\begin{equation}\label{eq:dynamic2341}
\begin{split}
     &\sum_{\omega}e^{ix\omega}\frac{\PPb_{23}^{\lambda_2+\lambda_3+\omega-1}}{\Gamma(\lambda_2+\lambda_3+\omega)}\frac{\PPb_{41}^{\lambda_4+\lambda_1-\omega-1}}{\Gamma(\lambda_4+\lambda_1-\omega)}\tr(\Phi_2\Phi_3\Phi_4\Phi_1)\\
     &\times\Bigg[e^{ix\omega}(\lambda_2-\lambda_3)\PPb_{41}+e^{-ix\omega}(\lambda_4-\lambda_1)\PPb_{23}+\omega\Big(e^{ix\omega}\frac{\beta_2-\beta_3}{\beta_2+\beta_3}\PPb_{41}+e^{-ix\omega}\frac{\beta_4-\beta_1}{\beta_2+\beta_3}\PPb_{23}\Big)\Bigg]\,.
     \end{split}
\end{equation}
Now, as we noted, $[1,2,3,4]$ should vanish by itself. This is only possible if $x=\pi$ or $\theta_{\omega}=(-)^{\omega}$. In this case, the expressions given above get simplified and one finally obtains
\small
\begin{equation}\label{eq:DU(N)1234constraint}
\begin{split}
    [1,2,3,4]=&\tr(\Phi_1\Phi_2\Phi_3\Phi_4)(\PPb_{12}-\PPb_{23}+\PPb_{34}-\PPb_{41})\times \frac{(\PPb_{12}+\PPb_{34})^{\Lambda_4-3}}{\Gamma(\Lambda_4-1)}\\
    &\times\Big[\lambda_1(\PPb_{23}+\PPb_{34})-\lambda_2(\PPb_{34}+\PPb_{41})+\lambda_3(\PPb_{41}+\PPb_{12})-\lambda_4(\PPb_{12}+\PPb_{23})\Big] =0\,.
    \end{split}
\end{equation}
\normalsize
In order to obtain the above result we used  momentum conservation and the identity $\PPb_{12}+\PPb_{34}=\PPb_{23}+\PPb_{41}$. Without having a common factor $(\PPb_{12}+\PPb_{34})^{\Lambda_4-4}=(\PPb_{23}+\PPb_{41})^{\Lambda_4-4}$, one cannot make another choice for $\theta_{\omega}$ to have (\ref{eq:magicalcoupling}) as the solution of $[1,2,3,4]=0$. For other partial contributions in (\ref{eq:colormodulo}), one can also see that they  vanish if $\theta_{\omega}=(-)^{\omega}$. Hence, $\theta_{\omega}=(-)^{\omega}$ is the unique solution of (\ref{eq:Dconstraint}) for $U(N)$ Chiral HiSGRA that has (\ref{eq:magicalcoupling}) as the coupling constants.

\paragraph{$\boldsymbol{SO(N) \ \text{and} \ USp(N)}$ gauging.}
In the case where fields have $SO(N)/USp(N)$ color indices, the trace is understood as
\begin{equation}
    \tr(\Phi_{\pvec_1}^{\lambda_1}\, ...\, \Phi^{\lambda_n}_{\pvec_n})=\Phi^1_{AB_1}\Phi^2_{B_1B_2}...\Phi^n_{B_nA}\,, \quad \Phi^i\equiv \Phi^{\lambda_i}_{\pvec_i}\,.
\end{equation} 
For the $SO(N)$-case the invariant tensor is $\delta_{AB}$ and the most general Poisson brackets read
\begin{equation}\label{eq:PoissonO(N)}
    [(\Phi^{\lambda}_{p})_{AB},(\Phi^{\mu}_{q})_{CD}]=\frac{\delta^{\lambda,-\mu}\delta^3(p+q)}{2q^+}\times [\delta_{AC}\delta_{BD}+\theta_{\lambda}\delta_{AD}\delta_{BC}]\,.
\end{equation}
Here, $\theta_{\lambda}$ is a phase factor that enters the Poisson brackets. 
The constraint (\ref{eq:Dconstraint})  reads
\begin{equation}\label{eq:O(N)Dynamic}
    \begin{split}
        0=&\sum_{\omega}\text{Sym}(-)^{\omega}\Big[\theta_{\lambda_3}\theta_{\lambda_4}\tr(\Phi_1\Phi_2\Phi_4\Phi_3)+\theta_{\omega}\tr(\Phi_1\Phi_2\Phi_3\Phi_4)\Big]\\
        &\times \Big[\frac{(\lambda_1+\omega-\lambda_2)\beta_1-(\lambda_2+\omega-\lambda_1)\beta_2}{\beta_1+\beta_2}C^{\lambda_1,\lambda_2,\omega}C^{\lambda_3,\lambda_4,-\omega}\PPb_{12}^{\lambda_1+\lambda_2+\omega-1}\PPb_{34}^{\lambda_3+\lambda_4-\omega}\Big]\,.
    \end{split}
\end{equation}
Now, we shall repeat the same procedure as for the
$U(N)$-case in order to determine the values of the phase factor $\theta_{\lambda_i}=e^{ix\lambda_i}$. However, unlike the $U(N)$-case, the $SO(N)$-case contains an extra trace that comes from the \textit{M\"obius twist} in the Poisson brackets (\ref{eq:PoissonO(N)}). As a consequence, there will be mixing between different $[i,j,k,l]$ partial contributions. First, let us look at $\{1,2,3,4\}\xrightarrow{\omega\rightarrow -\omega}\{3,4,1,2\}$ in $[1,2,3,4]$
\small
\begin{equation}\label{eq:O(N)1234}
\begin{split}
    &\sum_{\omega}\frac{e^{i\pi\omega}\,\PPb_{12}^{\lambda_1+\lambda_2+\omega-1}\PPb_{34}^{\lambda_3+\lambda_4-\omega-1}}{\Gamma(\lambda_1+\lambda_2+\omega)\Gamma(\lambda_3+\lambda_4-\omega)}\\
    \times&\Bigg[\tr(1234)\Big[e^{ix\omega}(\lambda_1-\lambda_2)\PPb_{34}+e^{-ix\omega}(\lambda_3-\lambda_4)\PPb_{12}+\omega\frac{e^{ix\omega}(\beta_1-\beta_2)\PPb_{34}+e^{-ix\omega}(\beta_3-\beta_4)\PPb_{12}}{\beta_1+\beta_2}\Big]\\
    &+\tr(1243)e^{ix(\lambda_3+\lambda_4)}\Big[(\lambda_1-\lambda_2)\PPb_{34}+(\lambda_3-\lambda_4)\PPb_{12}+\omega\frac{(\beta_1-\beta_2)\PPb_{34}+(\beta_3-\beta_4)\PPb_{12}}{\beta_1+\beta_2}\Big]\Bigg]\,,
    \end{split}
\end{equation}
\normalsize
where we denote $\tr(ijkl)\equiv\tr(\Phi_i\Phi_j\Phi_l\Phi_k)$ for simplicity. Similarly, the permutation $\{2,3,4,1\}\xrightarrow{\omega\rightarrow -\omega}\{4,1,2,3\}$ in $[1,2,3,4]$ reads
 \small
\begin{equation}\label{eq:O(N)2341}
\begin{split}
    &\sum_{\omega}\frac{e^{i\pi\omega}\,\PPb_{23}^{\lambda_2+\lambda_3+\omega-1}\PPb_{41}^{\lambda_4+\lambda_1-\omega-1}}{\Gamma(\lambda_2+\lambda_3+\omega)\Gamma(\lambda_4+\lambda_1-\omega)}\\
    \times&\Bigg[\tr(2341)\Big[e^{ix\omega}(\lambda_2-\lambda_3)\PPb_{41}+e^{-ix\omega}(\lambda_4-\lambda_1)\PPb_{12}+\omega\frac{e^{ix\omega}(\beta_2-\beta_3)\PPb_{41}+e^{-ix\omega}(\beta_4-\beta_1)\PPb_{23}}{\beta_2+\beta_3}\Big]\\
    &+\tr(2314)e^{ix(\lambda_1+\lambda_4)}\Big[(\lambda_2-\lambda_3)\PPb_{41}+(\lambda_4-\lambda_1)\PPb_{23}+\omega\frac{(\beta_2-\beta_3)\PPb_{41}+(\beta_4-\beta_1)\PPb_{23}}{\beta_2+\beta_3}\Big]\Bigg]
    \end{split}
\end{equation}
\normalsize
One can notice that there are additional contributions 
(compared to the $U(N)$ case) in the equation (\ref{eq:O(N)2341})
that combine two traces inside $[1,2,3,4]$: namely $\tr(2314)$ and an "exotic" one
$\tr(1243)$. Hence, $[1,2,3,4]$ cannot vanish by itself and we need to include some contributions from other $[i,j,k,l]$ in order to satisfy (\ref{eq:O(N)Dynamic}). Consider the permutation $\{1,3,2,4\}\xrightarrow{\omega\rightarrow -\omega}\{2,4,1,3\}$ in $[1,3,2,4]$
 \small
\begin{equation}\label{eq:O(N)1324}
\begin{split}
    &\sum_{\omega}\frac{e^{i\pi\omega}\,\PPb_{13}^{\lambda_1+\lambda_3+\omega-1}\PPb_{24}^{\lambda_2+\lambda_4-\omega-1}}{\Gamma(\lambda_1+\lambda_3+\omega)\Gamma(\lambda_2+\lambda_4-\omega)}\\
    \times&\Bigg[\tr(1324)\Big[e^{ix\omega}(\lambda_1-\lambda_3)\PPb_{24}+e^{-ix\omega}(\lambda_2-\lambda_4)\PPb_{13}+\omega\frac{e^{ix\omega}(\beta_1-\beta_3)\PPb_{24}+e^{-ix\omega}(\beta_2-\beta_4)\PPb_{13}}{\beta_1+\beta_3}\Big]\\
    &+\tr(3124)e^{ix(\lambda_1+\lambda_3)}\Big[(\lambda_1-\lambda_3)\PPb_{24}+(\lambda_2-\lambda_4)\PPb_{13}+\omega\frac{(\beta_1-\beta_3)\PPb_{24}+(\beta_2-\beta_4)\PPb_{13}}{\beta_1+\beta_3}\Big]\Bigg]
    \end{split}
\end{equation}
\normalsize
Then, we have in total six different color-ordered terms. Considering the combination of permutations $\{1,2,3,4\}\xrightarrow{\omega\rightarrow -\omega}\{3,4,1,2\}$ and $\{2,3,4,1\}\xrightarrow{\omega\rightarrow -\omega}\{4,1,2,3\}$ 
one can see, that
we need to set $x=\pi$ or $\theta_{\lambda_i}=(-)^{\lambda_i}$
to get (\ref{eq:DU(N)1234constraint}) for $\tr(1234)$ color ordering. Next, the contributions coming from $\tr(1243)(-)^{\omega}\theta_{\lambda_4}\theta_{\lambda_3}$ and $\tr(1324)$ also cancel each others with this choice of the phase factors in (\ref{eq:PoissonO(N)}). Similar argument applies for $\tr(2314)(-)^{\omega}\theta_{\lambda_4}\theta_{\lambda_1}$ and $\tr(3124)(-)^{\omega}\theta_{\lambda_3}\theta_{\lambda_1}$. Hence, even though $[i,j,k,l]$ can not vanish by themselves in the case of $SO(N)$-gauging, the total contribution does vanish by combining all the partial contributions together. This indicates that $\theta_{\omega}=(-)^{\omega}$ is the right choice for the phase factors in Poisson bracket (\ref{eq:PoissonO(N)}).

Finally, in the case of $USp(N)$-gauging, the Poisson bracket reads
\begin{equation}\label{eq:PoissonUsp(N)}
    [(\Phi^{\lambda}_{p})_{AB},(\Phi^{\mu}_{q})_{CD}]=\frac{\delta^{\lambda,-\mu}\delta^3(p+q)}{2q^+}\times [C_{AC}C_{BD}+\theta_{\lambda}C_{AD}C_{BC}]\,.
\end{equation}
where $C_{AB}$ is the anti-symmetric invariant tensor:
\begin{equation}
    C_{AB}=-C_{BA}\,, \qquad\qquad C_{AB}C^{CB}=\delta^C_A\,.
\end{equation}
The $C$-matrices are used to raise and lower indices as $V^A=C^{AB}V_B$, $V^B C_{BA}=V_A$. Finally, the trace for the $USp(N)$-case can be understood as
\begin{equation}
    \tr(\Phi \Phi...)=\Phi\fdu{A}{B}\Phi\fdu{B}{C}...
\end{equation}
The commutator (\ref{eq:Dconstraint}) reads
\begin{equation}\label{eq:USp(N)constraint}
\begin{split}
    0&=\sum_{\omega}\text{Sym}(-)^{\omega+1}\Big[\theta_{\lambda_3}\theta_{\lambda_4}\tr(\Phi_1\Phi_2\Phi_4\Phi_3)+\theta_{\omega}\tr(\Phi_1\Phi_2\Phi_3\Phi_4)\Big]\\
        &\times \Big[\frac{(\lambda_1+\omega-\lambda_2)\beta_1-(\lambda_2+\omega-\lambda_1)\beta_2}{\beta_1+\beta_2}C^{\lambda_1,\lambda_2,\omega}C^{\lambda_3,\lambda_4,-\omega}\PPb_{12}^{\lambda_1+\lambda_2+\omega-1}\PPb_{34}^{\lambda_3+\lambda_4-\omega}\Big]
    \end{split}
\end{equation}
Repeating the same procedure as in the $SO(N)$ case with the requirement that (\ref{eq:magicalcoupling}) is the solution of (\ref{eq:USp(N)constraint}), one obtains $\theta_{\omega}=(-)^{\omega+1}$. 

To summarize the $SO(N)/USp(N)$-valued fields have the following properties  under interchanging $SO(N)/USp(N)$ indices,
\begin{align}
    SO(N)&:\,(\Phi^{\lambda}_{\pvec})_{AB}=(-)^{\lambda}(\Phi^{\lambda}_{\pvec})_{BA}\,,\\ USp(N)&:(\Phi^{\lambda}_{\pvec})_{AB}=(-)^{\lambda+1}(\Phi^{\lambda}_{\pvec})_{BA}\,.
\end{align}
Here, fields with odd-spin in $SO(N)/USp(N)$ cases have odd/even parity, while fields with even-spin have even/odd parity. Fields with odd spins always take values in the adjoint representation.  

It is important to stress that, the constraint (\ref{eq:Dconstraint}) with the coupling constants (\ref{eq:magicalcoupling}) can only be satisfied with the above choices of $\theta_{\omega}$ for $U(N)$- and $SO(N)/USp(N)$-gauged Chiral HiSGRA. Interestingly enough the allowed gauge groups as well as the allowed representations coincide with the allowed Chan-Paton symmetry groups and the representations in open string theory \cite{Marcus:1982fr}.

%%%%%%%%%%%%%%%%%%%%%%%%%%%%%%%%%%%%%%%%%%%%%%%%%%%%%%%%%%
\section{Worldsheet-Friendly Regularization}
\label{app:Thornregulator}
\setcounter{equation}{0}
%%%%%%%%%%%%%%%%%%%%%%%%%%%%%%%%%%%%%%%%%%%%%%%%%%%%%%%%%%
Systematic calculations in the light-cone gauge have been performed in \cite{Chakrabarti:2005ny,Chakrabarti:2006mb} for the pure QCD case and we borrow most of this technology for the Chiral Theory case. In practice we face integrals of the following type:
\begin{align}\label{appendix-int}
    \int \frac{d^4 q}{(2\pi)^4} F(\beta, q^a) \frac{1}{\prod_i (\qvec-\kvec_i)^2}\,,
\end{align}
where the polynomial prefactor $F$ depends on external momenta (not shown here), and the loop momentum $\qvec$. Importantly, the $q^-$-component does not enter the vertex
and therefore is not present in the integral
(\ref{appendix-int}). Also, in practice $F$ is such that the
integration over angular variables in the $q^a$-plane vanishes. The regularization proposed in \cite{Chakrabarti:2005ny,Chakrabarti:2006mb} is to introduce the Gaussian cutoff in the transverse part of the loop momentum $q$, i.e. $q_\perp\equiv q^a$:
\begin{align}
    I=\int \frac{d^4 q}{(2\pi)^4} F(\beta, q_{\perp}) \frac{1}{\prod_i (\qvec-\kvec_i)^2} e^{-\xi q_\perp^2}
\end{align}
The integral can be performed by 
using the Schwinger trick with parameters $T_i$
as the first step and then doing the Gaussian integral over $q_\perp$. Integration over $q^-$ gives a delta function:
\begin{equation}
\begin{split}
    I= &\frac{2\pi^2}{2(2\pi)^4} \int d\beta \,F(\beta, \sum_i T_ik_i/(T+\xi)) \delta(\sum T_i \beta -\sum T_i \beta_i^+) \\
    &\times\exp{\Big[ 2\sum T_i(\beta-\beta_i^+) k_i^- -\sum T_i k_{i\perp}^2 +\frac{1}{\sum T_i+\xi} (\sum T_i k_{i\perp})^2\Big]}
    \end{split}
\end{equation}
If there are no IR divergences, we can safely solve for $\beta$. It is also convenient to change variables as $T_i=Tx_i$, $\sum x_i=1$, which gives Jacobian $T^{n-1}$. This way we get
\begin{align}
    I=  \int \frac{T^{n-1} dT\, \prod dx_i }{T^2\,(4\pi)^2}&\, F(\beta=\sum x_i \beta_i^+, q_{\perp}=\tfrac{T}{T+\xi}\sum x_ik_{i\perp})\, \delta(\sum x_i-1) \frac{\pi}{T+\xi}\\
    &\times\exp{\Big[ -T \sum_{i\leq j} x_i x_j (\kvec_i-\kvec_j)^2 -\frac{T \xi}{T+\xi} (\sum x_i k_{i\perp})^2\Big]}
\end{align}
In a lucky case when the integral is not divergent at all, we simply find
\small
\begin{eqnarray} 
    I&&= \frac{1}{(4\pi)^2}\int {T^{n-3}dT\, \prod dx_i } \times \\ \nonumber
    &&\times  F\left(\beta=\sum x_i \beta_i, q^a=\sum x_ik_i^a\right) \delta(\sum x_i-1) 
    \exp{\Big[ -\frac12 T \sum_{i,j} x_i x_j (\kvec_i-\kvec_j)^2 \Big]} 
\end{eqnarray}
\normalsize

Sometimes we use dual momenta. For example, consider the self-energy diagram: 
\begin{figure}[h]
    \centering
    \includegraphics[scale=0.3]{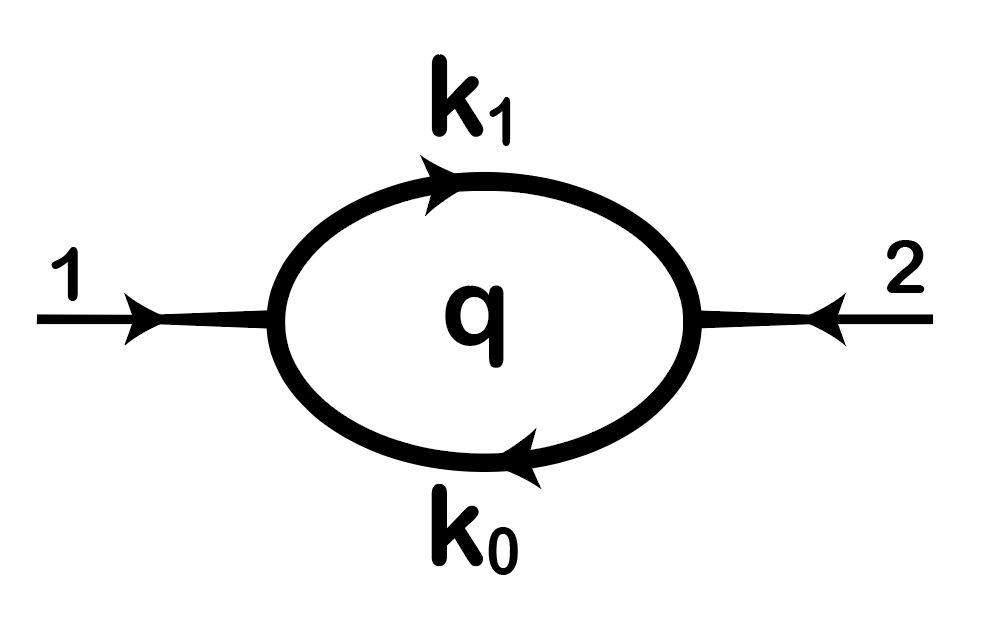}
\end{figure}\\
We choose the direction of the dual loop momentum $\kvec_i$ to be clockwise. A dual momentum is related to the original momentum as  follows. Take the first external leg and represent the corresponding 
four momentum $\pvec_1$ as
$\pvec_1=\kvec_1-\kvec_0$, then follow the same pattern for
the other external momenta by defining $\pvec_i=\kvec_i-\kvec_{i-1}$ at each of the vertices. The loop momentum
$\pvec$
is defined as the difference between $\qvec$ with its nearest dual regional momentum $\kvec_i$, where $\qvec$ is the dual momentum that is bounded by a loop. In our example, $\pvec=\qvec-\kvec_0$. 
We often use these rules of labeling dual momenta,
for  computations of the quantum correction at one-loop 
in the Section \ref{sec:leggedloop}.

%%%%%%%%%%%%%%%%%%%%%%%%%%%%%%%%%%%%%%%%%%%%%%%%%%%%%%%%%%
\section{Six-point Amplitude}
\label{app:sixpoint}
\setcounter{equation}{0}
%%%%%%%%%%%%%%%%%%%%%%%%%%%%%%%%%%%%%%%%%%%%%%%%%%%%%%%%%%
The combinatorics of Feynman graphs grows rapidly with the number of external legs. Nevertheless, the six-point tree-level amplitude can be dealt with directly, which provides an additional check of our recursive formula in the main text. We find the following topologies
\begin{figure}[h]
    \centering
    \includegraphics[scale=0.5]{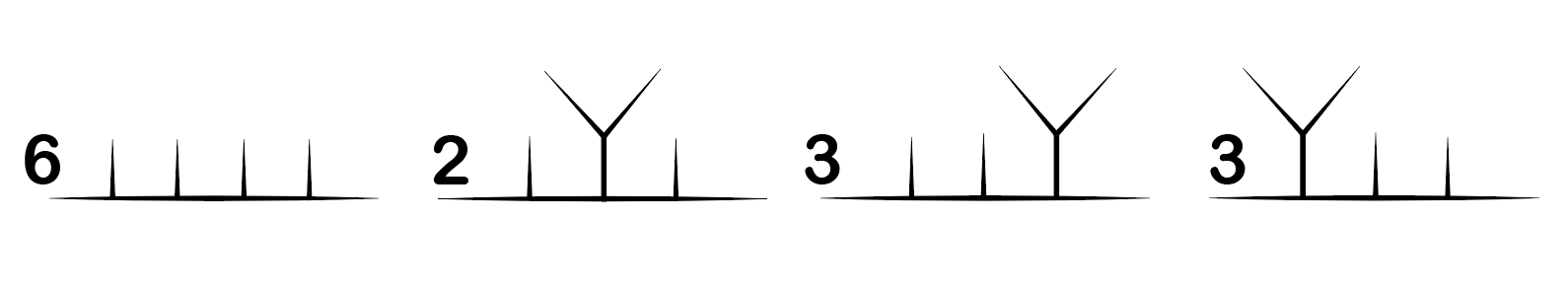}
\end{figure}
where the number in front of each topologies account for how many diagrams are there. These four topologies give:
\begin{align*}
     \hat{A}_1(123456)&=\frac{\PPb_{12}(\PPb_{13}+\PPb_{23})(\PPb_{14}+\PPb_{24}+\PPb_{34})\PPb_{56}}{\Gamma(\Lambda_6-3)\prod_{i=1}^6\beta_i^{\lambda_i}s_{12}s_{123}s_{56}}\alpha_6^{\Lambda_6-4}\\
     \hat{A}_2(123456)&=\frac{\PPb_{12}\PPb_{34}(\PPb_{61}+\PPb_{62}+\PPb_{51}+\PPb_{52})\PPb_{56}}{\Gamma(\Lambda_6-3)\prod_{i=1}^6\beta_i^{\lambda_i}s_{12}s_{34}s_{56}}\alpha_6^{\Lambda_6-4}\\
     \hat{A}_3(123456)&=\frac{\PPb_{12}(\PPb_{13}+\PPb_{23})(\PPb_{61}+\PPb_{62}+\PPb_{63})\PPb_{45}}{\Gamma(\Lambda_6-3)\prod_{i=1}^6\beta_i^{\lambda_i}s_{12}s_{123}s_{45}}\alpha_6^{\Lambda_6-4}\\
     \hat{A}_4(456123)&=\frac{\PPb_{23}(\PPb_{13}+\PPb_{12})(\PPb_{45}+\PPb_{46})\PPb_{56}}{\Gamma(\Lambda_6-3)\prod_{i=1}^6\beta_i^{\lambda_i}s_{23}s_{456}s_{56}}\alpha_6^{\Lambda_6-4}
\end{align*}
Let us omit $\alpha_6^{\Lambda_6-4}/\Gamma(\Lambda_6-3)\prod_{i=1}^6\beta_i^{\lambda_i}$ for a moment and focus on the prefactors. A short computation shows that
\begin{align*}
    A_1(123456)+A_4(456123)&=\frac{\beta_1\beta_2^2\beta_3\PPb_{56}(\PPb_{14}+\PPb_{24}+\PPb_{34})}{4s_{56}\PP_{12}\PP_{23}}=\frac{\beta_1\beta_2^2\beta_3\PPb_{56}(\PPb_{45}+\PPb_{46})}{4s_{56}\PP_{12}\PP_{23}}
\end{align*}
and similarly for other permutations. Together with the contribution from diagrams of the second topology
\begin{align*}
    \hat{A}_2(123456)&=\frac{\beta_1\beta_2\beta_3\beta_4(\PPb_{61}+\PPb_{62}+\PPb_{51}+\PPb_{52})\PPb_{56}}{4\PP_{12}\PP_{34}s_{56}}=\frac{\beta_1...\beta_4(\PPb_{13}+\PPb_{14}+\PPb_{23}+\PPb_{24})\PPb_{56}}{4\PP_{12}\PP_{34}s_{56}} \\
    \hat{A}_2(234561)&=\frac{\beta_2\beta_3\beta_4\beta_5(\PPb_{12}+\PPb_{13}+\PPb_{62}+\PPb_{63})\PPb_{61}}{4\PP_{23}\PP_{45}s_{61}}=\frac{\beta_2...\beta_5(\PPb_{24}+\PPb_{25}+\PPb_{34}+\PPb_{35})\PPb_{61}}{4\PP_{23}\PP_{45}s_{61}} 
\end{align*}
Grouping terms proportional to $\PPb_{56}/s_{56}$, one gets
\begin{align*}
    &\beta_1...\beta_4\frac{\PPb_{56}}{4s_{56}}\Big[\frac{\beta_2(\PPb_{45}+\PPb_{46})}{\beta_4\PP_{12}\PP_{23}}+\frac{\beta_3(\PPb_{51}+\PPb_{61})}{\beta_1\PP_{23}\PP_{34}}+\frac{\PPb_{61}+\PPb_{51}+\PPb_{62}+\PPb_{52}}{\PP_{12}\PP_{34}}\Big]\\
   =& \beta_1...\beta_4\frac{\PPb_{56}}{4s_{56}}\frac{\beta_2\beta_3}{\PP_{12}\PP_{23}\PP_{34}}\Big[-\frac{(\beta_5+\beta_6)\pvec_6^2}{2\beta_6}+\frac{\PPb_{56}\PP_{56}}{\beta_5\beta_6}\Big]=-\frac{\beta_1\beta_2^2\beta_3^2\beta_4\PPb_{56}}{8\PP_{12}\PP_{23}\PP_{34}}
\end{align*}
Similarly, for terms proportional to $\PPb_{61}/s_{61}$, we get
\begin{align*}
    &\beta_2...\beta_5\frac{\PPb_{61}}{4s_{61}}\Big[\frac{\beta_3(\PPb_{51}+\PPb_{56})}{\beta_5\PP_{23}\PP_{34}}+\frac{\beta_4(\PPb_{62}+\PPb_{12})}{\beta_2\PP_{34}\PP_{45}}+\frac{\PPb_{12}+\PPb_{13}+\PPb_{62}+\PPb_{63}}{\PP_{23}\PP_{45}}\Big]\\
    =&\beta_2...\beta_5\frac{\PPb_{61}}{4s_{61}}\frac{\beta_3\beta_4}{\PP_{23}\PP_{34}\PP_{45}}\Big[-\frac{(\beta_6+\beta_1)\pvec_6^2}{2\beta_6}+\frac{\PPb_{61}\PP_{61}}{\beta_6\beta_1}\Big]=-\frac{\beta_2\beta_3^2\beta_4^2\beta_5\PPb_{61}}{8\PP_{23}\PP_{34}\PP_{45}}
\end{align*}
The remaining terms combine into
\begin{equation*}
    -\frac{\beta_1...\beta_5}{8\PP_{12}\PP_{45}}\Big[\frac{\beta_4(\PPb_{61}+\PPb_{62})}{\PP_{34}}+\frac{\beta_2(\PPb_{46}+\PPb_{56})}{\PP_{23}}\Big]=\frac{\beta_1...\beta_5}{8\PP_{12}\PP_{45}}\frac{\beta_2\beta_4}{\PP_{23}\PP_{34}}\Big[\frac{\beta_6\beta_3\pvec_6^2}{2\beta_6}+\frac{\beta_3\PPb_{61}\PP_{12}}{\beta_1\beta_2}+\frac{\beta_3\PPb_{56}\PP_{45}}{\beta_4\beta_5}\Big]
\end{equation*}
Summing all of the above partial results together and we get an concise expression for $6$-point amplitude:
\begin{equation}
    A(123456)=\frac{\alpha_6^{\Lambda_6-4}}{16\Gamma(\Lambda_6-3)\prod_{i=1}^6\beta_i^{\lambda_i-1}}\frac{\beta_2\beta_3\beta_4\,\pvec_6^2 }{\beta_6\PP_{12}\PP_{23}\PP_{34}\PP_{45}}
\end{equation}

%%%%%%%%%%%%%%%%%%%%%%%%%%%%%%%%%%%%%%%%%%%%%%%%%%%%%%%%%%
\end{appendix}
%%%%%%%%%%%%%%%%%%%%%%%%%%%%%%%%%%%%%%%%%%%%%%%%%%%%%%%%%%

\footnotesize
\providecommand{\href}[2]{#2}\begingroup\raggedright\endgroup

\end{document}